\documentclass[aps,prx,twocolumn,superscriptaddress,amssymb,amsmath,floatfix,longbibliography,english,nofootinbib]{revtex4-2}

\usepackage[T1]{fontenc}
\usepackage{mathtools} %
\usepackage{braket}
\usepackage{natbib}
\usepackage{graphicx}%
\usepackage{hyperref}%
\usepackage{multirow}%
\usepackage{xcolor}%
\usepackage{newtxtext,newtxmath}%
\definecolor{prxblue}{HTML}{2E358F}
\definecolor{BLUE}{rgb}{0,0,1} %
\definecolor{RED}{rgb}{1,0,0} %
\definecolor{myora}{HTML}{F28C28}
\definecolor{supplmatcolour}{HTML}{C12903}
\usepackage{lastpage}
\usepackage{fancyhdr}
\usepackage{orcidlink}

\usepackage[capitalise,nameinlink]{cleveref}%

\makeatletter %
\@ifpackageloaded{gitinfo2}{%
    \newcommand*{\gitpdfsubject}{{Git Commit: \gitHash{}}}
}{%
    \newcommand*{\gitpdfsubject}{{Git Commit: NULL}}
}
\makeatother

\hypersetup{
    colorlinks = true,
    citecolor = prxblue,
    linkcolor = prxblue,
    urlcolor =  prxblue,
    pdfsubject = {\gitpdfsubject{}}, %
}
\begin{document}

\newcommand*\ketdownz{{\ket{\downarrow}_{Z_1}}}
\newcommand*\ketupz{{\ket{\uparrow}_{Z_1}}}
\newcommand*\ketdowny{{\ket{\downarrow}_{Y_1}}}
\newcommand*\ketupy{{\ket{\uparrow}_{Y_1}}}

\newcommand*\textManifoldG{{\textsuperscript{4}\textit{I}\textsubscript{15/2}}}
\newcommand*\textManifoldE{{\textsuperscript{4}\textit{I}\textsubscript{13/2}}}

\newcommand*\textCaWO{{CaWO\textsubscript{4}}}
\newcommand*\textEr{{Er\textsuperscript{3+}}}
\newcommand*\textErNuclear{{\textsuperscript{167}Er\textsuperscript{3+}}}
\newcommand*\textErCaWO{{\textEr{}:\textCaWO{}}}
\newcommand*\textYSO{{Y\textsubscript{2}SiO\textsubscript{5}}}
\newcommand*\textErYSO{{\textEr{}:\textYSO{}}}
\newcommand*\textErYO{{\textEr{}:Y\textsubscript{2}O\textsubscript{3}}}

\newcommand*\textLN{{LiNbO\textsubscript{3}}}

\newcommand*\textQuartz{{SiO\textsubscript{2}}}

\newcommand*\textTungstenNuclear{{\textsuperscript{183}W}}

\newcommand*\textAmmonia{{NH\textsubscript{4}OH}}

\newcommand*\textSupplMat{{\textcolor{supplmatcolour}{Suppl}}}

\newcommand*{\pavel}[1]{\textcolor{blue}{[Pavel: #1]}}
\newcommand*{\kjwo}[1]{\textcolor{myora}{[Bern: #1]}}
\newcommand*{\steven}[1]{\textcolor{cyan}{[Steven: #1]}}

\newcommand*\GammaSD{{\Gamma_{\mkern-2.5mu\text{SD}}}}
\newcommand*\SSD{{S_{\mkern-1mu\text{SD}}}}
\newcommand*\Gammaeff{{\Gamma_{\mkern-2.5mu\text{eff}}}}
\newcommand*\Gammainh{{\Gamma_{\mkern-2.5mu\text{inh}}}}
\newcommand*\GammaeffZ{{\Gamma_{\mkern-2.5mu\text{eff,0}}}}
\newcommand*\GammaZ{{\Gamma_{\mkern-2.5mu0}}}
\newcommand*\Gammaff{{\Gamma_{\mkern-2.5mu\text{ff}}}}
\newcommand*\GammaD{{\Gamma_{\mkern-2.5mu\text{D}}}}

\newcommand*\theoGammaeff{{\Gammaeff{}}}
\newcommand*\empiGammaeff{{\tilde{\Gamma}_{\mkern-2.5mu\text{eff}}}}

\newcommand*\geff{{g_{\mkern-1mu\text{eff}}}}
\newcommand*\bohrM{{\mu_{\mkern-1mu\text{B}}}}
\newcommand*\kB{{k_{\mkern-1mu\text{B}}}}

\newcommand*{\iu}{{i\mkern0.6mu}}

\newcommand*{\unitSpacing}{\mkern1.5mu}

\newcommand*{\triExp}{{
        I_{\text{inv}}\propto
        [1-
            (
                p_1\exp{(-{\tau_{\text{inv}}/T_1})}
                +
                p_{\text{Z}}\exp{(-{\tau_{\text{inv}}/T_\text{Z}})}
                +
                p_{\text{W}}\exp{(-{(\tau_{\text{inv}}/T_{\text{W}})^{{1/2}}})}
            )
]^2}}
\newcommand*{\oscillatorStrength}{
    {f=4\pi\epsilon_0\frac{m_e c}{\pi e^{{2}}}
        \frac{1}{\rho}
        \frac{n}{{\chi}_{L}{\mkern-2mu}^{2}}
    \!\int\!\!\alpha(\mkern-1mu\nu\mkern-1mu)\,\mathrm{d}\nu}
}
\newcommand*{\transitionDipoleMoment}{
    {\mu=\sqrt{{\hbar e^2 f/(2m_e \omega)}}}
}
\newcommand*{\visibility}{{V=(I_\text{max}-I_\text{min})/(I_\text{max}+I_\text{min})}}
\newcommand*{\theoGammaeffExpression}{{\theoGammaeff{}=\GammaZ{}+\frac{1}{2}\GammaSD{}(R\tau_{12}+1-\exp{(-R\tau_{23})})}}

\newcommand*\absB{{|\mathbf{B}|}}
\newcommand*{\absBequal}[1]{{|\mathbf{B}|={#1}\unitSpacing{}}}

\makeatletter
\newcommand*{\raisemath}[1]{\mathpalette{\raisem@th{#1}}}
\newcommand*{\raisem@th}[3]{\raisebox{#1}{$#2#3$}}
\makeatother

\newcommand*\textSampleA{{\textsc{main~sample}}}
\newcommand*\textSampleB{{\textsc{sample~b}}}
\newcommand*\textSampleACap{{\textsc{Main~sample}}}
\newcommand*\textSampleBCap{{\textsc{Sample~b}}}

\newcommand*\hollowSquare{{\setlength{\fboxsep}{0pt}\setlength{\fboxrule}{0.4pt}\fbox{\rule{0pt}{0.8ex}\rule{0.8ex}{0pt}}}}
\newcommand*\ringMarker{{\raisebox{0.0ex}{\rotatebox{45}{\hollowSquare{}}}}}
\newcommand*\busMarker{{\raisebox{0.2ex}{\hollowSquare{}}}}

\newcommand{\nocontentsline}[3]{}
\newcommand\stoptoc{%
    \let\origcontentsline\addcontentsline
    \let\addcontentsline\nocontentsline
}
\newcommand\resumetoc{%
    \let\addcontentsline\origcontentsline
}
 
\title{
    Millisecond optical coherence and strong collective coupling in an integrated telecom rare-earth photonic platform
}
\date{\today}
\author{Kah Jen Wo\,\orcidlink{0000-0003-3806-9233}}\email{kjwo@u.nus.edu}
\affiliation{Centre for Quantum Technologies, Queenstown 117543, Singapore}

\author{Pavel A. Dmitriev\,\orcidlink{0009-0006-9681-9409}}\email{pavel.a.dmitriev@nus.edu.sg}
\affiliation{Centre for Quantum Technologies, Queenstown 117543, Singapore}
\affiliation{National University of Singapore, Department of Materials Science and Engineering, Singapore}

\author{Karthik Dasigi\,\orcidlink{0009-0007-8127-2292}}%
\affiliation{Centre for Quantum Technologies, Queenstown 117543, Singapore}

\author{Fumiya Hanamura\,\orcidlink{0000-0002-2382-4593}}%
\affiliation{Centre for Quantum Technologies, Queenstown 117543, Singapore}

\author{Steven Touzard\,\orcidlink{0000-0002-3475-2839}}\email{steven.touzard@nus.edu.sg}
\affiliation{Centre for Quantum Technologies, Queenstown 117543, Singapore}
\affiliation{National University of Singapore, Department of Physics, Singapore}
 
\begin{abstract}
    Long-range quantum network nodes require the combination of strong light-matter coupling, long coherence times and in-situ spectral control at telecom wavelengths.
    The coherence of erbium in integrated devices is held back by its hosts, which do not simultaneously provide the weakly magnetic nuclear-spin environment and the well-defined substitutional sites found in coherence-optimised bulk crystals.
    Here we bring such an optimised crystal onto a photonic chip, by bonding an \textErCaWO{} host without an adhesive interlayer to a high-${Q}$ electro-optically tuneable thin-film lithium niobate microring resonator.
    At an effective temperature of 75${\unitSpacing{}}$mK and a field of only 0.2${\unitSpacing{}}$T, ions probed through a waveguide show an effective homogeneous linewidth of ${289\pm5\unitSpacing{}}$Hz (${T_{\text{M}}=1.10\pm0.02\unitSpacing{}}$ms), with spectral diffusion proceeding at ${86\pm18\unitSpacing{}}$Hz and saturating at ${1.5\pm0.2\unitSpacing{}}$kHz.
    Electro-optically tuning the resonator through the erbium optical transition resolves an avoided crossing with a collective cooperativity of ${C=6.7\pm0.4}$ at zero field and ${C=4.9\pm0.2}$ at 0.4${\unitSpacing{}}$T, where ${T_\text{M}}$ exceeds 1${\unitSpacing{}}$ms.
    Exploiting superhyperfine coupling to the host's \textTungstenNuclear{} nuclear spins, we store and retrieve optical phase information over 5${\unitSpacing{}}$s with a visibility of ${0.935\pm0.015}$.
    Strong collective coupling, millisecond coherence and in-situ spectral tuning in a single device thus establish heterogeneous integration leveraging coherence-optimised hosts as a route to scalable telecom quantum networks.
\end{abstract}

\maketitle

\section{Introduction}\label{sec:intro}

Scalable quantum networks require coherent interfaces between travelling optical photons and stationary matter systems~\cite{Kimble2008,Sangouard2011,Wehner2018}.
Rare-earth ions in crystalline hosts are among the most mature solid-state candidates, combining narrow optical transitions with long-lived spin and shelving states~\cite{Tittel2010,thiel2011,Afzelius2015}, and erbium is singular among them because its \textManifoldG{}${\leftrightarrow}$\textManifoldE{} transition lies in the telecom C-band~\cite{Dieke1968} and interfaces directly with existing fibre networks.
Deploying such interfaces at manufacturing scale requires combining erbium ensembles with photonic circuits that confine the optical mode and provide in-situ tuneable resonators, thereby enhancing light-matter coupling and reducing the optical power required at millikelvin temperatures.

This effort has seen substantial progress: integrated photonics with direct erbium doping has demonstrated quantum memories, with atomic-frequency-comb (AFC) storage of single-photon-level telecom pulses~\cite{duttaAtomicFrequencyComb2023,Barya2026}~and efficient resonator-enhanced storage of entangled photons with electro-optically routed retrieval~\cite{TFLNprog2026}.
However, the erbium coherence in these devices is limited by their choice of materials.
Each doped photonic material, such as silicon~\cite{Weiss2021,gritschNarrowOpticalTransitions2022,Rinner2023,berkmanLongOpticalElectron2025}, silicon carbide~\cite{Lyasota2026}, silicon nitride~\cite{Gupta2025} and thin-film lithium niobate (TFLN)~\cite{Wang2020,duttaIntegratedPhotonicPlatform2020}, was adopted because it is available as large wafers of thin-film on insulator with scalable fabrication techniques, and not because it suits erbium.
These choices can lead to a high level of magnetic noise, ill-defined substitution sites or implantation damage. For example, although the effective homogeneous linewidth in erbium-doped TFLN reaches 1.8${\unitSpacing{}}$kHz~\cite{wangErLiNb2022}, in hosts where the spin bath environment causes spectral diffusion and static spin broadening~\cite{Kanai2022,Guo2026}, the storage time is currently limited to the 0.1-1${\unitSpacing{}}$µs range~\cite{TFLNprog2026,Barya2026}.

In contrast, bulk rare-earth hosts offer a large coherence advantage over direct doping of photonic circuits. \textErYSO{}, a host combining clear substitution sites with moderate magnetic noise, reaches 73${\unitSpacing{}}$Hz optical linewidths~\cite{Boettger2009} and second-long hyperfine coherence~\cite{Rancic2017}, but only near an extreme field of 7${\unitSpacing{}}$T.
Even though the optical effective linewidth itself can be of order 100${\unitSpacing{}}$Hz at lower temperature and field~\cite{Fukumori2020, Guo2026}, the 100\%-abundant \textsuperscript{89}Y bath caps the spin coherence that underpins long-lived memories, and imposes superhyperfine structure obstructing precise spectral tailoring~\cite{Guo2026}.
Optimised hosts such as calcium tungstate (\textCaWO{}) were predicted to relax this constraint because their constituent nuclear spins are dilute and weakly magnetic~\cite{Kanai2022,Ferrenti2020}.
Indeed, \textErCaWO{} yields 23${\unitSpacing{}}$ms electron-spin coherence~\cite{LeDantec2021}, week-long narrow spectral holes~\cite{Wang2025}, indistinguishable photons~\cite{ourariIndistinguishableTelecomBand2023a} and spin-photon entanglement from single erbium ions~\cite{uysalSpinPhotonEntanglementSingle2025}.
However, because optimised hosts are not available as large thin films, scalable memories built from them require heterogeneous integration.
This route has already been explored for \textErYSO{}, whose mechanical robustness and low refractive index make it convenient to integrate.
These heterogeneous integration attempts achieved resonator tuning but weak coupling in lithium niobate~\cite{yangPhotonicIntegrationEr2021}, or strong coupling without resonator tuning in silicon carbide~\cite{Kolar2026}.
For optimised hosts, heterogeneous integration has been attempted for CeO\textsubscript{2} via deposition~\cite{Zhang2024}, without reaching bulk-level coherence.
Can the coherence of erbium in optimised hosts be combined with the scalability of integrated photonics?
\begin{figure*}[!ht]
    \centering
    \includegraphics{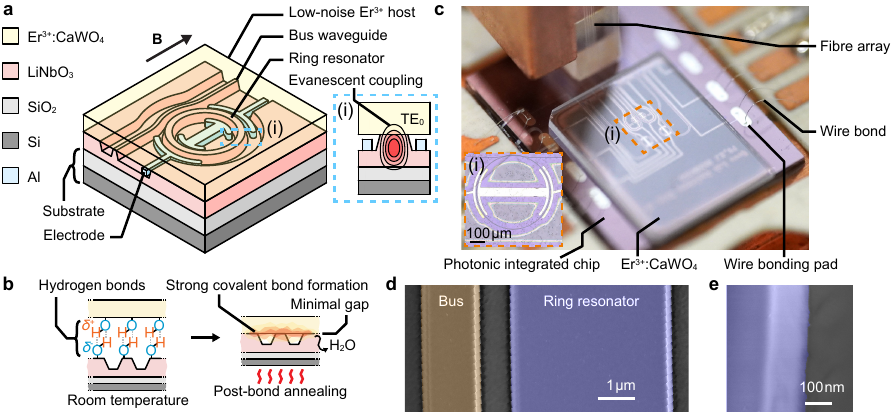}
    \caption{
        \textbf{Heterogeneously integrated photonic platform for strong light-matter coupling.}
        \textbf{a},~Schematic of the heterogeneously integrated platform.
        The photonic platform is thin-film lithium niobate (TFLN), with the \textQuartz{} and Si layers forming its substrate.
        An external magnetic field ${\mathbf{B}}$ is applied in the plane of the chip.
        A bus waveguide couples light in and out of the ring resonator via a pulley coupler,
        while aluminium electrodes enable electro-optic tuning of the ring resonator.
        Inset:~cross-section of the ring resonator with its single-mode profile evanescently coupling to the \textEr{} ions in the bonded \textCaWO{}.
        \textbf{b},~Schematic of the direct bonding procedure.
        At room temperature, hydrogen bonds form at the \textErCaWO{}--TFLN interface via surface activation (see \textSupplMat{}).
        A subsequent post-bond anneal drives the formation of covalent bonds with the release of H\textsubscript{2}O~\cite{Kang2024}.
        \textbf{c},~Photograph of the experimental setup, with a zoomed-in view of the ring resonator.
        \textbf{d},
        \textbf{e},~False-coloured scanning electron microscope (SEM) images of the TFLN waveguides prior to direct bonding, showing the bus waveguide (orange) and ring resonator (blue) with low sidewall roughness.
    }
    \label{fig:fig_01}
\end{figure*}%

Here we integrate \textErCaWO{} with thin-film lithium niobate (TFLN) (\cref{fig:fig_01}).
We bond a 50${\unitSpacing{}}$ppm \textErCaWO{} crystal directly to a high-${Q}$ (pre-bonding ${Q_\text{int}=5\times10^6}$ and post-bonding ${Q_\text{int}=1\times10^6}$; \cref{fig:fig_01}d-e and \textSupplMat{}), electro-optically tuneable microring resonator without an adhesive interlayer (\cref{fig:fig_01}b), placing the ions well within the evanescent field of the resonator mode (\cref{fig:fig_01}a; inset).
The electro-optic effect of TFLN enables spectral alignment of the resonator with a spin-preserving optical transition at millikelvin temperatures, where thermal and gas-condensation tuning are impractical.
Operating the device in a dilution refrigerator (\cref{fig:fig_01}c), we measure, through a bus waveguide, an effective homogeneous linewidth of ${289\pm5\unitSpacing{}}$Hz (${T_{\text{M}}=1.10\pm0.02\unitSpacing{}}$ms) at 0.2${\unitSpacing{}}$T, with spectral diffusion proceeding at a rate ${R=86\pm18\unitSpacing{}}$Hz and saturating at an amplitude ${\GammaSD{}=1.5\pm0.2\unitSpacing{}}$kHz.
Tuning the resonator through the optical transition resolves an avoided crossing with collective cooperativity ${C=6.7\pm0.4}$ at zero field.
At 0.4${\unitSpacing{}}$T, where ${T_\text{M}=1.07\pm0.01\unitSpacing{}}$ms, the same device yields ${C=4.9\pm0.2}$ (\textSupplMat{}).
Superhyperfine coupling to host \textTungstenNuclear{} nuclear spins allows storage of optical phases encoded in spectral gratings for 5${\unitSpacing{}}$s, retrieved with a visibility of ${0.935\pm0.015}$.
Strong collective coupling, millisecond optical coherence, slow spectral diffusion and electro-optic tuneability are thus obtained in one integrated device, establishing an optimal platform for scalable quantum networks.

\section{Results}\label{sec:results}
\subsection{High coherence and low spectral diffusion in an integrated chip}\label{subsec:fig_02}%
\begin{figure}[!htp]
    \centering
    \includegraphics{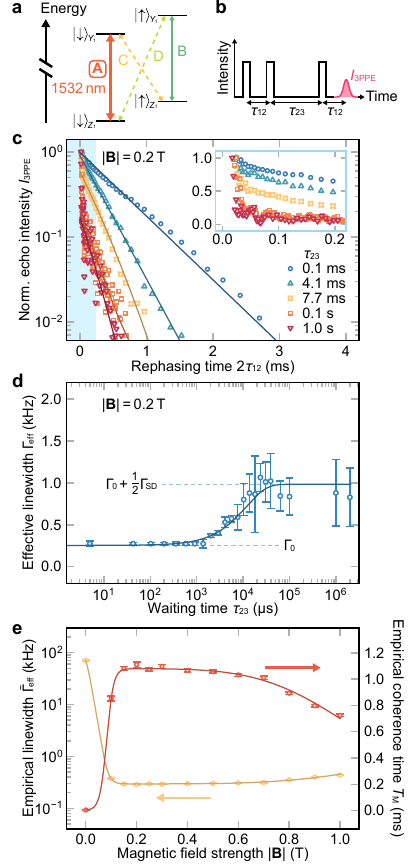}
    \caption{
        \textbf{Spectral diffusion and coherence of Er\textsuperscript{3+} ions.}
        \textbf{a},~\textEr{} energy level structure in the presence of magnetic field, depicting four distinct optical transitions (A--D) between the Kramers doublets of the ${Z_1}$~(\textManifoldG{}) and ${Y_1}$~(\textManifoldE{}) manifolds.
        Transition~A (${{\ketdownz{}}{\leftrightarrow}{\ketdowny{}}}$) is driven here.
        \textbf{b},~Pulse sequence for the three-pulse photon echo (3PPE) measurement, defining the pulse separation ${\tau_{12}}$ and waiting time ${\tau_{23}}$~\cite{Boettger2006}.
        \textbf{c},~Normalised echo intensity ${I_\text{3PPE}}$ as a function of rephasing time ${2\tau_{12}}$, obtained for various waiting times $\tau_{23}$ (different markers) at ${\absB{}=0.2\unitSpacing{}}$T.
        Each curve is fitted to a spectral diffusion model (solid line).
        Inset: Zoom-in on the shaded region revealing electron spin-echo envelope modulation (ESEEM).
        \textbf{d},~Effective linewidth $\theoGammaeff{}$ as a function of waiting time $\tau_{23}$ at ${\absB{}=0.2\unitSpacing{}}$T.
        The curve is fitted to the same spectral diffusion model (solid line).
        $\GammaZ{}$ is the homogeneous linewidth and $\GammaSD{}$ is the FWHM of the frequency distribution explored by \textEr{} ions.
        \textbf{e},~Empirical linewidth ${\empiGammaeff{}}$ (left axis) and empirical coherence time ${T_\text{M}}$ (right axis) as a function of magnetic field strength ${\absB{}}$ up to 1${\unitSpacing{}}$T.
        Solid line:~fit combining electronic spin flip-flop and one-phonon direct process contributions (see \textSupplMat{}), yielding a temperature of 75${\unitSpacing{}}$mK.
        The error bars represent one-standard-deviation confidence intervals from least-squares fits.
    }
    \label{fig:fig_02}
\end{figure}

We characterise the optical coherence of the \textEr{} ions using the bus waveguide, thereby probing the ions free from resonator-induced Purcell modification, and from non-linear effects due to strong collective coupling.
The erbium dopant is present in natural isotopic abundance, comprising both the hyperfine-active \textErNuclear{} isotope (${I={7/2}}$,~22.9\%) and hyperfine-free even isotopes (${I=0}$,~77.1\%).
Of the four optical transitions (A--D) between the Kramers doublets (\cref{fig:fig_02}a), we drive transition~A (${\ketdownz{}}{\leftrightarrow}{\ketdowny{}}$) in all experiments reported in this work. This corresponds to a spin-preserving optical transition, robust to magnetic noise.

To probe both the spectral diffusion and the homogeneous linewidth, we employ the three-pulse photon echo (3PPE) technique (\cref{fig:fig_02}b)~\cite{Levenson1988,Boettger2006}.
The first two pulses, separated by time ${\tau_{12}}$, imprint a spectral population grating on the ensemble.
After a waiting time ${\tau_{23}}$, a third pulse converts this grating back into optical coherence, which rephases as an echo at a time ${\tau_{12}}$ after the third pulse. For the measurement setup, see \textSupplMat{}.

As long as ${\tau_{23}}$ is less than the optical lifetime ${T_1=6.5\unitSpacing{}}$ms (\textSupplMat{}), the population grating is stored in the optical transition and the decay of the echo intensity follows ${I_{\text{3PPE}}\propto\exp{\mkern-1.5mu(-4\pi\tau_{12}\theoGammaeff{})}}$, where ${\theoGammaeff{}}$ is the effective linewidth. This time-dependence is shown in~\cref{fig:fig_02}c at the field of ${\absB{}=0.2\unitSpacing{}}$T.
In this regime, the effective linewidth remains close to 260${\unitSpacing{}}$Hz (\cref{fig:fig_02}d). For ${\tau_{23}\gg T_1}$, the optical excited state has fully decayed and the population grating is stored in shelving states with much longer lifetimes (see \cref{subsec:fig_04}). At short ${\tau_{12}}$, the resulting echoes are modulated by electron spin-echo envelope modulation (ESEEM) due to superhyperfine coupling~\cite{Probst2020}.
At ${\tau_{12}>20\unitSpacing{}}$µs, we fit the decay of the intensity following the same model, from which we deduce ${\theoGammaeff{}}$.
We see that the effective linewidth rises around ${\tau_{23}=1\unitSpacing{}}$ms and plateaus near 1${\unitSpacing{}}$kHz after ${\tau_{23}=10\unitSpacing{}}$ms verified up to ${\tau_{23}=2\unitSpacing{}}$s.

The evolution of the effective linewidth shown in \cref{fig:fig_02}d matches the ubiquitous spectral diffusion model well, for which ${\theoGammaeffExpression{}}$~\cite{Boettger2006}, where ${\GammaZ{}}$ is the homogeneous linewidth, ${R}$ is the characteristic rate at which neighbouring perturbers flip, and ${\GammaSD{}}$ is the full-width half-maximum (FWHM) of the frequency distribution explored by \textEr{} ions.
We find that the homogeneous linewidth is ${\GammaZ{}=254\pm7\unitSpacing{}}$Hz, the spectral diffusion rate is ${R=86\pm18\unitSpacing{}}$Hz, and the process saturates the linewidth at ${\GammaSD{}=1.5\pm0.2\unitSpacing{}}$kHz.
This spectral diffusion proceeds more slowly than the optical lifetime (${1/R=11.6\unitSpacing{}}$ms~${>T_1}$), and does not significantly affect the coherence of the optical transition.
Indeed, the empirical coherence time calculated from these parameters as ${T_\text{M}=\frac{2\GammaZ{}}{\GammaSD{}R}(-1+\surd{}(1+\frac{\GammaSD{}R}{\pi\GammaZ{}^{\mkern-5mu\smash{2}}}))}$~\cite{Boettger2006}, gives ${T_\text{M}=1.10\pm0.02\unitSpacing{}}$ms, which is close to ${1/\pi\GammaZ{}=1.25\pm0.03\unitSpacing{}}$ms.
Correspondingly, the empirical linewidth is ${\empiGammaeff{}=1/\pi T_\text{M}=289\pm5\unitSpacing{}}$Hz.
In comparison, the empirical linewidth at millikelvin temperatures with direct doping is limited to 1.8${\unitSpacing{}}$kHz in TFLN~\cite{wangErLiNb2022} and 441${\unitSpacing{}}$kHz in silicon carbide on insulator~\cite{Lyasota2026}, and with heterogeneous integration 1.3${\unitSpacing{}}$kHz for \textErYO{} on silicon~\cite{guptaDualEpitaxial2025} and 5${\unitSpacing{}}$kHz for \textEr{}:TiO\textsubscript{2} on silicon nitride~\cite{Gupta2025}.

For different magnetic fields, the empirical coherence time exceeds one millisecond between ${0.15\unitSpacing{}}$T and ${0.7\unitSpacing{}}$T, falling sharply at lower fields and declining gradually at higher fields, as shown in~\cref{fig:fig_02}e, where each point is obtained from a linearised fit of the spectral diffusion model (\textSupplMat{}).
We attribute the fall at low fields to the \textEr{} electronic spin flip-flop interactions in the $Z_1$ manifold~\cite{Boettger2006}, and the decline at high fields to the one-phonon direct process that involves an increased density of phonon states whose energies are degenerate with the Zeeman splitting \cite{Boettger2009}.
A fit combining both contributions (solid line) yields an effective spin temperature of 75${\unitSpacing{}}$mK (\textSupplMat{}).
This temperature is limited by thermalisation and residual heating induced by optical measurement pulses~\cite{Rochman2023}.

\subsection{Strong optical collective coupling with an integrated {\textEr{}} ensemble}\label{subsec:fig_03}%
{%
    Having established millisecond optical coherence at telecom wavelengths, we now demonstrate that the same integrated platform achieves strong collective coupling between the ring resonator and the \textEr{} ensemble.

    By electro-optically tuning the ring resonator across transition~A, we map the optical field transmission as a function of resonator-ensemble detuning (\cref{fig:fig_03}a).
    For all magnetic fields between ${0\unitSpacing{}}$T and ${1\unitSpacing{}}$T, we observe a clear avoided crossing, which is a spectroscopic signature of strong collective coupling.
    For the extracted parameters across all magnetic fields, see \textSupplMat{}.
    We model this resonator-ensemble coupling with a Lorentzian spectral distribution for the inhomogeneous broadening of the \textEr{} ensemble.
    This choice is physically justified by the strain caused by dilute defects in rare-earth crystals~\cite{thiel2011,Boettger2006a,Orth1994}, and is consistent with prior observation~\cite{yangPhotonicIntegrationEr2021}.

    For an inhomogeneously broadened ensemble with linewidth ${\Gammainh{}}$ coupled to a resonator of total linewidth ${\kappa_\text{tot}}$, the collective cooperativity is defined as ${C=\tfrac{4G^2}{\kappa_\text{tot}\Gammainh{}}}$, where $G$ is the collective coupling strength~\cite{Diniz2011,Gorshkov2007,Afzelius2010}.
    Fitting the transmission yields ${C=6.7\pm0.4}$ with ${\Gammainh{}/2\pi=332\pm21\unitSpacing{}}$MHz, ${G/2\pi=331\pm6\unitSpacing{}}$MHz at zero field, and a loaded resonator linewidth of ${\kappa_{\text{tot}}/2\pi \approx 200\unitSpacing{}}$MHz.
    At $\absBequal{1}$T, we extract ${C=5.8\pm0.8}$ with ${\Gammainh{}/2\pi=387\pm29\unitSpacing{}}$MHz and ${G/2\pi=353\pm26\unitSpacing{}}$MHz.
    One-dimensional slices across the avoided crossings confirm the fit in both amplitude and phase (\cref{fig:fig_03}b): (i) on resonance, the transmitted power splits into two polariton branches accompanied by a sharp dispersive phase feature, and (ii) both signatures vanish once the resonator is detuned.

    In comparison,
    \textErYSO{} direct-bonded to a TFLN microring reached ${C=0.36}$~\cite{yangPhotonicIntegrationEr2021}, and \textsuperscript{167}\textErYSO{} on silicon carbide microrings reached ${C=1.9}$~\cite{Kolar2026}.
    We attribute our improvement in cooperativity to annealed bonding and absence of an adhesive layer.
    In TFLN integration, the electro-optic effect enables tuning compatible with millikelvin temperatures to the spin-preserving transition~A.
    Without this tuning capability, integrated memories
    rely on spin-flipping transitions that can be tuned via magnetic field bias, at the cost of increased sensitivity to magnetic noise.

    \begin{figure}[!htp]
        \centering
        \includegraphics{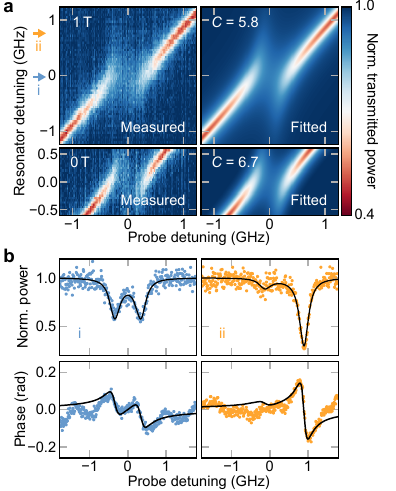}
        \caption{
            \textbf{Strong collective light-matter coupling.}
            \textbf{a},~Normalised transmission spectrum of the ring resonator as it is electro-optically tuned across
            the ${Z_1}{\leftrightarrow}{Y_1}$ transition (transition~A at finite field).
            An avoided crossing is resolved at both ${\absB{}=1\unitSpacing{}}$T (top; ${C=5.8\pm0.8}$) and ${\absB{}=0\unitSpacing{}}$T (bottom; ${C=6.7\pm0.4}$).
            The measured transmission spectra are fitted to a coupled resonator--ensemble model assuming a Lorentzian inhomogeneous broadening profile (see \textSupplMat{}).
            \textbf{b},~Transmitted power (top) and phase response (bottom) when the ring resonator is on resonance~(i; left) or detuned~(ii; right) relative to transition~A, at ${\absB{}=1\unitSpacing{}}$T.
            Markers:~measured data; solid lines:~fit.
            The power and phase response were acquired simultaneously using a phase-sensitive heterodyne setup (see \textSupplMat{} and Ref.~\cite{Dasigi2026}).
        }
        \label{fig:fig_03}
    \end{figure}
}

\subsection{Optical phase storage at seconds timescale}\label{subsec:fig_04}
\begin{figure*}[!htp]
    \centering
    \includegraphics{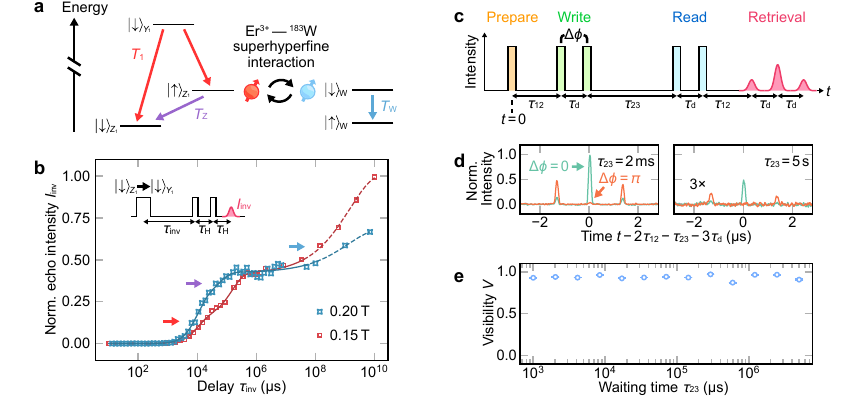}
    \caption{
        \textbf{Optical phase storage.}
        \textbf{a},~Decay mechanisms of \textEr{} ions populated in the $\ket{\downarrow}_{Y_1}$ state following optical excitation.
        \textbf{b},~Inversion-recovery echo intensity ${I_\text{inv}}$ as a function of waiting time $\tau_{\text{inv}}$ between the first pulse (see inset for sequence) and the subsequent Hahn echo sequence with time $\tau_{\text{H}}$ at different fields (markers).
        The echo intensities are normalised to their common plateau at ${T_1,T_{\text{Z}}<\tau_{\text{inv}}<T_{\text{W}}}$.
        Lines: fit to a triple exponential ${\triExp{}}$, from which $T_1$, $T_\text{Z}$, and $T_\text{W}$ are extracted.
        The dashed line section indicates an estimate for the ultra-long-lived states.
        Arrows indicate the echo features corresponding to each of the three decay channels shown in~\textbf{a}: optical ($T_1$), electronic spin ($T_\text{Z}$), and nuclear spin ($T_\text{W}$), from left to right.
        \textbf{c}, The pulse sequence for optical phase storage protocol, based on the 3PPE sequence~\cite{Staudt2007}.
        A preparation pulse is followed by two write pulses after a delay $\tau_{12}$.
        The write pulses are separated by $\tau_\text{d}$ and carry a relative phase $\Delta\phi$.
        After a time $\tau_{23}$, two read pulses retrieve the stored relative phase as an interference pattern of the emitted echoes.
        \textbf{d},~Emitted echoes for an input pulse pair with relative phase ${\Delta\phi=0}$ (constructive interference, green) and ${\Delta\phi=\pi}$ (destructive interference, orange), at delays ${\tau_{23}=2\unitSpacing{}}$ms (left) and ${\tau_{23}=5\unitSpacing{}}$s (right) at ${\absB{}=0.2\unitSpacing{}}$T.
        The traces for ${\tau_{23}=5\unitSpacing{}}$s are scaled up by a factor of 3${\times}$ for clarity.
        \textbf{e},~Visibility $V$ of the retrieved interference pattern as a function of waiting time $\tau_{23}$.
        The error bars are estimated from the baseline of the intensity trace.
    }
    \label{fig:fig_04}
\end{figure*}

In \textErCaWO{}, the \textTungstenNuclear{} isotope (${I={1/2}}$, 14.3\% natural abundance) provides access to ultra-long-lived shelving states.
The mechanism relies on superhyperfine coupling between the \textEr{} electronic spin and the surrounding \textTungstenNuclear{} nuclear spins~(\cref{fig:fig_04}a), which has been shown to enable spectral hole lifetimes on the order of minutes to hours~\cite{Wang2025}.
We demonstrate this shelving timescale through an inversion-recovery measurement as shown in~\cref{fig:fig_04}b:
at a delay time $\tau_\text{inv}$ after an initial optical excitation to $\ketdowny{}$,
we measure the ground state population at the excitation frequency with a Hahn echo sequence.
The excited \textEr{} ions relax through different channels with three well-separated timescales.
The first timescale involves the optical decay from the ${Y_1}$ to the ${Z_1}$ manifold with lifetime ${T_1=6.34\pm0.25\unitSpacing{}}$ms (${T_1=6.42\pm0.25\unitSpacing{}}$ms) at 0.15${\unitSpacing{}}$T (0.2${\unitSpacing{}}$T) (measured independently via fluorescence \textSupplMat{}).
The second timescale is the lifetime ${T_\text{Z}=126\pm8\unitSpacing{}}$ms (${T_\text{Z}=45\pm7\unitSpacing{}}$ms) of the upper Zeeman sublevel ${\ketupz{}}$ at 0.15${\unitSpacing{}}$T (0.2${\unitSpacing{}}$T).
At longer delays, the intensity of the recovered echo continuously increases for delays up to 10,000${\unitSpacing{}}$s.
We fit this increase with a stretched exponential with a stretching factor of 0.5, which suggests a timescale beyond 1,000${\unitSpacing{}}$s (a better estimate would require observing a clear plateau).
We attribute this third timescale ${T_\text{W}\gg T_\text{Z}}$ to the flips of the tungsten \textTungstenNuclear{} nuclear spins, which are coupled to erbium ions via superhyperfine interaction~\cite{Wang2025} and can serve as long-lived shelving states.

These shelving states are critical for many quantum memory protocols such as the atomic frequency comb (AFC). In AFCs, the phase information of time-bin qubits is typically measured using the interference pattern through a dual-comb interferometer~\cite{Kolar2026,Liu2022}. A similar capability is available for an optical phase using the 3PPE sequence shown in \cref{fig:fig_04}c. The 3PPE protocol is analogous to an AFC for which the finesse of the comb is intrinsically insufficient for efficient quantum storage~\cite{Bonarota2010}.
An initial preparation pulse first excites a frequency band within the inhomogeneous line.
After a delay ${\tau_{12}}$, two write pulses, separated by ${\tau_{\text{d}}}$, create a pair of spectral population gratings within this band.
The relative phase ${\Delta\phi}$ between the write pulses is thereby encoded into the frequency offset between these two gratings.
After a time ${\tau_{23}}$, two read pulses separated by ${\tau_{\text{d}}}$ convert the spectral population gratings back into optical coherence, generating three echoes. If the period of both population gratings and their relative frequency offset are preserved, the central echo displays consistent interference between the retrieved pulses encoding the original phase information ${\Delta\phi}$.
The delay time ${\tau_\text{d}=6.15\unitSpacing{}}$µs is chosen for robustness against the phase drift of our 2${\unitSpacing{}}$kHz linewidth source laser.
For the pulse parameters, see \textSupplMat{}.
We observe high contrast interference for delays up to ${\tau_{23}=5\unitSpacing{}}$s as shown in \cref{fig:fig_04}d.
We quantify this contrast using the visibility ${\visibility{}}$, where $I_\text{max}$ and $I_\text{min}$ are the integrated intensities of the central interfered echo for constructive and destructive interference, respectively.
The visibility remains constant across all investigated storage times with an average value ${V = 0.935 \pm 0.015}$, as shown in \cref{fig:fig_04}e.
This confirms that the relative optical phase encoded in spectral gratings with periods of 0.13~--~0.75${\unitSpacing{}}$MHz is retained over 5${\unitSpacing{}}$s, and validates that the \textTungstenNuclear{} nuclear spins act as shelving states.

\section{Discussion and conclusion}\label{sec:conclusion}

We have integrated an erbium host optimised for coherence into scalable photonics, with an optical coherence time that exceeds those of other integrated erbium platforms and an effective linewidth of the same order as in bulk \textsuperscript{167}\textErYSO{} at the same field and comparable temperature~\cite{Guo2026}.
Bonding an \textErCaWO{} crystal directly to a high-${Q}$ thin-film lithium niobate microring places the ions in the evanescent field of the resonator, where their optical coherence time reaches the millisecond scale.

Electro-optic control of the resonator allows the coupling to be measured on the highly coherent spin-preserving transition on which a quantum memory would operate, whereas neither thermal nor gas-condensation tuning is practical at millikelvin temperatures.
We find a collective cooperativity of up to ${C=6.7}$, and ${C=4.9}$ at 0.4${\unitSpacing{}}$T where ${T_\text{M}>1\unitSpacing{}}$ms.
The device therefore simultaneously offers strong collective coupling, millisecond optical coherence and on-chip resonator tuning.

The optical transition frequency is stable over long times, as it varies more slowly than the optical lifetime and remains within 1${\unitSpacing{}}$kHz from its original value even after seconds.
We attribute this stability to the sparse bath of nuclear spins in \textCaWO{} and to the high crystallinity of the host.
This stability is paramount for quantum memory protocols based on spectral-hole burning such as AFC~\cite{Afzelius2009}.
Sustaining a sub-kilohertz effective linewidth over the preparation time of an AFC makes kilohertz tooth spacings plausible, corresponding to delays two orders of magnitude beyond the microsecond range demonstrated so far~\cite{Barya2026,TFLNprog2026}.
Our demonstration of 3PPE-based interference provides a necessary primitive towards this goal.
Preparing a high-efficiency AFC and operating it on quantum signals remains to be demonstrated.

The tight mode confinement of our integrated photonics lowers the optical power needed to drive the \textEr{} ions and makes millikelvin operation possible.
At this operating temperature, a field as small as 0.2${\unitSpacing{}}$T polarises the electronic spins and suppresses spin flip-flops, leading to our narrow optical effective linewidth.
Compared to the multi-tesla fields required to reach comparable effective linewidths at higher temperature~\cite{Boettger2009,Rancic2017}, here a permanent magnet would suffice to operate our devices as a scalable quantum network node.

The combination of high cooperativity and optical coherence also enables efficient echo-based quantum memories~\cite{Damon2011,AWilliamson2014}.
In our device, the cooperativity ${C=6.7}$ indicates that the collective coupling of the erbium ensemble to the resonator far exceeds its coupling to the environment.
The same architecture can be designed to operate in the overcoupled regime, enabling a high-efficiency impedance-matched quantum memory with ${C=1}$~\cite{Afzelius2010,Moiseev2010}.
Identifying and removing the loss introduced by bonding would provide a straightforward path towards further improvements in efficiency.
Such an echo-based memory would allow on-demand storage of order millisecond, limited by the optical effective linewidth.

Our rare-earth host also provides a route towards far longer storage,
through hyperfine shelving in erbium enriched in the \textsuperscript{167}Er isotope.
The dilute nuclear spin background, together with the zero first-order-Zeeman points identified in \textCaWO{}~\cite{Marsh2026}, should lift the hyperfine coherence restriction of about a second previously measured in \textsuperscript{167}\textErYSO{} at 7${\unitSpacing{}}$T~\cite{Rancic2017}.
Together with the low temperatures that integrated devices enable, this platform opens a path towards spin-wave quantum storage on the minute timescales.
The platform thus supports a long-term programme, from high-efficiency nodes for metropolitan quantum networks to the far longer storage that intercontinental and satellite links require.

\section*{Data availability}
Data will be made available upon reasonable request.

\section*{Acknowledgements}
We thank Di~Zhu, Jian-Rui~Soh, Mikhael~Sayat, John~Bartholomew, and Milo{\v{s}}~Ran\v{c}i\'{c} for helpful discussions.
This research is supported by the Ministry of Education, Singapore, the National Research Foundation, Singapore, under grants ID NRFF14-2022-0002 and CRP33-2025-0072, and through the National Quantum Office, hosted in A\texttt{*}STAR, under its Centre for Quantum Technologies Funding Initiative (S24Q2d0009).
We also acknowledge the funding support from The University of Sydney~--~National University of Singapore 2026 Ignition Grants, and from E6Nanofab facilities. K.J.W.~and~K.D.~acknowledge the support of the Singapore National Quantum Scholarship Scheme (NQSS).

\section*{Competing interests}
The authors declare no competing interests.

\section*{Author contributions}
K.J.W.~carried out the measurements and developed the direct bonding procedure. P.A.D.~fabricated and packaged the TFLN samples and assisted in developing the direct bonding procedure. K.J.W.,~P.A.D.~and~K.D. designed the cryogenic measurement setup. K.D.~developed the setup for the heterodyne measurements, assisted with nanofabrication and measurements. F.H.~assisted with measurements and data interpretation. S.T.~developed the concept, coordinated the work and assisted with data interpretation. All authors participated in the writing and reviewing of the manuscript.


\begin{thebibliography}{52}%
\makeatletter
\providecommand \@ifxundefined [1]{%
 \@ifx{#1\undefined}
}%
\providecommand \@ifnum [1]{%
 \ifnum #1\expandafter \@firstoftwo
 \else \expandafter \@secondoftwo
 \fi
}%
\providecommand \@ifx [1]{%
 \ifx #1\expandafter \@firstoftwo
 \else \expandafter \@secondoftwo
 \fi
}%
\providecommand \natexlab [1]{#1}%
\providecommand \enquote  [1]{``#1''}%
\providecommand \bibnamefont  [1]{#1}%
\providecommand \bibfnamefont [1]{#1}%
\providecommand \citenamefont [1]{#1}%
\providecommand \href@noop [0]{\@secondoftwo}%
\providecommand \href [0]{\begingroup \@sanitize@url \@href}%
\providecommand \@href[1]{\@@startlink{#1}\@@href}%
\providecommand \@@href[1]{\endgroup#1\@@endlink}%
\providecommand \@sanitize@url [0]{\catcode `\\12\catcode `\$12\catcode
  `\&12\catcode `\#12\catcode `\^12\catcode `\_12\catcode `\%12\relax}%
\providecommand \@@startlink[1]{}%
\providecommand \@@endlink[0]{}%
\providecommand \url  [0]{\begingroup\@sanitize@url \@url }%
\providecommand \@url [1]{\endgroup\@href {#1}{\urlprefix }}%
\providecommand \urlprefix  [0]{URL }%
\providecommand \Eprint [0]{\href }%
\providecommand \doibase [0]{https://doi.org/}%
\providecommand \selectlanguage [0]{\@gobble}%
\providecommand \bibinfo  [0]{\@secondoftwo}%
\providecommand \bibfield  [0]{\@secondoftwo}%
\providecommand \translation [1]{[#1]}%
\providecommand \BibitemOpen [0]{}%
\providecommand \bibitemStop [0]{}%
\providecommand \bibitemNoStop [0]{.\EOS\space}%
\providecommand \EOS [0]{\spacefactor3000\relax}%
\providecommand \BibitemShut  [1]{\csname bibitem#1\endcsname}%
\let\auto@bib@innerbib\@empty
%
\bibitem [{\citenamefont {Kimble}(2008)}]{Kimble2008}%
  \BibitemOpen
  \bibfield  {author} {\bibinfo {author} {\bibfnamefont {H.~J.}\ \bibnamefont
  {Kimble}},\ }\bibfield  {title} {\bibinfo {title} {The quantum internet},\
  }\href {https://doi.org/10.1038/nature07127} {\bibfield  {journal} {\bibinfo
  {journal} {Nature}\ }\textbf {\bibinfo {volume} {453}},\ \bibinfo {pages}
  {1023} (\bibinfo {year} {2008})}\BibitemShut {NoStop}%
\bibitem [{\citenamefont {Sangouard}\ \emph {et~al.}(2011)\citenamefont
  {Sangouard}, \citenamefont {Simon}, \citenamefont {de~Riedmatten},\ and\
  \citenamefont {Gisin}}]{Sangouard2011}%
  \BibitemOpen
  \bibfield  {author} {\bibinfo {author} {\bibfnamefont {N.}~\bibnamefont
  {Sangouard}}, \bibinfo {author} {\bibfnamefont {C.}~\bibnamefont {Simon}},
  \bibinfo {author} {\bibfnamefont {H.}~\bibnamefont {de~Riedmatten}},\ and\
  \bibinfo {author} {\bibfnamefont {N.}~\bibnamefont {Gisin}},\ }\bibfield
  {title} {\bibinfo {title} {Quantum repeaters based on atomic ensembles and
  linear optics},\ }\href {https://doi.org/10.1103/revmodphys.83.33} {\bibfield
   {journal} {\bibinfo  {journal} {Reviews of Modern Physics}\ }\textbf
  {\bibinfo {volume} {83}},\ \bibinfo {pages} {33} (\bibinfo {year}
  {2011})}\BibitemShut {NoStop}%
\bibitem [{\citenamefont {Wehner}\ \emph {et~al.}(2018)\citenamefont {Wehner},
  \citenamefont {Elkouss},\ and\ \citenamefont {Hanson}}]{Wehner2018}%
  \BibitemOpen
  \bibfield  {author} {\bibinfo {author} {\bibfnamefont {S.}~\bibnamefont
  {Wehner}}, \bibinfo {author} {\bibfnamefont {D.}~\bibnamefont {Elkouss}},\
  and\ \bibinfo {author} {\bibfnamefont {R.}~\bibnamefont {Hanson}},\
  }\bibfield  {title} {\bibinfo {title} {Quantum internet: A vision for the
  road ahead},\ }\bibfield  {journal} {\bibinfo  {journal} {Science}\ }\textbf
  {\bibinfo {volume} {362}},\ \href {https://doi.org/10.1126/science.aam9288}
  {10.1126/science.aam9288} (\bibinfo {year} {2018})\BibitemShut {NoStop}%
\bibitem [{\citenamefont {Tittel}\ \emph {et~al.}(2010)\citenamefont {Tittel},
  \citenamefont {Afzelius}, \citenamefont {Chaneliére}, \citenamefont {Cone},
  \citenamefont {Kr{\"{o}}ll}, \citenamefont {Moiseev},\ and\ \citenamefont
  {Sellars}}]{Tittel2010}%
  \BibitemOpen
  \bibfield  {author} {\bibinfo {author} {\bibfnamefont {W.}~\bibnamefont
  {Tittel}}, \bibinfo {author} {\bibfnamefont {M.}~\bibnamefont {Afzelius}},
  \bibinfo {author} {\bibfnamefont {T.}~\bibnamefont {Chaneliére}}, \bibinfo
  {author} {\bibfnamefont {R.}~\bibnamefont {Cone}}, \bibinfo {author}
  {\bibfnamefont {S.}~\bibnamefont {Kr{\"{o}}ll}}, \bibinfo {author}
  {\bibfnamefont {S.}~\bibnamefont {Moiseev}},\ and\ \bibinfo {author}
  {\bibfnamefont {M.}~\bibnamefont {Sellars}},\ }\bibfield  {title} {\bibinfo
  {title} {Photon--echo quantum memory in solid state systems},\ }\href
  {https://doi.org/10.1002/lpor.200810056} {\bibfield  {journal} {\bibinfo
  {journal} {Laser \& Photonics Reviews}\ }\textbf {\bibinfo {volume} {4}},\
  \bibinfo {pages} {244} (\bibinfo {year} {2010})}\BibitemShut {NoStop}%
\bibitem [{\citenamefont {Thiel}\ \emph {et~al.}(2011)\citenamefont {Thiel},
  \citenamefont {Böttger},\ and\ \citenamefont {Cone}}]{thiel2011}%
  \BibitemOpen
  \bibfield  {author} {\bibinfo {author} {\bibfnamefont {C.}~\bibnamefont
  {Thiel}}, \bibinfo {author} {\bibfnamefont {T.}~\bibnamefont {Böttger}},\
  and\ \bibinfo {author} {\bibfnamefont {R.}~\bibnamefont {Cone}},\ }\bibfield
  {title} {\bibinfo {title} {Rare-earth-doped materials for applications in
  quantum information storage and signal processing},\ }\href
  {https://doi.org/10.1016/j.jlumin.2010.12.015} {\bibfield  {journal}
  {\bibinfo  {journal} {Journal of Luminescence}\ }\textbf {\bibinfo {volume}
  {131}},\ \bibinfo {pages} {353} (\bibinfo {year} {2011})}\BibitemShut
  {NoStop}%
\bibitem [{\citenamefont {Afzelius}\ \emph {et~al.}(2015)\citenamefont
  {Afzelius}, \citenamefont {Gisin},\ and\ \citenamefont
  {de~Riedmatten}}]{Afzelius2015}%
  \BibitemOpen
  \bibfield  {author} {\bibinfo {author} {\bibfnamefont {M.}~\bibnamefont
  {Afzelius}}, \bibinfo {author} {\bibfnamefont {N.}~\bibnamefont {Gisin}},\
  and\ \bibinfo {author} {\bibfnamefont {H.}~\bibnamefont {de~Riedmatten}},\
  }\bibfield  {title} {\bibinfo {title} {Quantum memory for photons},\ }\href
  {https://doi.org/10.1063/pt.3.3021} {\bibfield  {journal} {\bibinfo
  {journal} {Physics Today}\ }\textbf {\bibinfo {volume} {68}},\ \bibinfo
  {pages} {42} (\bibinfo {year} {2015})}\BibitemShut {NoStop}%
\bibitem [{\citenamefont {Dieke}(1968)}]{Dieke1968}%
  \BibitemOpen
  \bibfield  {author} {\bibinfo {author} {\bibfnamefont {G.~H.}\ \bibnamefont
  {Dieke}},\ }\href@noop {} {\emph {\bibinfo {title} {Spectra and energy levels
  of rare earth ions in crystals}}},\ edited by\ \bibinfo {editor}
  {\bibfnamefont {H.~M.}\ \bibnamefont {Crosswhite}}\ and\ \bibinfo {editor}
  {\bibfnamefont {H.}~\bibnamefont {Crosswhite}}\ (\bibinfo  {publisher}
  {Interscience Publishers},\ \bibinfo {address} {New York},\ \bibinfo {year}
  {1968})\BibitemShut {NoStop}%
\bibitem [{\citenamefont {Dutta}\ \emph {et~al.}(2023)\citenamefont {Dutta},
  \citenamefont {Zhao}, \citenamefont {Saha}, \citenamefont {Farfurnik},
  \citenamefont {Goldschmidt},\ and\ \citenamefont
  {Waks}}]{duttaAtomicFrequencyComb2023}%
  \BibitemOpen
  \bibfield  {author} {\bibinfo {author} {\bibfnamefont {S.}~\bibnamefont
  {Dutta}}, \bibinfo {author} {\bibfnamefont {Y.}~\bibnamefont {Zhao}},
  \bibinfo {author} {\bibfnamefont {U.}~\bibnamefont {Saha}}, \bibinfo {author}
  {\bibfnamefont {D.}~\bibnamefont {Farfurnik}}, \bibinfo {author}
  {\bibfnamefont {E.~A.}\ \bibnamefont {Goldschmidt}},\ and\ \bibinfo {author}
  {\bibfnamefont {E.}~\bibnamefont {Waks}},\ }\bibfield  {title} {\bibinfo
  {title} {An {{Atomic Frequency Comb Memory}} in {{Rare-Earth-Doped Thin-Film
  Lithium Niobate}}},\ }\href {https://doi.org/10.1021/acsphotonics.2c01835}
  {\bibfield  {journal} {\bibinfo  {journal} {ACS Photonics}\ }\textbf
  {\bibinfo {volume} {10}},\ \bibinfo {pages} {1104} (\bibinfo {year}
  {2023})}\BibitemShut {NoStop}%
\bibitem [{\citenamefont {Barya}\ \emph {et~al.}(2026)\citenamefont {Barya},
  \citenamefont {Chen}, \citenamefont {Prabhu}, \citenamefont {Heller},
  \citenamefont {Chow}, \citenamefont {Kim}, \citenamefont {Akin},
  \citenamefont {Niaouris}, \citenamefont {Zhang}, \citenamefont {Dibos},
  \citenamefont {Wang},\ and\ \citenamefont {Goldschmidt}}]{Barya2026}%
  \BibitemOpen
  \bibfield  {author} {\bibinfo {author} {\bibfnamefont {P.}~\bibnamefont
  {Barya}}, \bibinfo {author} {\bibfnamefont {D.}~\bibnamefont {Chen}},
  \bibinfo {author} {\bibfnamefont {A.}~\bibnamefont {Prabhu}}, \bibinfo
  {author} {\bibfnamefont {L.}~\bibnamefont {Heller}}, \bibinfo {author}
  {\bibfnamefont {E.}~\bibnamefont {Chow}}, \bibinfo {author} {\bibfnamefont
  {H.}~\bibnamefont {Kim}}, \bibinfo {author} {\bibfnamefont {J.}~\bibnamefont
  {Akin}}, \bibinfo {author} {\bibfnamefont {V.}~\bibnamefont {Niaouris}},
  \bibinfo {author} {\bibfnamefont {J.}~\bibnamefont {Zhang}}, \bibinfo
  {author} {\bibfnamefont {A.~M.}\ \bibnamefont {Dibos}}, \bibinfo {author}
  {\bibfnamefont {P.}~\bibnamefont {Wang}},\ and\ \bibinfo {author}
  {\bibfnamefont {E.~A.}\ \bibnamefont {Goldschmidt}},\ }\href
  {https://doi.org/10.48550/ARXIV.2605.11588} {\bibinfo {title} {Telecom
  quantum memory over one microsecond in nanophotonic lithium niobate}}
  (\bibinfo {year} {2026})\BibitemShut {NoStop}%
\bibitem [{\citenamefont {Yang}\ \emph {et~al.}(2026)\citenamefont {Yang},
  \citenamefont {Guo}, \citenamefont {An}, \citenamefont {He}, \citenamefont
  {Lu}, \citenamefont {Jiang}, \citenamefont {Lu}, \citenamefont {Zhu},\ and\
  \citenamefont {Ma}}]{TFLNprog2026}%
  \BibitemOpen
  \bibfield  {author} {\bibinfo {author} {\bibfnamefont {C.}~\bibnamefont
  {Yang}}, \bibinfo {author} {\bibfnamefont {H.}~\bibnamefont {Guo}}, \bibinfo
  {author} {\bibfnamefont {Y.-Y.}\ \bibnamefont {An}}, \bibinfo {author}
  {\bibfnamefont {Q.}~\bibnamefont {He}}, \bibinfo {author} {\bibfnamefont
  {C.}~\bibnamefont {Lu}}, \bibinfo {author} {\bibfnamefont {Z.}~\bibnamefont
  {Jiang}}, \bibinfo {author} {\bibfnamefont {Y.-Q.}\ \bibnamefont {Lu}},
  \bibinfo {author} {\bibfnamefont {S.}~\bibnamefont {Zhu}},\ and\ \bibinfo
  {author} {\bibfnamefont {X.-S.}\ \bibnamefont {Ma}},\ }\href
  {https://doi.org/10.48550/ARXIV.2605.14777} {\bibinfo {title} {Programmable
  cavity-enhanced telecom quantum memory in thin-film lithium niobate}}
  (\bibinfo {year} {2026})\BibitemShut {NoStop}%
\bibitem [{\citenamefont {Weiss}\ \emph {et~al.}(2021)\citenamefont {Weiss},
  \citenamefont {Gritsch}, \citenamefont {Merkel},\ and\ \citenamefont
  {Reiserer}}]{Weiss2021}%
  \BibitemOpen
  \bibfield  {author} {\bibinfo {author} {\bibfnamefont {L.}~\bibnamefont
  {Weiss}}, \bibinfo {author} {\bibfnamefont {A.}~\bibnamefont {Gritsch}},
  \bibinfo {author} {\bibfnamefont {B.}~\bibnamefont {Merkel}},\ and\ \bibinfo
  {author} {\bibfnamefont {A.}~\bibnamefont {Reiserer}},\ }\bibfield  {title}
  {\bibinfo {title} {Erbium dopants in nanophotonic silicon waveguides},\
  }\href {https://doi.org/10.1364/optica.413330} {\bibfield  {journal}
  {\bibinfo  {journal} {Optica}\ }\textbf {\bibinfo {volume} {8}},\ \bibinfo
  {pages} {40} (\bibinfo {year} {2021})}\BibitemShut {NoStop}%
\bibitem [{\citenamefont {Gritsch}\ \emph {et~al.}(2022)\citenamefont
  {Gritsch}, \citenamefont {Weiss}, \citenamefont {Fr{\"u}h}, \citenamefont
  {Rinner},\ and\ \citenamefont
  {Reiserer}}]{gritschNarrowOpticalTransitions2022}%
  \BibitemOpen
  \bibfield  {author} {\bibinfo {author} {\bibfnamefont {A.}~\bibnamefont
  {Gritsch}}, \bibinfo {author} {\bibfnamefont {L.}~\bibnamefont {Weiss}},
  \bibinfo {author} {\bibfnamefont {J.}~\bibnamefont {Fr{\"u}h}}, \bibinfo
  {author} {\bibfnamefont {S.}~\bibnamefont {Rinner}},\ and\ \bibinfo {author}
  {\bibfnamefont {A.}~\bibnamefont {Reiserer}},\ }\bibfield  {title} {\bibinfo
  {title} {Narrow {{Optical Transitions}} in {{Erbium-Implanted Silicon
  Waveguides}}},\ }\href {https://doi.org/10.1103/PhysRevX.12.041009}
  {\bibfield  {journal} {\bibinfo  {journal} {Physical Review X}\ }\textbf
  {\bibinfo {volume} {12}},\ \bibinfo {pages} {041009} (\bibinfo {year}
  {2022})}\BibitemShut {NoStop}%
\bibitem [{\citenamefont {Rinner}\ \emph {et~al.}(2023)\citenamefont {Rinner},
  \citenamefont {Burger}, \citenamefont {Gritsch}, \citenamefont {Schmitt},\
  and\ \citenamefont {Reiserer}}]{Rinner2023}%
  \BibitemOpen
  \bibfield  {author} {\bibinfo {author} {\bibfnamefont {S.}~\bibnamefont
  {Rinner}}, \bibinfo {author} {\bibfnamefont {F.}~\bibnamefont {Burger}},
  \bibinfo {author} {\bibfnamefont {A.}~\bibnamefont {Gritsch}}, \bibinfo
  {author} {\bibfnamefont {J.}~\bibnamefont {Schmitt}},\ and\ \bibinfo {author}
  {\bibfnamefont {A.}~\bibnamefont {Reiserer}},\ }\bibfield  {title} {\bibinfo
  {title} {Erbium emitters in commercially fabricated nanophotonic silicon
  waveguides},\ }\href {https://doi.org/10.1515/nanoph-2023-0287} {\bibfield
  {journal} {\bibinfo  {journal} {Nanophotonics}\ }\textbf {\bibinfo {volume}
  {12}},\ \bibinfo {pages} {3455} (\bibinfo {year} {2023})}\BibitemShut
  {NoStop}%
\bibitem [{\citenamefont {Berkman}\ \emph {et~al.}(2025)\citenamefont
  {Berkman}, \citenamefont {Lyasota}, \citenamefont {{de Boo}}, \citenamefont
  {Bartholomew}, \citenamefont {Lim}, \citenamefont {Johnson}, \citenamefont
  {McCallum}, \citenamefont {Xu}, \citenamefont {Xie}, \citenamefont
  {Abrosimov}, \citenamefont {Pohl}, \citenamefont {Ahlefeldt}, \citenamefont
  {Sellars}, \citenamefont {Yin},\ and\ \citenamefont
  {Rogge}}]{berkmanLongOpticalElectron2025}%
  \BibitemOpen
  \bibfield  {author} {\bibinfo {author} {\bibfnamefont {I.~R.}\ \bibnamefont
  {Berkman}}, \bibinfo {author} {\bibfnamefont {A.}~\bibnamefont {Lyasota}},
  \bibinfo {author} {\bibfnamefont {G.~G.}\ \bibnamefont {{de Boo}}}, \bibinfo
  {author} {\bibfnamefont {J.~G.}\ \bibnamefont {Bartholomew}}, \bibinfo
  {author} {\bibfnamefont {S.~Q.}\ \bibnamefont {Lim}}, \bibinfo {author}
  {\bibfnamefont {B.~C.}\ \bibnamefont {Johnson}}, \bibinfo {author}
  {\bibfnamefont {J.~C.}\ \bibnamefont {McCallum}}, \bibinfo {author}
  {\bibfnamefont {B.-B.}\ \bibnamefont {Xu}}, \bibinfo {author} {\bibfnamefont
  {S.}~\bibnamefont {Xie}}, \bibinfo {author} {\bibfnamefont {N.~V.}\
  \bibnamefont {Abrosimov}}, \bibinfo {author} {\bibfnamefont {H.-J.}\
  \bibnamefont {Pohl}}, \bibinfo {author} {\bibfnamefont {R.~L.}\ \bibnamefont
  {Ahlefeldt}}, \bibinfo {author} {\bibfnamefont {M.~J.}\ \bibnamefont
  {Sellars}}, \bibinfo {author} {\bibfnamefont {C.}~\bibnamefont {Yin}},\ and\
  \bibinfo {author} {\bibfnamefont {S.}~\bibnamefont {Rogge}},\ }\bibfield
  {title} {\bibinfo {title} {Long optical and electron spin coherence times for
  erbium ions in silicon},\ }\href {https://doi.org/10.1038/s41534-025-01008-x}
  {\bibfield  {journal} {\bibinfo  {journal} {npj Quantum Information}\
  }\textbf {\bibinfo {volume} {11}},\ \bibinfo {pages} {66} (\bibinfo {year}
  {2025})}\BibitemShut {NoStop}%
\bibitem [{\citenamefont {Lyasota}\ \emph {et~al.}(2026)\citenamefont
  {Lyasota}, \citenamefont {Bader}, \citenamefont {Lim}, \citenamefont
  {Johnson}, \citenamefont {McCallum}, \citenamefont {Li}, \citenamefont
  {Rogge},\ and\ \citenamefont {Castelletto}}]{Lyasota2026}%
  \BibitemOpen
  \bibfield  {author} {\bibinfo {author} {\bibfnamefont {A.}~\bibnamefont
  {Lyasota}}, \bibinfo {author} {\bibfnamefont {J.}~\bibnamefont {Bader}},
  \bibinfo {author} {\bibfnamefont {S.~Q.}\ \bibnamefont {Lim}}, \bibinfo
  {author} {\bibfnamefont {B.~C.}\ \bibnamefont {Johnson}}, \bibinfo {author}
  {\bibfnamefont {J.~C.}\ \bibnamefont {McCallum}}, \bibinfo {author}
  {\bibfnamefont {Q.}~\bibnamefont {Li}}, \bibinfo {author} {\bibfnamefont
  {S.}~\bibnamefont {Rogge}},\ and\ \bibinfo {author} {\bibfnamefont
  {S.}~\bibnamefont {Castelletto}},\ }\bibfield  {title} {\bibinfo {title}
  {Narrow magneto-optical transitions of erbium implanted into silicon
  carbide-on-insulator},\ }\bibfield  {journal} {\bibinfo  {journal}
  {Communications Materials}\ }\textbf {\bibinfo {volume} {7}},\ \href
  {https://doi.org/10.1038/s43246-026-01154-5} {10.1038/s43246-026-01154-5}
  (\bibinfo {year} {2026})\BibitemShut {NoStop}%
\bibitem [{\citenamefont {Gupta}\ \emph
  {et~al.}(2025{\natexlab{a}})\citenamefont {Gupta}, \citenamefont {Pettit},
  \citenamefont {Sundaresh}, \citenamefont {Niaouris}, \citenamefont
  {Deckoff-Jones}, \citenamefont {Crowley}, \citenamefont {Carpenter},
  \citenamefont {Dibos}, \citenamefont {Singh},\ and\ \citenamefont
  {Sullivan}}]{Gupta2025}%
  \BibitemOpen
  \bibfield  {author} {\bibinfo {author} {\bibfnamefont {S.}~\bibnamefont
  {Gupta}}, \bibinfo {author} {\bibfnamefont {R.~M.}\ \bibnamefont {Pettit}},
  \bibinfo {author} {\bibfnamefont {A.}~\bibnamefont {Sundaresh}}, \bibinfo
  {author} {\bibfnamefont {V.}~\bibnamefont {Niaouris}}, \bibinfo {author}
  {\bibfnamefont {S.}~\bibnamefont {Deckoff-Jones}}, \bibinfo {author}
  {\bibfnamefont {D.~P.}\ \bibnamefont {Crowley}}, \bibinfo {author}
  {\bibfnamefont {L.~G.}\ \bibnamefont {Carpenter}}, \bibinfo {author}
  {\bibfnamefont {A.~M.}\ \bibnamefont {Dibos}}, \bibinfo {author}
  {\bibfnamefont {M.~K.}\ \bibnamefont {Singh}},\ and\ \bibinfo {author}
  {\bibfnamefont {S.~E.}\ \bibnamefont {Sullivan}},\ }\bibfield  {title}
  {\bibinfo {title} {Erbium quantum memory platform with long optical coherence
  via back-end-of-line deposition on foundry-fabricated photonics},\ }\bibfield
   {journal} {\bibinfo  {journal} {Physical Review Applied}\ }\textbf {\bibinfo
  {volume} {24}},\ \href {https://doi.org/10.1103/xj8y-b6sl}
  {10.1103/xj8y-b6sl} (\bibinfo {year} {2025}{\natexlab{a}})\BibitemShut
  {NoStop}%
\bibitem [{\citenamefont {Wang}\ \emph {et~al.}(2020)\citenamefont {Wang},
  \citenamefont {Yang}, \citenamefont {Cheng}, \citenamefont {Xu},
  \citenamefont {Shen}, \citenamefont {Cone}, \citenamefont {Thiel},\ and\
  \citenamefont {Tang}}]{Wang2020}%
  \BibitemOpen
  \bibfield  {author} {\bibinfo {author} {\bibfnamefont {S.}~\bibnamefont
  {Wang}}, \bibinfo {author} {\bibfnamefont {L.}~\bibnamefont {Yang}}, \bibinfo
  {author} {\bibfnamefont {R.}~\bibnamefont {Cheng}}, \bibinfo {author}
  {\bibfnamefont {Y.}~\bibnamefont {Xu}}, \bibinfo {author} {\bibfnamefont
  {M.}~\bibnamefont {Shen}}, \bibinfo {author} {\bibfnamefont {R.~L.}\
  \bibnamefont {Cone}}, \bibinfo {author} {\bibfnamefont {C.~W.}\ \bibnamefont
  {Thiel}},\ and\ \bibinfo {author} {\bibfnamefont {H.~X.}\ \bibnamefont
  {Tang}},\ }\bibfield  {title} {\bibinfo {title} {Incorporation of erbium ions
  into thin-film lithium niobate integrated photonics},\ }\bibfield  {journal}
  {\bibinfo  {journal} {Applied Physics Letters}\ }\textbf {\bibinfo {volume}
  {116}},\ \href {https://doi.org/10.1063/1.5142631} {10.1063/1.5142631}
  (\bibinfo {year} {2020})\BibitemShut {NoStop}%
\bibitem [{\citenamefont {Dutta}\ \emph {et~al.}(2020)\citenamefont {Dutta},
  \citenamefont {Goldschmidt}, \citenamefont {Barik}, \citenamefont {Saha},\
  and\ \citenamefont {Waks}}]{duttaIntegratedPhotonicPlatform2020}%
  \BibitemOpen
  \bibfield  {author} {\bibinfo {author} {\bibfnamefont {S.}~\bibnamefont
  {Dutta}}, \bibinfo {author} {\bibfnamefont {E.~A.}\ \bibnamefont
  {Goldschmidt}}, \bibinfo {author} {\bibfnamefont {S.}~\bibnamefont {Barik}},
  \bibinfo {author} {\bibfnamefont {U.}~\bibnamefont {Saha}},\ and\ \bibinfo
  {author} {\bibfnamefont {E.}~\bibnamefont {Waks}},\ }\bibfield  {title}
  {\bibinfo {title} {Integrated {{Photonic Platform}} for {{Rare-Earth Ions}}
  in {{Thin Film Lithium Niobate}}},\ }\href
  {https://doi.org/10.1021/acs.nanolett.9b04679} {\bibfield  {journal}
  {\bibinfo  {journal} {Nano Letters}\ }\textbf {\bibinfo {volume} {20}},\
  \bibinfo {pages} {741} (\bibinfo {year} {2020})}\BibitemShut {NoStop}%
\bibitem [{\citenamefont {Wang}\ \emph {et~al.}(2022)\citenamefont {Wang},
  \citenamefont {Yang}, \citenamefont {Shen}, \citenamefont {Fu}, \citenamefont
  {Xu}, \citenamefont {Cone}, \citenamefont {Thiel},\ and\ \citenamefont
  {Tang}}]{wangErLiNb2022}%
  \BibitemOpen
  \bibfield  {author} {\bibinfo {author} {\bibfnamefont {S.}~\bibnamefont
  {Wang}}, \bibinfo {author} {\bibfnamefont {L.}~\bibnamefont {Yang}}, \bibinfo
  {author} {\bibfnamefont {M.}~\bibnamefont {Shen}}, \bibinfo {author}
  {\bibfnamefont {W.}~\bibnamefont {Fu}}, \bibinfo {author} {\bibfnamefont
  {Y.}~\bibnamefont {Xu}}, \bibinfo {author} {\bibfnamefont {R.~L.}\
  \bibnamefont {Cone}}, \bibinfo {author} {\bibfnamefont {C.~W.}\ \bibnamefont
  {Thiel}},\ and\ \bibinfo {author} {\bibfnamefont {H.~X.}\ \bibnamefont
  {Tang}},\ }\bibfield  {title} {\bibinfo {title} {Er:{LiNbO\textsubscript{3}}
  with {{High Optical Coherence Enabling Optical Thickness Control}}},\ }\href
  {https://doi.org/10.1103/PhysRevApplied.18.014069} {\bibfield  {journal}
  {\bibinfo  {journal} {Physical Review Applied}\ }\textbf {\bibinfo {volume}
  {18}},\ \bibinfo {pages} {014069} (\bibinfo {year} {2022})}\BibitemShut
  {NoStop}%
\bibitem [{\citenamefont {Kanai}\ \emph {et~al.}(2022)\citenamefont {Kanai},
  \citenamefont {Heremans}, \citenamefont {Seo}, \citenamefont {Wolfowicz},
  \citenamefont {Anderson}, \citenamefont {Sullivan}, \citenamefont {Onizhuk},
  \citenamefont {Galli}, \citenamefont {Awschalom},\ and\ \citenamefont
  {Ohno}}]{Kanai2022}%
  \BibitemOpen
  \bibfield  {author} {\bibinfo {author} {\bibfnamefont {S.}~\bibnamefont
  {Kanai}}, \bibinfo {author} {\bibfnamefont {F.~J.}\ \bibnamefont {Heremans}},
  \bibinfo {author} {\bibfnamefont {H.}~\bibnamefont {Seo}}, \bibinfo {author}
  {\bibfnamefont {G.}~\bibnamefont {Wolfowicz}}, \bibinfo {author}
  {\bibfnamefont {C.~P.}\ \bibnamefont {Anderson}}, \bibinfo {author}
  {\bibfnamefont {S.~E.}\ \bibnamefont {Sullivan}}, \bibinfo {author}
  {\bibfnamefont {M.}~\bibnamefont {Onizhuk}}, \bibinfo {author} {\bibfnamefont
  {G.}~\bibnamefont {Galli}}, \bibinfo {author} {\bibfnamefont {D.~D.}\
  \bibnamefont {Awschalom}},\ and\ \bibinfo {author} {\bibfnamefont
  {H.}~\bibnamefont {Ohno}},\ }\bibfield  {title} {\bibinfo {title}
  {Generalized scaling of spin qubit coherence in over 12,000 host materials},\
  }\bibfield  {journal} {\bibinfo  {journal} {Proceedings of the National
  Academy of Sciences}\ }\textbf {\bibinfo {volume} {119}},\ \href
  {https://doi.org/10.1073/pnas.2121808119} {10.1073/pnas.2121808119} (\bibinfo
  {year} {2022})\BibitemShut {NoStop}%
\bibitem [{\citenamefont {Guo}\ \emph {et~al.}(2026)\citenamefont {Guo},
  \citenamefont {Li}, \citenamefont {Xu}, \citenamefont {Liu}, \citenamefont
  {Wang},\ and\ \citenamefont {Zhong}}]{Guo2026}%
  \BibitemOpen
  \bibfield  {author} {\bibinfo {author} {\bibfnamefont {M.}~\bibnamefont
  {Guo}}, \bibinfo {author} {\bibfnamefont {Q.}~\bibnamefont {Li}}, \bibinfo
  {author} {\bibfnamefont {Z.}~\bibnamefont {Xu}}, \bibinfo {author}
  {\bibfnamefont {S.}~\bibnamefont {Liu}}, \bibinfo {author} {\bibfnamefont
  {F.}~\bibnamefont {Wang}},\ and\ \bibinfo {author} {\bibfnamefont
  {M.}~\bibnamefont {Zhong}},\ }\bibfield  {title} {\bibinfo {title}
  {Decoherence characterization and quantum memory design in
  \textsuperscript{167}{Er}\textsuperscript{3+}:{Y}\textsubscript{2}{SiO}\textsubscript{5}},\
  }\href {https://doi.org/10.15302/frontphys.2026.033201} {\bibfield  {journal}
  {\bibinfo  {journal} {Frontiers of Physics}\ }\textbf {\bibinfo {volume}
  {21}},\ \bibinfo {pages} {033201} (\bibinfo {year} {2026})}\BibitemShut
  {NoStop}%
\bibitem [{\citenamefont {B\"ottger}\ \emph {et~al.}(2009)\citenamefont
  {B\"ottger}, \citenamefont {Thiel}, \citenamefont {Cone},\ and\ \citenamefont
  {Sun}}]{Boettger2009}%
  \BibitemOpen
  \bibfield  {author} {\bibinfo {author} {\bibfnamefont {T.}~\bibnamefont
  {B\"ottger}}, \bibinfo {author} {\bibfnamefont {C.~W.}\ \bibnamefont
  {Thiel}}, \bibinfo {author} {\bibfnamefont {R.~L.}\ \bibnamefont {Cone}},\
  and\ \bibinfo {author} {\bibfnamefont {Y.}~\bibnamefont {Sun}},\ }\bibfield
  {title} {\bibinfo {title} {Effects of magnetic field orientation on optical
  decoherence in
  {Er\textsuperscript{3+}}:{Y\textsubscript{2}SiO\textsubscript{5}}},\ }\href
  {https://doi.org/10.1103/PhysRevB.79.115104} {\bibfield  {journal} {\bibinfo
  {journal} {Phys. Rev. B}\ }\textbf {\bibinfo {volume} {79}},\ \bibinfo
  {pages} {115104} (\bibinfo {year} {2009})}\BibitemShut {NoStop}%
\bibitem [{\citenamefont {Rančić}\ \emph {et~al.}(2017)\citenamefont
  {Rančić}, \citenamefont {Hedges}, \citenamefont {Ahlefeldt},\ and\
  \citenamefont {Sellars}}]{Rancic2017}%
  \BibitemOpen
  \bibfield  {author} {\bibinfo {author} {\bibfnamefont {M.}~\bibnamefont
  {Rančić}}, \bibinfo {author} {\bibfnamefont {M.~P.}\ \bibnamefont
  {Hedges}}, \bibinfo {author} {\bibfnamefont {R.~L.}\ \bibnamefont
  {Ahlefeldt}},\ and\ \bibinfo {author} {\bibfnamefont {M.~J.}\ \bibnamefont
  {Sellars}},\ }\bibfield  {title} {\bibinfo {title} {Coherence time of over a
  second in a telecom-compatible quantum memory storage material},\ }\href
  {https://doi.org/10.1038/nphys4254} {\bibfield  {journal} {\bibinfo
  {journal} {Nature Physics}\ }\textbf {\bibinfo {volume} {14}},\ \bibinfo
  {pages} {50} (\bibinfo {year} {2017})}\BibitemShut {NoStop}%
\bibitem [{\citenamefont {Fukumori}\ \emph {et~al.}(2020)\citenamefont
  {Fukumori}, \citenamefont {Huang}, \citenamefont {Yang}, \citenamefont
  {Zhang},\ and\ \citenamefont {Zhong}}]{Fukumori2020}%
  \BibitemOpen
  \bibfield  {author} {\bibinfo {author} {\bibfnamefont {R.}~\bibnamefont
  {Fukumori}}, \bibinfo {author} {\bibfnamefont {Y.}~\bibnamefont {Huang}},
  \bibinfo {author} {\bibfnamefont {J.}~\bibnamefont {Yang}}, \bibinfo {author}
  {\bibfnamefont {H.}~\bibnamefont {Zhang}},\ and\ \bibinfo {author}
  {\bibfnamefont {T.}~\bibnamefont {Zhong}},\ }\bibfield  {title} {\bibinfo
  {title} {Subkilohertz optical homogeneous linewidth and dephasing mechanisms
  in {Er\textsuperscript{3+}}:{Y\textsubscript{2}O\textsubscript{3}}
  ceramics},\ }\href {https://doi.org/10.1103/physrevb.101.214202} {\bibfield
  {journal} {\bibinfo  {journal} {Physical Review B}\ }\textbf {\bibinfo
  {volume} {101}},\ \bibinfo {pages} {214202} (\bibinfo {year}
  {2020})}\BibitemShut {NoStop}%
\bibitem [{\citenamefont {Ferrenti}\ \emph {et~al.}(2020)\citenamefont
  {Ferrenti}, \citenamefont {de~Leon}, \citenamefont {Thompson},\ and\
  \citenamefont {Cava}}]{Ferrenti2020}%
  \BibitemOpen
  \bibfield  {author} {\bibinfo {author} {\bibfnamefont {A.~M.}\ \bibnamefont
  {Ferrenti}}, \bibinfo {author} {\bibfnamefont {N.~P.}\ \bibnamefont
  {de~Leon}}, \bibinfo {author} {\bibfnamefont {J.~D.}\ \bibnamefont
  {Thompson}},\ and\ \bibinfo {author} {\bibfnamefont {R.~J.}\ \bibnamefont
  {Cava}},\ }\bibfield  {title} {\bibinfo {title} {Identifying candidate hosts
  for quantum defects via data mining},\ }\bibfield  {journal} {\bibinfo
  {journal} {npj Computational Materials}\ }\textbf {\bibinfo {volume} {6}},\
  \href {https://doi.org/10.1038/s41524-020-00391-7}
  {10.1038/s41524-020-00391-7} (\bibinfo {year} {2020})\BibitemShut {NoStop}%
\bibitem [{\citenamefont {Le~Dantec}\ \emph {et~al.}(2021)\citenamefont
  {Le~Dantec}, \citenamefont {Ran{\v{c}}i{\'{c}}}, \citenamefont {Lin},
  \citenamefont {Billaud}, \citenamefont {Ranjan}, \citenamefont {Flanigan},
  \citenamefont {Bertaina}, \citenamefont {Chaneli{\`{e}}re}, \citenamefont
  {Goldner}, \citenamefont {Erb}, \citenamefont {Liu}, \citenamefont {Estève},
  \citenamefont {Vion}, \citenamefont {Flurin},\ and\ \citenamefont
  {Bertet}}]{LeDantec2021}%
  \BibitemOpen
  \bibfield  {author} {\bibinfo {author} {\bibfnamefont {M.}~\bibnamefont
  {Le~Dantec}}, \bibinfo {author} {\bibfnamefont {M.}~\bibnamefont
  {Ran{\v{c}}i{\'{c}}}}, \bibinfo {author} {\bibfnamefont {S.}~\bibnamefont
  {Lin}}, \bibinfo {author} {\bibfnamefont {E.}~\bibnamefont {Billaud}},
  \bibinfo {author} {\bibfnamefont {V.}~\bibnamefont {Ranjan}}, \bibinfo
  {author} {\bibfnamefont {D.}~\bibnamefont {Flanigan}}, \bibinfo {author}
  {\bibfnamefont {S.}~\bibnamefont {Bertaina}}, \bibinfo {author}
  {\bibfnamefont {T.}~\bibnamefont {Chaneli{\`{e}}re}}, \bibinfo {author}
  {\bibfnamefont {P.}~\bibnamefont {Goldner}}, \bibinfo {author} {\bibfnamefont
  {A.}~\bibnamefont {Erb}}, \bibinfo {author} {\bibfnamefont {R.~B.}\
  \bibnamefont {Liu}}, \bibinfo {author} {\bibfnamefont {D.}~\bibnamefont
  {Estève}}, \bibinfo {author} {\bibfnamefont {D.}~\bibnamefont {Vion}},
  \bibinfo {author} {\bibfnamefont {E.}~\bibnamefont {Flurin}},\ and\ \bibinfo
  {author} {\bibfnamefont {P.}~\bibnamefont {Bertet}},\ }\bibfield  {title}
  {\bibinfo {title} {Twenty-three--millisecond electron spin coherence of
  erbium ions in a natural-abundance crystal},\ }\bibfield  {journal} {\bibinfo
   {journal} {Science Advances}\ }\textbf {\bibinfo {volume} {7}},\ \href
  {https://doi.org/10.1126/sciadv.abj9786} {10.1126/sciadv.abj9786} (\bibinfo
  {year} {2021})\BibitemShut {NoStop}%
\bibitem [{\citenamefont {Wang}\ \emph {et~al.}(2025)\citenamefont {Wang},
  \citenamefont {Lin}, \citenamefont {Le~Dantec}, \citenamefont
  {Ran\v{c}i\'{c}}, \citenamefont {Goldner}, \citenamefont {Bertaina},
  \citenamefont {Chaneliere}, \citenamefont {Liu}, \citenamefont {Esteve},
  \citenamefont {Vion}, \citenamefont {Flurin},\ and\ \citenamefont
  {Bertet}}]{Wang2025}%
  \BibitemOpen
  \bibfield  {author} {\bibinfo {author} {\bibfnamefont {Z.}~\bibnamefont
  {Wang}}, \bibinfo {author} {\bibfnamefont {S.}~\bibnamefont {Lin}}, \bibinfo
  {author} {\bibfnamefont {M.}~\bibnamefont {Le~Dantec}}, \bibinfo {author}
  {\bibfnamefont {M.}~\bibnamefont {Ran\v{c}i\'{c}}}, \bibinfo {author}
  {\bibfnamefont {P.}~\bibnamefont {Goldner}}, \bibinfo {author} {\bibfnamefont
  {S.}~\bibnamefont {Bertaina}}, \bibinfo {author} {\bibfnamefont
  {T.}~\bibnamefont {Chaneliere}}, \bibinfo {author} {\bibfnamefont
  {R.}~\bibnamefont {Liu}}, \bibinfo {author} {\bibfnamefont {D.}~\bibnamefont
  {Esteve}}, \bibinfo {author} {\bibfnamefont {D.}~\bibnamefont {Vion}},
  \bibinfo {author} {\bibfnamefont {E.}~\bibnamefont {Flurin}},\ and\ \bibinfo
  {author} {\bibfnamefont {P.}~\bibnamefont {Bertet}},\ }\bibfield  {title}
  {\bibinfo {title} {Week-long-lifetime microwave spectral holes in an
  erbium-doped scheelite crystal at millikelvin temperature},\ }\bibfield
  {journal} {\bibinfo  {journal} {Nature Communications}\ }\textbf {\bibinfo
  {volume} {16}},\ \href {https://doi.org/10.1038/s41467-025-64087-6}
  {10.1038/s41467-025-64087-6} (\bibinfo {year} {2025})\BibitemShut {NoStop}%
\bibitem [{\citenamefont {Ourari}\ \emph {et~al.}(2023)\citenamefont {Ourari},
  \citenamefont {Dusanowski}, \citenamefont {Horvath}, \citenamefont {Uysal},
  \citenamefont {Phenicie}, \citenamefont {Stevenson}, \citenamefont {Raha},
  \citenamefont {Chen}, \citenamefont {Cava}, \citenamefont {{de Leon}},\ and\
  \citenamefont {Thompson}}]{ourariIndistinguishableTelecomBand2023a}%
  \BibitemOpen
  \bibfield  {author} {\bibinfo {author} {\bibfnamefont {S.}~\bibnamefont
  {Ourari}}, \bibinfo {author} {\bibfnamefont {{\L}.}~\bibnamefont
  {Dusanowski}}, \bibinfo {author} {\bibfnamefont {S.~P.}\ \bibnamefont
  {Horvath}}, \bibinfo {author} {\bibfnamefont {M.~T.}\ \bibnamefont {Uysal}},
  \bibinfo {author} {\bibfnamefont {C.~M.}\ \bibnamefont {Phenicie}}, \bibinfo
  {author} {\bibfnamefont {P.}~\bibnamefont {Stevenson}}, \bibinfo {author}
  {\bibfnamefont {M.}~\bibnamefont {Raha}}, \bibinfo {author} {\bibfnamefont
  {S.}~\bibnamefont {Chen}}, \bibinfo {author} {\bibfnamefont {R.~J.}\
  \bibnamefont {Cava}}, \bibinfo {author} {\bibfnamefont {N.~P.}\ \bibnamefont
  {{de Leon}}},\ and\ \bibinfo {author} {\bibfnamefont {J.~D.}\ \bibnamefont
  {Thompson}},\ }\bibfield  {title} {\bibinfo {title} {Indistinguishable
  telecom band photons from a single {{Er}} ion in the solid state},\ }\href
  {https://doi.org/10.1038/s41586-023-06281-4} {\bibfield  {journal} {\bibinfo
  {journal} {Nature}\ }\textbf {\bibinfo {volume} {620}},\ \bibinfo {pages}
  {977} (\bibinfo {year} {2023})}\BibitemShut {NoStop}%
\bibitem [{\citenamefont {Uysal}\ \emph {et~al.}(2025)\citenamefont {Uysal},
  \citenamefont {Dusanowski}, \citenamefont {Xu}, \citenamefont {Horvath},
  \citenamefont {Ourari}, \citenamefont {Cava}, \citenamefont {{de Leon}},\
  and\ \citenamefont {Thompson}}]{uysalSpinPhotonEntanglementSingle2025}%
  \BibitemOpen
  \bibfield  {author} {\bibinfo {author} {\bibfnamefont {M.~T.}\ \bibnamefont
  {Uysal}}, \bibinfo {author} {\bibfnamefont {{\L}.}~\bibnamefont
  {Dusanowski}}, \bibinfo {author} {\bibfnamefont {H.}~\bibnamefont {Xu}},
  \bibinfo {author} {\bibfnamefont {S.~P.}\ \bibnamefont {Horvath}}, \bibinfo
  {author} {\bibfnamefont {S.}~\bibnamefont {Ourari}}, \bibinfo {author}
  {\bibfnamefont {R.~J.}\ \bibnamefont {Cava}}, \bibinfo {author}
  {\bibfnamefont {N.~P.}\ \bibnamefont {{de Leon}}},\ and\ \bibinfo {author}
  {\bibfnamefont {J.~D.}\ \bibnamefont {Thompson}},\ }\bibfield  {title}
  {\bibinfo {title} {Spin-{{Photon Entanglement}} of a {{Single}}
  {Er\textsuperscript{3+}} {{Ion}} in the {{Telecom Band}}},\ }\href
  {https://doi.org/10.1103/PhysRevX.15.011071} {\bibfield  {journal} {\bibinfo
  {journal} {Physical Review X}\ }\textbf {\bibinfo {volume} {15}},\ \bibinfo
  {pages} {011071} (\bibinfo {year} {2025})}\BibitemShut {NoStop}%
\bibitem [{\citenamefont {Yang}\ \emph {et~al.}(2021)\citenamefont {Yang},
  \citenamefont {Shen}, \citenamefont {Xu}, \citenamefont {Xie},\ and\
  \citenamefont {Tang}}]{yangPhotonicIntegrationEr2021}%
  \BibitemOpen
  \bibfield  {author} {\bibinfo {author} {\bibfnamefont {L.}~\bibnamefont
  {Yang}}, \bibinfo {author} {\bibfnamefont {M.}~\bibnamefont {Shen}}, \bibinfo
  {author} {\bibfnamefont {Y.}~\bibnamefont {Xu}}, \bibinfo {author}
  {\bibfnamefont {J.}~\bibnamefont {Xie}},\ and\ \bibinfo {author}
  {\bibfnamefont {H.~X.}\ \bibnamefont {Tang}},\ }\bibfield  {title} {\bibinfo
  {title} {Photonic integration of
  {{Er}}{\textsuperscript{3+}}:{{Y}}{\textsubscript{2}}{{SiO}}{\textsubscript{5}}
  with thin-film lithium niobate by flip chip bonding},\ }\href
  {https://doi.org/10.1364/OE.423659} {\bibfield  {journal} {\bibinfo
  {journal} {Optics Express}\ }\textbf {\bibinfo {volume} {29}},\ \bibinfo
  {pages} {15497} (\bibinfo {year} {2021})}\BibitemShut {NoStop}%
\bibitem [{\citenamefont {Kolar}\ \emph {et~al.}(2026)\citenamefont {Kolar},
  \citenamefont {Chin}, \citenamefont {Fong}, \citenamefont {Lukin},
  \citenamefont {Guidry}, \citenamefont {Palei}, \citenamefont {Vučković},\
  and\ \citenamefont {Zhong}}]{Kolar2026}%
  \BibitemOpen
  \bibfield  {author} {\bibinfo {author} {\bibfnamefont {A.}~\bibnamefont
  {Kolar}}, \bibinfo {author} {\bibfnamefont {I.}~\bibnamefont {Chin}},
  \bibinfo {author} {\bibfnamefont {C.}~\bibnamefont {Fong}}, \bibinfo {author}
  {\bibfnamefont {D.~M.}\ \bibnamefont {Lukin}}, \bibinfo {author}
  {\bibfnamefont {M.~A.}\ \bibnamefont {Guidry}}, \bibinfo {author}
  {\bibfnamefont {M.}~\bibnamefont {Palei}}, \bibinfo {author} {\bibfnamefont
  {J.}~\bibnamefont {Vučković}},\ and\ \bibinfo {author} {\bibfnamefont
  {T.}~\bibnamefont {Zhong}},\ }\href
  {https://doi.org/10.48550/ARXIV.2607.01324} {\bibinfo {title} {Integrated
  photon-memory entanglement generation using dual photonic resonators}}
  (\bibinfo {year} {2026})\BibitemShut {NoStop}%
\bibitem [{\citenamefont {Zhang}\ \emph {et~al.}(2024)\citenamefont {Zhang},
  \citenamefont {Grant}, \citenamefont {Masiulionis}, \citenamefont {Solomon},
  \citenamefont {Marcks}, \citenamefont {Bindra}, \citenamefont {Niklas},
  \citenamefont {Dibos}, \citenamefont {Poluektov}, \citenamefont {Heremans},
  \citenamefont {Guha},\ and\ \citenamefont {Awschalom}}]{Zhang2024}%
  \BibitemOpen
  \bibfield  {author} {\bibinfo {author} {\bibfnamefont {J.}~\bibnamefont
  {Zhang}}, \bibinfo {author} {\bibfnamefont {G.~D.}\ \bibnamefont {Grant}},
  \bibinfo {author} {\bibfnamefont {I.}~\bibnamefont {Masiulionis}}, \bibinfo
  {author} {\bibfnamefont {M.~T.}\ \bibnamefont {Solomon}}, \bibinfo {author}
  {\bibfnamefont {J.~C.}\ \bibnamefont {Marcks}}, \bibinfo {author}
  {\bibfnamefont {J.~K.}\ \bibnamefont {Bindra}}, \bibinfo {author}
  {\bibfnamefont {J.}~\bibnamefont {Niklas}}, \bibinfo {author} {\bibfnamefont
  {A.~M.}\ \bibnamefont {Dibos}}, \bibinfo {author} {\bibfnamefont {O.~G.}\
  \bibnamefont {Poluektov}}, \bibinfo {author} {\bibfnamefont {F.~J.}\
  \bibnamefont {Heremans}}, \bibinfo {author} {\bibfnamefont {S.}~\bibnamefont
  {Guha}},\ and\ \bibinfo {author} {\bibfnamefont {D.~D.}\ \bibnamefont
  {Awschalom}},\ }\bibfield  {title} {\bibinfo {title} {Optical and spin
  coherence of er spin qubits in epitaxial cerium dioxide on silicon},\
  }\bibfield  {journal} {\bibinfo  {journal} {npj Quantum Information}\
  }\textbf {\bibinfo {volume} {10}},\ \href
  {https://doi.org/10.1038/s41534-024-00903-z} {10.1038/s41534-024-00903-z}
  (\bibinfo {year} {2024})\BibitemShut {NoStop}%
\bibitem [{\citenamefont {Kang}\ \emph {et~al.}(2024)\citenamefont {Kang},
  \citenamefont {Yan}, \citenamefont {Niu}, \citenamefont {Liu}, \citenamefont
  {Suga},\ and\ \citenamefont {Wang}}]{Kang2024}%
  \BibitemOpen
  \bibfield  {author} {\bibinfo {author} {\bibfnamefont {Q.}~\bibnamefont
  {Kang}}, \bibinfo {author} {\bibfnamefont {H.}~\bibnamefont {Yan}}, \bibinfo
  {author} {\bibfnamefont {F.}~\bibnamefont {Niu}}, \bibinfo {author}
  {\bibfnamefont {K.}~\bibnamefont {Liu}}, \bibinfo {author} {\bibfnamefont
  {T.}~\bibnamefont {Suga}},\ and\ \bibinfo {author} {\bibfnamefont
  {C.}~\bibnamefont {Wang}},\ }\bibfield  {title} {\bibinfo {title} {Inp/linbo3
  covalent heterointerface construction via an asymmetric plasma activation
  strategy for hybrid integrated quantum systems},\ }\href
  {https://doi.org/10.1021/acsami.4c08823} {\bibfield  {journal} {\bibinfo
  {journal} {ACS Applied Materials \& Interfaces}\ }\textbf {\bibinfo {volume}
  {16}},\ \bibinfo {pages} {48502} (\bibinfo {year} {2024})},\ \bibinfo {note}
  {pMID: 39193874}\BibitemShut {NoStop}%
\bibitem [{\citenamefont {B\"ottger}\ \emph
  {et~al.}(2006{\natexlab{a}})\citenamefont {B\"ottger}, \citenamefont {Thiel},
  \citenamefont {Sun},\ and\ \citenamefont {Cone}}]{Boettger2006}%
  \BibitemOpen
  \bibfield  {author} {\bibinfo {author} {\bibfnamefont {T.}~\bibnamefont
  {B\"ottger}}, \bibinfo {author} {\bibfnamefont {C.~W.}\ \bibnamefont
  {Thiel}}, \bibinfo {author} {\bibfnamefont {Y.}~\bibnamefont {Sun}},\ and\
  \bibinfo {author} {\bibfnamefont {R.~L.}\ \bibnamefont {Cone}},\ }\bibfield
  {title} {\bibinfo {title} {Optical decoherence and spectral diffusion at 1.5
  µm in {Er\textsuperscript{3+}}:{Y\textsubscript{2}SiO\textsubscript{5}}
  versus magnetic field, temperature, and {Er\textsuperscript{3+}}
  concentration},\ }\href {https://doi.org/10.1103/PhysRevB.73.075101}
  {\bibfield  {journal} {\bibinfo  {journal} {Phys. Rev. B}\ }\textbf {\bibinfo
  {volume} {73}},\ \bibinfo {pages} {075101} (\bibinfo {year}
  {2006}{\natexlab{a}})}\BibitemShut {NoStop}%
\bibitem [{\citenamefont {Levenson}\ and\ \citenamefont
  {Kano}(1988)}]{Levenson1988}%
  \BibitemOpen
  \bibfield  {author} {\bibinfo {author} {\bibfnamefont {M.~D.}\ \bibnamefont
  {Levenson}}\ and\ \bibinfo {author} {\bibfnamefont {S.~S.}\ \bibnamefont
  {Kano}},\ }\href {https://doi.org/10.1016/b978-0-12-444722-6.x5001-9} {\emph
  {\bibinfo {title} {{I}ntroduction to {N}onlinear {L}aser {S}pectroscopy}}}\
  (\bibinfo  {publisher} {Academic Press},\ \bibinfo {address} {New York},\
  \bibinfo {year} {1988})\BibitemShut {NoStop}%
\bibitem [{\citenamefont {Probst}\ \emph {et~al.}(2020)\citenamefont {Probst},
  \citenamefont {Zhang}, \citenamefont {Ran\v{c}i\'c}, \citenamefont {Ranjan},
  \citenamefont {Le~Dantec}, \citenamefont {Zhang}, \citenamefont {Albanese},
  \citenamefont {Doll}, \citenamefont {Liu}, \citenamefont {Morton},
  \citenamefont {Chaneli\`ere}, \citenamefont {Goldner}, \citenamefont {Vion},
  \citenamefont {Esteve},\ and\ \citenamefont {Bertet}}]{Probst2020}%
  \BibitemOpen
  \bibfield  {author} {\bibinfo {author} {\bibfnamefont {S.}~\bibnamefont
  {Probst}}, \bibinfo {author} {\bibfnamefont {G.}~\bibnamefont {Zhang}},
  \bibinfo {author} {\bibfnamefont {M.}~\bibnamefont {Ran\v{c}i\'c}}, \bibinfo
  {author} {\bibfnamefont {V.}~\bibnamefont {Ranjan}}, \bibinfo {author}
  {\bibfnamefont {M.}~\bibnamefont {Le~Dantec}}, \bibinfo {author}
  {\bibfnamefont {Z.}~\bibnamefont {Zhang}}, \bibinfo {author} {\bibfnamefont
  {B.}~\bibnamefont {Albanese}}, \bibinfo {author} {\bibfnamefont
  {A.}~\bibnamefont {Doll}}, \bibinfo {author} {\bibfnamefont {R.~B.}\
  \bibnamefont {Liu}}, \bibinfo {author} {\bibfnamefont {J.}~\bibnamefont
  {Morton}}, \bibinfo {author} {\bibfnamefont {T.}~\bibnamefont
  {Chaneli\`ere}}, \bibinfo {author} {\bibfnamefont {P.}~\bibnamefont
  {Goldner}}, \bibinfo {author} {\bibfnamefont {D.}~\bibnamefont {Vion}},
  \bibinfo {author} {\bibfnamefont {D.}~\bibnamefont {Esteve}},\ and\ \bibinfo
  {author} {\bibfnamefont {P.}~\bibnamefont {Bertet}},\ }\bibfield  {title}
  {\bibinfo {title} {Hyperfine spectroscopy in a quantum-limited
  spectrometer},\ }\href {https://doi.org/10.5194/mr-1-315-2020} {\bibfield
  {journal} {\bibinfo  {journal} {Magnetic Resonance}\ }\textbf {\bibinfo
  {volume} {1}},\ \bibinfo {pages} {315} (\bibinfo {year} {2020})}\BibitemShut
  {NoStop}%
\bibitem [{\citenamefont {Gupta}\ \emph
  {et~al.}(2025{\natexlab{b}})\citenamefont {Gupta}, \citenamefont {Huang},
  \citenamefont {Liu}, \citenamefont {Pei}, \citenamefont {Gao}, \citenamefont
  {Yang}, \citenamefont {Tomm}, \citenamefont {Warburton},\ and\ \citenamefont
  {Zhong}}]{guptaDualEpitaxial2025}%
  \BibitemOpen
  \bibfield  {author} {\bibinfo {author} {\bibfnamefont {S.}~\bibnamefont
  {Gupta}}, \bibinfo {author} {\bibfnamefont {Y.}~\bibnamefont {Huang}},
  \bibinfo {author} {\bibfnamefont {S.}~\bibnamefont {Liu}}, \bibinfo {author}
  {\bibfnamefont {Y.}~\bibnamefont {Pei}}, \bibinfo {author} {\bibfnamefont
  {Q.}~\bibnamefont {Gao}}, \bibinfo {author} {\bibfnamefont {S.}~\bibnamefont
  {Yang}}, \bibinfo {author} {\bibfnamefont {N.}~\bibnamefont {Tomm}}, \bibinfo
  {author} {\bibfnamefont {R.~J.}\ \bibnamefont {Warburton}},\ and\ \bibinfo
  {author} {\bibfnamefont {T.}~\bibnamefont {Zhong}},\ }\bibfield  {title}
  {\bibinfo {title} {Dual epitaxial telecom spin-photon interfaces with
  long-lived coherence},\ }\href {https://doi.org/10.1038/s41467-025-64780-6}
  {\bibfield  {journal} {\bibinfo  {journal} {Nature Communications}\ }\textbf
  {\bibinfo {volume} {16}},\ \bibinfo {pages} {9814} (\bibinfo {year}
  {2025}{\natexlab{b}})}\BibitemShut {NoStop}%
\bibitem [{\citenamefont {Rochman}\ \emph {et~al.}(2023)\citenamefont
  {Rochman}, \citenamefont {Xie}, \citenamefont {Bartholomew}, \citenamefont
  {Schwab},\ and\ \citenamefont {Faraon}}]{Rochman2023}%
  \BibitemOpen
  \bibfield  {author} {\bibinfo {author} {\bibfnamefont {J.}~\bibnamefont
  {Rochman}}, \bibinfo {author} {\bibfnamefont {T.}~\bibnamefont {Xie}},
  \bibinfo {author} {\bibfnamefont {J.~G.}\ \bibnamefont {Bartholomew}},
  \bibinfo {author} {\bibfnamefont {K.~C.}\ \bibnamefont {Schwab}},\ and\
  \bibinfo {author} {\bibfnamefont {A.}~\bibnamefont {Faraon}},\ }\bibfield
  {title} {\bibinfo {title} {Microwave-to-optical transduction with erbium ions
  coupled to planar photonic and superconducting resonators},\ }\bibfield
  {journal} {\bibinfo  {journal} {Nature Communications}\ }\textbf {\bibinfo
  {volume} {14}},\ \href {https://doi.org/10.1038/s41467-023-36799-0}
  {10.1038/s41467-023-36799-0} (\bibinfo {year} {2023})\BibitemShut {NoStop}%
\bibitem [{\citenamefont {B\"ottger}\ \emph
  {et~al.}(2006{\natexlab{b}})\citenamefont {B\"ottger}, \citenamefont {Sun},
  \citenamefont {Thiel},\ and\ \citenamefont {Cone}}]{Boettger2006a}%
  \BibitemOpen
  \bibfield  {author} {\bibinfo {author} {\bibfnamefont {T.}~\bibnamefont
  {B\"ottger}}, \bibinfo {author} {\bibfnamefont {Y.}~\bibnamefont {Sun}},
  \bibinfo {author} {\bibfnamefont {C.~W.}\ \bibnamefont {Thiel}},\ and\
  \bibinfo {author} {\bibfnamefont {R.~L.}\ \bibnamefont {Cone}},\ }\bibfield
  {title} {\bibinfo {title} {Spectroscopy and dynamics of
  {Er\textsuperscript{3+}}:{Y\textsubscript{2}SiO\textsubscript{5}} at 1.5
  µm},\ }\href {https://doi.org/10.1103/PhysRevB.74.075107} {\bibfield
  {journal} {\bibinfo  {journal} {Phys. Rev. B}\ }\textbf {\bibinfo {volume}
  {74}},\ \bibinfo {pages} {075107} (\bibinfo {year}
  {2006}{\natexlab{b}})}\BibitemShut {NoStop}%
\bibitem [{\citenamefont {Orth}\ and\ \citenamefont
  {Skinner}(1994)}]{Orth1994}%
  \BibitemOpen
  \bibfield  {author} {\bibinfo {author} {\bibfnamefont {D.~L.}\ \bibnamefont
  {Orth}}\ and\ \bibinfo {author} {\bibfnamefont {J.~L.}\ \bibnamefont
  {Skinner}},\ }\bibfield  {title} {\bibinfo {title} {Lattice model of
  inhomogeneous broadening in crystals: Correlation of frequency distributions
  for different transitions},\ }\href {https://doi.org/10.1021/j100081a018}
  {\bibfield  {journal} {\bibinfo  {journal} {The Journal of Physical
  Chemistry}\ }\textbf {\bibinfo {volume} {98}},\ \bibinfo {pages} {7342}
  (\bibinfo {year} {1994})}\BibitemShut {NoStop}%
\bibitem [{\citenamefont {Diniz}\ \emph {et~al.}(2011)\citenamefont {Diniz},
  \citenamefont {Portolan}, \citenamefont {Ferreira}, \citenamefont {Gérard},
  \citenamefont {Bertet},\ and\ \citenamefont {Auffèves}}]{Diniz2011}%
  \BibitemOpen
  \bibfield  {author} {\bibinfo {author} {\bibfnamefont {I.}~\bibnamefont
  {Diniz}}, \bibinfo {author} {\bibfnamefont {S.}~\bibnamefont {Portolan}},
  \bibinfo {author} {\bibfnamefont {R.}~\bibnamefont {Ferreira}}, \bibinfo
  {author} {\bibfnamefont {J.~M.}\ \bibnamefont {Gérard}}, \bibinfo {author}
  {\bibfnamefont {P.}~\bibnamefont {Bertet}},\ and\ \bibinfo {author}
  {\bibfnamefont {A.}~\bibnamefont {Auffèves}},\ }\bibfield  {title} {\bibinfo
  {title} {Strongly coupling a cavity to inhomogeneous ensembles of emitters:
  Potential for long-lived solid-state quantum memories},\ }\href
  {https://doi.org/10.1103/physreva.84.063810} {\bibfield  {journal} {\bibinfo
  {journal} {Physical Review A}\ }\textbf {\bibinfo {volume} {84}},\ \bibinfo
  {pages} {063810} (\bibinfo {year} {2011})}\BibitemShut {NoStop}%
\bibitem [{\citenamefont {Gorshkov}\ \emph {et~al.}(2007)\citenamefont
  {Gorshkov}, \citenamefont {André}, \citenamefont {Lukin},\ and\
  \citenamefont {Sørensen}}]{Gorshkov2007}%
  \BibitemOpen
  \bibfield  {author} {\bibinfo {author} {\bibfnamefont {A.~V.}\ \bibnamefont
  {Gorshkov}}, \bibinfo {author} {\bibfnamefont {A.}~\bibnamefont {André}},
  \bibinfo {author} {\bibfnamefont {M.~D.}\ \bibnamefont {Lukin}},\ and\
  \bibinfo {author} {\bibfnamefont {A.~S.}\ \bibnamefont {Sørensen}},\
  }\bibfield  {title} {\bibinfo {title} {Photon storage in {$\Lambda$}-type
  optically dense atomic media. i. cavity model},\ }\href
  {https://doi.org/10.1103/physreva.76.033804} {\bibfield  {journal} {\bibinfo
  {journal} {Physical Review A}\ }\textbf {\bibinfo {volume} {76}},\ \bibinfo
  {pages} {033804} (\bibinfo {year} {2007})}\BibitemShut {NoStop}%
\bibitem [{\citenamefont {Afzelius}\ and\ \citenamefont
  {Simon}(2010)}]{Afzelius2010}%
  \BibitemOpen
  \bibfield  {author} {\bibinfo {author} {\bibfnamefont {M.}~\bibnamefont
  {Afzelius}}\ and\ \bibinfo {author} {\bibfnamefont {C.}~\bibnamefont
  {Simon}},\ }\bibfield  {title} {\bibinfo {title} {Impedance-matched cavity
  quantum memory},\ }\href {https://doi.org/10.1103/physreva.82.022310}
  {\bibfield  {journal} {\bibinfo  {journal} {Physical Review A}\ }\textbf
  {\bibinfo {volume} {82}},\ \bibinfo {pages} {022310} (\bibinfo {year}
  {2010})}\BibitemShut {NoStop}%
\bibitem [{\citenamefont {Dasigi}\ \emph {et~al.}(2026)\citenamefont {Dasigi},
  \citenamefont {Dmitriev}, \citenamefont {Wo}, \citenamefont {Hanamura},
  \citenamefont {Kong},\ and\ \citenamefont {Touzard}}]{Dasigi2026}%
  \BibitemOpen
  \bibfield  {author} {\bibinfo {author} {\bibfnamefont {K.}~\bibnamefont
  {Dasigi}}, \bibinfo {author} {\bibfnamefont {P.~A.}\ \bibnamefont
  {Dmitriev}}, \bibinfo {author} {\bibfnamefont {K.~J.}\ \bibnamefont {Wo}},
  \bibinfo {author} {\bibfnamefont {F.}~\bibnamefont {Hanamura}}, \bibinfo
  {author} {\bibfnamefont {L.}~\bibnamefont {Kong}},\ and\ \bibinfo {author}
  {\bibfnamefont {S.}~\bibnamefont {Touzard}},\ }\bibfield  {title} {\bibinfo
  {title} {Quantum-limited optical vector analysis},\ }\href
  {https://doi.org/10.1109/tim.2026.3699774} {\bibfield  {journal} {\bibinfo
  {journal} {IEEE Transactions on Instrumentation and Measurement}\ ,\ \bibinfo
  {pages} {1}} (\bibinfo {year} {2026})}\BibitemShut {NoStop}%
\bibitem [{\citenamefont {Staudt}\ \emph {et~al.}(2007)\citenamefont {Staudt},
  \citenamefont {Hastings-Simon}, \citenamefont {Nilsson}, \citenamefont
  {Afzelius}, \citenamefont {Scarani}, \citenamefont {Ricken}, \citenamefont
  {Suche}, \citenamefont {Sohler}, \citenamefont {Tittel},\ and\ \citenamefont
  {Gisin}}]{Staudt2007}%
  \BibitemOpen
  \bibfield  {author} {\bibinfo {author} {\bibfnamefont {M.~U.}\ \bibnamefont
  {Staudt}}, \bibinfo {author} {\bibfnamefont {S.~R.}\ \bibnamefont
  {Hastings-Simon}}, \bibinfo {author} {\bibfnamefont {M.}~\bibnamefont
  {Nilsson}}, \bibinfo {author} {\bibfnamefont {M.}~\bibnamefont {Afzelius}},
  \bibinfo {author} {\bibfnamefont {V.}~\bibnamefont {Scarani}}, \bibinfo
  {author} {\bibfnamefont {R.}~\bibnamefont {Ricken}}, \bibinfo {author}
  {\bibfnamefont {H.}~\bibnamefont {Suche}}, \bibinfo {author} {\bibfnamefont
  {W.}~\bibnamefont {Sohler}}, \bibinfo {author} {\bibfnamefont
  {W.}~\bibnamefont {Tittel}},\ and\ \bibinfo {author} {\bibfnamefont
  {N.}~\bibnamefont {Gisin}},\ }\bibfield  {title} {\bibinfo {title} {Fidelity
  of an optical memory based on stimulated photon echoes},\ }\href
  {https://doi.org/10.1103/physrevlett.98.113601} {\bibfield  {journal}
  {\bibinfo  {journal} {Physical Review Letters}\ }\textbf {\bibinfo {volume}
  {98}},\ \bibinfo {pages} {113601} (\bibinfo {year} {2007})}\BibitemShut
  {NoStop}%
\bibitem [{\citenamefont {Liu}\ \emph {et~al.}(2022)\citenamefont {Liu},
  \citenamefont {Li}, \citenamefont {Zhu}, \citenamefont {Zheng}, \citenamefont
  {Huang}, \citenamefont {Zhou}, \citenamefont {Li},\ and\ \citenamefont
  {Guo}}]{Liu2022}%
  \BibitemOpen
  \bibfield  {author} {\bibinfo {author} {\bibfnamefont {D.-C.}\ \bibnamefont
  {Liu}}, \bibinfo {author} {\bibfnamefont {P.-Y.}\ \bibnamefont {Li}},
  \bibinfo {author} {\bibfnamefont {T.-X.}\ \bibnamefont {Zhu}}, \bibinfo
  {author} {\bibfnamefont {L.}~\bibnamefont {Zheng}}, \bibinfo {author}
  {\bibfnamefont {J.-Y.}\ \bibnamefont {Huang}}, \bibinfo {author}
  {\bibfnamefont {Z.-Q.}\ \bibnamefont {Zhou}}, \bibinfo {author}
  {\bibfnamefont {C.-F.}\ \bibnamefont {Li}},\ and\ \bibinfo {author}
  {\bibfnamefont {G.-C.}\ \bibnamefont {Guo}},\ }\bibfield  {title} {\bibinfo
  {title} {On-demand storage of photonic qubits at telecom wavelengths},\
  }\href {https://doi.org/10.1103/physrevlett.129.210501} {\bibfield  {journal}
  {\bibinfo  {journal} {Physical Review Letters}\ }\textbf {\bibinfo {volume}
  {129}},\ \bibinfo {pages} {210501} (\bibinfo {year} {2022})}\BibitemShut
  {NoStop}%
\bibitem [{\citenamefont {Bonarota}\ \emph {et~al.}(2010)\citenamefont
  {Bonarota}, \citenamefont {Ruggiero}, \citenamefont {Gouët},\ and\
  \citenamefont {Chanelière}}]{Bonarota2010}%
  \BibitemOpen
  \bibfield  {author} {\bibinfo {author} {\bibfnamefont {M.}~\bibnamefont
  {Bonarota}}, \bibinfo {author} {\bibfnamefont {J.}~\bibnamefont {Ruggiero}},
  \bibinfo {author} {\bibfnamefont {J.~L.~L.}\ \bibnamefont {Gouët}},\ and\
  \bibinfo {author} {\bibfnamefont {T.}~\bibnamefont {Chanelière}},\
  }\bibfield  {title} {\bibinfo {title} {Efficiency optimization for atomic
  frequency comb storage},\ }\href {https://doi.org/10.1103/physreva.81.033803}
  {\bibfield  {journal} {\bibinfo  {journal} {Physical Review A}\ }\textbf
  {\bibinfo {volume} {81}},\ \bibinfo {pages} {033803} (\bibinfo {year}
  {2010})}\BibitemShut {NoStop}%
\bibitem [{\citenamefont {Afzelius}\ \emph {et~al.}(2009)\citenamefont
  {Afzelius}, \citenamefont {Simon}, \citenamefont {de~Riedmatten},\ and\
  \citenamefont {Gisin}}]{Afzelius2009}%
  \BibitemOpen
  \bibfield  {author} {\bibinfo {author} {\bibfnamefont {M.}~\bibnamefont
  {Afzelius}}, \bibinfo {author} {\bibfnamefont {C.}~\bibnamefont {Simon}},
  \bibinfo {author} {\bibfnamefont {H.}~\bibnamefont {de~Riedmatten}},\ and\
  \bibinfo {author} {\bibfnamefont {N.}~\bibnamefont {Gisin}},\ }\bibfield
  {title} {\bibinfo {title} {Multimode quantum memory based on atomic frequency
  combs},\ }\href {https://doi.org/10.1103/physreva.79.052329} {\bibfield
  {journal} {\bibinfo  {journal} {Physical Review A}\ }\textbf {\bibinfo
  {volume} {79}},\ \bibinfo {pages} {052329} (\bibinfo {year}
  {2009})}\BibitemShut {NoStop}%
\bibitem [{\citenamefont {Damon}\ \emph {et~al.}(2011)\citenamefont {Damon},
  \citenamefont {Bonarota}, \citenamefont {Louchet-Chauvet}, \citenamefont
  {Chanelière},\ and\ \citenamefont {Le~Gouët}}]{Damon2011}%
  \BibitemOpen
  \bibfield  {author} {\bibinfo {author} {\bibfnamefont {V.}~\bibnamefont
  {Damon}}, \bibinfo {author} {\bibfnamefont {M.}~\bibnamefont {Bonarota}},
  \bibinfo {author} {\bibfnamefont {A.}~\bibnamefont {Louchet-Chauvet}},
  \bibinfo {author} {\bibfnamefont {T.}~\bibnamefont {Chanelière}},\ and\
  \bibinfo {author} {\bibfnamefont {J.-L.}\ \bibnamefont {Le~Gouët}},\
  }\bibfield  {title} {\bibinfo {title} {Revival of silenced echo and quantum
  memory for light},\ }\href {https://doi.org/10.1088/1367-2630/13/9/093031}
  {\bibfield  {journal} {\bibinfo  {journal} {New Journal of Physics}\ }\textbf
  {\bibinfo {volume} {13}},\ \bibinfo {pages} {093031} (\bibinfo {year}
  {2011})}\BibitemShut {NoStop}%
\bibitem [{\citenamefont {A~Williamson}\ and\ \citenamefont
  {J~Longdell}(2014)}]{AWilliamson2014}%
  \BibitemOpen
  \bibfield  {author} {\bibinfo {author} {\bibfnamefont {L.}~\bibnamefont
  {A~Williamson}}\ and\ \bibinfo {author} {\bibfnamefont {J.}~\bibnamefont
  {J~Longdell}},\ }\bibfield  {title} {\bibinfo {title} {Cavity enhanced
  rephased amplified spontaneous emission},\ }\href
  {https://doi.org/10.1088/1367-2630/16/7/073046} {\bibfield  {journal}
  {\bibinfo  {journal} {New Journal of Physics}\ }\textbf {\bibinfo {volume}
  {16}},\ \bibinfo {pages} {073046} (\bibinfo {year} {2014})}\BibitemShut
  {NoStop}%
\bibitem [{\citenamefont {Moiseev}\ \emph {et~al.}(2010)\citenamefont
  {Moiseev}, \citenamefont {Andrianov},\ and\ \citenamefont
  {Gubaidullin}}]{Moiseev2010}%
  \BibitemOpen
  \bibfield  {author} {\bibinfo {author} {\bibfnamefont {S.~A.}\ \bibnamefont
  {Moiseev}}, \bibinfo {author} {\bibfnamefont {S.~N.}\ \bibnamefont
  {Andrianov}},\ and\ \bibinfo {author} {\bibfnamefont {F.~F.}\ \bibnamefont
  {Gubaidullin}},\ }\bibfield  {title} {\bibinfo {title} {Efficient multimode
  quantum memory based on photon echo in an optimal qed cavity},\ }\href
  {https://doi.org/10.1103/physreva.82.022311} {\bibfield  {journal} {\bibinfo
  {journal} {Physical Review A}\ }\textbf {\bibinfo {volume} {82}},\ \bibinfo
  {pages} {022311} (\bibinfo {year} {2010})}\BibitemShut {NoStop}%
\bibitem [{\citenamefont {Marsh}\ \emph {et~al.}(2026)\citenamefont {Marsh},
  \citenamefont {Yikai}, \citenamefont {Mattiroli}, \citenamefont {Sayat},
  \citenamefont {Trí}, \citenamefont {Rønnow}, \citenamefont {Longdell},\
  and\ \citenamefont {Soh}}]{Marsh2026}%
  \BibitemOpen
  \bibfield  {author} {\bibinfo {author} {\bibfnamefont {L.}~\bibnamefont
  {Marsh}}, \bibinfo {author} {\bibfnamefont {Y.}~\bibnamefont {Yikai}},
  \bibinfo {author} {\bibfnamefont {C.}~\bibnamefont {Mattiroli}}, \bibinfo
  {author} {\bibfnamefont {M.~T.}\ \bibnamefont {Sayat}}, \bibinfo {author}
  {\bibfnamefont {{\fontencoding{T1}\selectfont\char208}.~M.}\ \bibnamefont
  {Trí}}, \bibinfo {author} {\bibfnamefont {H.~M.}\ \bibnamefont {Rønnow}},
  \bibinfo {author} {\bibfnamefont {J.~J.}\ \bibnamefont {Longdell}},\ and\
  \bibinfo {author} {\bibfnamefont {J.-R.}\ \bibnamefont {Soh}},\ }\bibfield
  {title} {\bibinfo {title} {Nuclear quadrupole interaction and zero
  first-order zeeman transitions of
  {\textsuperscript{167}}{{Er}}{\textsuperscript{3+}} in
  {CaWO\textsubscript{4}}},\ }\bibfield  {journal} {\bibinfo  {journal}
  {Physical Review B}\ }\textbf {\bibinfo {volume} {113}},\ \href
  {https://doi.org/10.1103/pg5m-hk2n} {10.1103/pg5m-hk2n} (\bibinfo {year}
  {2026})\BibitemShut {NoStop}%
\end{thebibliography}
\end{document}


\clearpage{}%

\newcommand*\ketdownz{{\ket{\downarrow}_{Z_1}}}
\newcommand*\ketupz{{\ket{\uparrow}_{Z_1}}}
\newcommand*\ketdowny{{\ket{\downarrow}_{Y_1}}}
\newcommand*\ketupy{{\ket{\uparrow}_{Y_1}}}

\newcommand*\textManifoldG{{\textsuperscript{4}\textit{I}\textsubscript{15/2}}}
\newcommand*\textManifoldE{{\textsuperscript{4}\textit{I}\textsubscript{13/2}}}

\newcommand*\textCaWO{{CaWO\textsubscript{4}}}
\newcommand*\textEr{{Er\textsuperscript{3+}}}
\newcommand*\textErNuclear{{\textsuperscript{167}Er\textsuperscript{3+}}}
\newcommand*\textErCaWO{{\textEr{}:\textCaWO{}}}
\newcommand*\textYSO{{Y\textsubscript{2}SiO\textsubscript{5}}}
\newcommand*\textErYSO{{\textEr{}:\textYSO{}}}
\newcommand*\textErYO{{\textEr{}:Y\textsubscript{2}O\textsubscript{3}}}

\newcommand*\textLN{{LiNbO\textsubscript{3}}}

\newcommand*\textQuartz{{SiO\textsubscript{2}}}

\newcommand*\textTungstenNuclear{{\textsuperscript{183}W}}

\newcommand*\textAmmonia{{NH\textsubscript{4}OH}}

\newcommand*\textSupplMat{{\textcolor{supplmatcolour}{Suppl}}}

\newcommand*{\pavel}[1]{\textcolor{blue}{[Pavel: #1]}}
\newcommand*{\kjwo}[1]{\textcolor{myora}{[Bern: #1]}}
\newcommand*{\steven}[1]{\textcolor{cyan}{[Steven: #1]}}

\newcommand*\GammaSD{{\Gamma_{\mkern-2.5mu\text{SD}}}}
\newcommand*\SSD{{S_{\mkern-1mu\text{SD}}}}
\newcommand*\Gammaeff{{\Gamma_{\mkern-2.5mu\text{eff}}}}
\newcommand*\Gammainh{{\Gamma_{\mkern-2.5mu\text{inh}}}}
\newcommand*\GammaeffZ{{\Gamma_{\mkern-2.5mu\text{eff,0}}}}
\newcommand*\GammaZ{{\Gamma_{\mkern-2.5mu0}}}
\newcommand*\Gammaff{{\Gamma_{\mkern-2.5mu\text{ff}}}}
\newcommand*\GammaD{{\Gamma_{\mkern-2.5mu\text{D}}}}

\newcommand*\theoGammaeff{{\Gammaeff{}}}
\newcommand*\empiGammaeff{{\tilde{\Gamma}_{\mkern-2.5mu\text{eff}}}}

\newcommand*\geff{{g_{\mkern-1mu\text{eff}}}}
\newcommand*\bohrM{{\mu_{\mkern-1mu\text{B}}}}
\newcommand*\kB{{k_{\mkern-1mu\text{B}}}}

\newcommand*{\iu}{{i\mkern0.6mu}}

\newcommand*{\unitSpacing}{\mkern1.5mu}

\newcommand*{\triExp}{{
        I_{\text{inv}}\propto
        [1-
            (
                p_1\exp{(-{\tau_{\text{inv}}/T_1})}
                +
                p_{\text{Z}}\exp{(-{\tau_{\text{inv}}/T_\text{Z}})}
                +
                p_{\text{W}}\exp{(-{(\tau_{\text{inv}}/T_{\text{W}})^{{1/2}}})}
            )
]^2}}
\newcommand*{\oscillatorStrength}{
    {f=4\pi\epsilon_0\frac{m_e c}{\pi e^{{2}}}
        \frac{1}{\rho}
        \frac{n}{{\chi}_{L}{\mkern-2mu}^{2}}
    \!\int\!\!\alpha(\mkern-1mu\nu\mkern-1mu)\,\mathrm{d}\nu}
}
\newcommand*{\transitionDipoleMoment}{
    {\mu=\sqrt{{\hbar e^2 f/(2m_e \omega)}}}
}
\newcommand*{\visibility}{{V=(I_\text{max}-I_\text{min})/(I_\text{max}+I_\text{min})}}
\newcommand*{\theoGammaeffExpression}{{\theoGammaeff{}=\GammaZ{}+\frac{1}{2}\GammaSD{}(R\tau_{12}+1-\exp{(-R\tau_{23})})}}

\newcommand*\absB{{|\mathbf{B}|}}
\newcommand*{\absBequal}[1]{{|\mathbf{B}|={#1}\unitSpacing{}}}

\makeatletter
\newcommand*{\raisemath}[1]{\mathpalette{\raisem@th{#1}}}
\newcommand*{\raisem@th}[3]{\raisebox{#1}{$#2#3$}}
\makeatother

\newcommand*\textSampleA{{\textsc{main~sample}}}
\newcommand*\textSampleB{{\textsc{sample~b}}}
\newcommand*\textSampleACap{{\textsc{Main~sample}}}
\newcommand*\textSampleBCap{{\textsc{Sample~b}}}

\newcommand*\hollowSquare{{\setlength{\fboxsep}{0pt}\setlength{\fboxrule}{0.4pt}\fbox{\rule{0pt}{0.8ex}\rule{0.8ex}{0pt}}}}
\newcommand*\ringMarker{{\raisebox{0.0ex}{\rotatebox{45}{\hollowSquare{}}}}}
\newcommand*\busMarker{{\raisebox{0.2ex}{\hollowSquare{}}}}

\newcommand{\nocontentsline}[3]{}
\newcommand\stoptoc{%
    \let\origcontentsline\addcontentsline
    \let\addcontentsline\nocontentsline
}
\newcommand\resumetoc{%
    \let\addcontentsline\origcontentsline
}
\clearpage{}%
\clearpage{}%
\newcommand*\footnoteAvoidedCrossing{{\footnote{Data for \textSampleB{} (${r_{\text{bend}}=140\unitSpacing{}}$µm) during its initial thermal cycles. Nominal mixing chamber temperature was ${\sim}$20${\unitSpacing{}}$mK.}}}
\newcommand*\extendedDataTableAvoidedCrossing{
    \begin{tabularx}{0.975\linewidth}{s@{\hspace{0.0em}}s@{\hspace{1.0em}}s@{\hspace{1.0em}}s@{\hspace{1.0em}}s@{\hspace{1.0em}}s@{\hspace{1.0em}}s} \toprule[1.3pt]
        Condition                                                                   & $\kappa_\text{ext}/2\pi$ (MHz) & $\kappa_\text{int}/2\pi$ (MHz) & ${\Gamma_{\mkern-2.5mu\text{inh}}/2\pi}$ (MHz) & $G/2\pi$ (MHz)     & ${C=\tfrac{4G^{\smash{2}}}{\kappa_\text{tot}\Gammainh{}}}$ & ${\lambda_\text{ions}}$ (nm) \\ \midrule \\ [-2.0ex]
        \multicolumn{7}{l}{\textbf{\textSampleACap{}}}                                                                                                                                                                                                                                                                  \\ \midrule
        0.0${\unitSpacing{}}$T                                                      & ${39.12\pm0.46}$               & ${158.68\pm8.44}$              & ${331.79\pm20.94}$                             & ${330.57\pm5.67}$  & ${6.66\pm0.42}$                                            & 1532.6350                    \\ \midrule %
        0.1${\unitSpacing{}}$T                                                      & ${31.38\pm0.24}$               & ${153.01\pm2.72}$              & ${195.00\pm10.22}$                             & ${211.00\pm2.60}$  & ${4.95\pm0.25}$                                            & 1532.6305                    \\ \midrule
        0.4${\unitSpacing{}}$T                                                      & ${36.55\pm0.17}$               & ${155.43\pm3.18}$              & ${226.22\pm10.75}$                             & ${231.34\pm2.56}$  & ${4.93\pm0.20}$                                            & 1532.6201                    \\ \midrule %
        1.0${\unitSpacing{}}$T                                                      & ${36.96\pm0.29}$               & ${186.97\pm7.65}$              & ${386.58\pm29.49}$                             & ${353.49\pm25.87}$ & ${5.77\pm0.77}$                                            & 1532.6197                    \\ \midrule %
        \textbf{\textSampleBCap{}} (0${\unitSpacing{}}$T)\footnoteAvoidedCrossing{} & ${93.69\pm1.71}$               & ${75.23\pm14.27}$              & ${264.40\pm13.30}$                             & ${318.02\pm3.51}$  & ${9.06\pm0.96}$                                            & 1532.6346                    \\ \bottomrule[1.3pt]
\end{tabularx}}%

\newcommand*\extendedDataTableSpectralDiffusion{
    \begin{tabularx}{0.99\linewidth}{l@{\hspace{2em}}X@{\hspace{0.5em}}X@{\hspace{0.5em}}X@{\hspace{0.5em}}X@{\hspace{0.5em}}X@{\hspace{0.5em}}X@{\hspace{0.5em}}X@{\hspace{0.5em}}X@{\hspace{0.5em}}X@{\hspace{0.5em}}X@{\hspace{0.5em}}X@{\hspace{0.5em}}X} \toprule[1.3pt]
        ${|\mathbf{B}|}$ (T)                             & 0.1                           & 0.15                          & 0.2                           & 0.25                          & 0.3                           & 0.4                           & 0.5                           & 0.6                           & 0.7                           & 0.8                           & 0.9                           & 1.0                           \\ \midrule \\ [-2.0ex]
        \multicolumn{13}{l}{\textbf{Full}}                                                                                                                                                                                                                                                                                                                                                                                                               \\ \midrule
        ${R}$ (Hz)                                       & ${<415}$                      & ${29}$\newline${\pm19}$       & ${86}$\newline${\pm18}$       & ${<31}$                       & ${<43}$                       & ${<28}$                       & ${<57}$                       & ${<46}$                       & ${<51}$                       & ${<74}$                       & ${<76}$                       & ${<88}$                       \\ \midrule
        ${\Gamma_{\mkern-2.5mu0}}$ (Hz)                  & ${317}$\newline${\pm10}$      & ${268}$\newline${\pm5}$       & ${254}$\newline${\pm7}$       & ${267}$\newline${\pm4}$       & ${262}$\newline${\pm5}$       & ${266}$\newline${\pm4}$       & ${267}$\newline${\pm5}$       & ${269}$\newline${\pm4}$       & ${274}$\newline${\pm6}$       & ${318}$\newline${\pm5}$       & ${364}$\newline${\pm5}$       & ${389}$\newline${\pm8}$       \\ \midrule
        ${\GammaSD{}R}$ (kHz\textsuperscript{2})         & ${0.285}$\newline${\pm0.063}$ & ${0.099}$\newline${\pm0.018}$ & ${0.125}$\newline${\pm0.015}$ & ${0.108}$\newline${\pm0.013}$ & ${0.103}$\newline${\pm0.014}$ & ${0.120}$\newline${\pm0.021}$ & ${0.127}$\newline${\pm0.018}$ & ${0.155}$\newline${\pm0.023}$ & ${0.164}$\newline${\pm0.029}$ & ${0.172}$\newline${\pm0.025}$ & ${0.176}$\newline${\pm0.031}$ & ${0.282}$\newline${\pm0.037}$ \\ \midrule
        ${T_\text{M}}$ (µs)                       & ${844}$\newline${\pm23}$      & ${1079}$\newline${\pm17}$     & ${1103}$\newline${\pm20}$     & ${1074}$\newline${\pm13}$     & ${1097}$\newline${\pm14}$     & ${1067}$\newline${\pm16}$     & ${1058}$\newline${\pm14}$     & ${1032}$\newline${\pm16}$     & ${1010}$\newline${\pm19}$     & ${893}$\newline${\pm12}$      & ${798}$\newline${\pm11}$      & ${724}$\newline${\pm11}$      \\ \midrule
        ${\tilde{\Gamma}_{\mkern-2.5mu\text{eff}}}$ (Hz) & ${377}$\newline${\pm10}$      & ${295}$\newline${\pm5}$       & ${289}$\newline${\pm5}$       & ${296}$\newline${\pm4}$       & ${290}$\newline${\pm4}$       & ${298}$\newline${\pm4}$       & ${301}$\newline${\pm4}$       & ${309}$\newline${\pm5}$       & ${315}$\newline${\pm6}$       & ${356}$\newline${\pm5}$       & ${399}$\newline${\pm6}$       & ${440}$\newline${\pm6}$       \\ \midrule
        \multicolumn{13}{l}{\textbf{Linear,} $\boldsymbol{{\tau_{23}\leq13}\unitSpacing{}}$\textbf{ms}}                                                                                                                                                                                                                                                                                                                                                  \\ \midrule
        ${\Gamma_{\mkern-2.5mu0}}$ (Hz)                  & ${319}$\newline${\pm9}$       & ${268}$\newline${\pm5}$       & ${258}$\newline${\pm7}$       & ${268}$\newline${\pm4}$       & ${261}$\newline${\pm5}$       & ${266}$\newline${\pm4}$       & ${267}$\newline${\pm4}$       & ${269}$\newline${\pm4}$       & ${274}$\newline${\pm5}$       & ${318}$\newline${\pm5}$       & ${364}$\newline${\pm5}$       & ${388}$\newline${\pm7}$       \\ \midrule
        ${\GammaSD{}R}$ (kHz\textsuperscript{2})         & ${0.258}$\newline${\pm0.039}$ & ${0.097}$\newline${\pm0.016}$ & ${0.104}$\newline${\pm0.010}$ & ${0.107}$\newline${\pm0.010}$ & ${0.106}$\newline${\pm0.010}$ & ${0.119}$\newline${\pm0.012}$ & ${0.126}$\newline${\pm0.011}$ & ${0.154}$\newline${\pm0.014}$ & ${0.163}$\newline${\pm0.014}$ & ${0.172}$\newline${\pm0.016}$ & ${0.176}$\newline${\pm0.022}$ & ${0.284}$\newline${\pm0.022}$ \\ \midrule
        ${T_\text{M}}$ (µs)                       & ${852}$\newline${\pm19}$      & ${1081}$\newline${\pm16}$     & ${1111}$\newline${\pm20}$     & ${1074}$\newline${\pm12}$     & ${1097}$\newline${\pm14}$     & ${1068}$\newline${\pm13}$     & ${1058}$\newline${\pm12}$     & ${1032}$\newline${\pm13}$     & ${1011}$\newline${\pm14}$     & ${893}$\newline${\pm10}$      & ${798}$\newline${\pm10}$      & ${724}$\newline${\pm10}$      \\ \midrule
        ${\tilde{\Gamma}_{\mkern-2.5mu\text{eff}}}$ (Hz) & ${374}$\newline${\pm8}$       & ${295}$\newline${\pm4}$       & ${287}$\newline${\pm5}$       & ${296}$\newline${\pm3}$       & ${290}$\newline${\pm4}$       & ${298}$\newline${\pm4}$       & ${301}$\newline${\pm3}$       & ${308}$\newline${\pm4}$       & ${315}$\newline${\pm4}$       & ${356}$\newline${\pm4}$       & ${399}$\newline${\pm5}$       & ${440}$\newline${\pm6}$       \\ \midrule
        \multicolumn{13}{l}{\textbf{Linear,} $\boldsymbol{{\tau_{23}\leq3}\unitSpacing{}}$\textbf{ms}}                                                                                                                                                                                                                                                                                                                                                   \\ \midrule
        ${\Gamma_{\mkern-2.5mu0}}$ (Hz)                  & ${316}$\newline${\pm10}$      & ${272}$\newline${\pm6}$       & ${261}$\newline${\pm8}$       & ${271}$\newline${\pm4}$       & ${268}$\newline${\pm5}$       & ${272}$\newline${\pm4}$       & ${268}$\newline${\pm5}$       & ${270}$\newline${\pm4}$       & ${276}$\newline${\pm6}$       & ${319}$\newline${\pm5}$       & ${365}$\newline${\pm5}$       & ${389}$\newline${\pm8}$       \\ \midrule
        ${\GammaSD{}R}$ (kHz\textsuperscript{2})         & ${0.289}$\newline${\pm0.052}$ & ${0.068}$\newline${\pm0.022}$ & ${0.092}$\newline${\pm0.015}$ & ${0.081}$\newline${\pm0.014}$ & ${0.074}$\newline${\pm0.015}$ & ${0.067}$\newline${\pm0.018}$ & ${0.124}$\newline${\pm0.017}$ & ${0.142}$\newline${\pm0.019}$ & ${0.147}$\newline${\pm0.021}$ & ${0.166}$\newline${\pm0.019}$ & ${0.162}$\newline${\pm0.032}$ & ${0.276}$\newline${\pm0.033}$ \\ \midrule
        ${T_\text{M}}$ (µs)                       & ${844}$\newline${\pm20}$      & ${1094}$\newline${\pm19}$     & ${1112}$\newline${\pm21}$     & ${1086}$\newline${\pm14}$     & ${1105}$\newline${\pm15}$     & ${1096}$\newline${\pm16}$     & ${1059}$\newline${\pm14}$     & ${1037}$\newline${\pm15}$     & ${1017}$\newline${\pm16}$     & ${895}$\newline${\pm10}$      & ${801}$\newline${\pm11}$      & ${725}$\newline${\pm10}$      \\ \midrule
        ${\tilde{\Gamma}_{\mkern-2.5mu\text{eff}}}$ (Hz) & ${377}$\newline${\pm9}$       & ${291}$\newline${\pm5}$       & ${286}$\newline${\pm5}$       & ${293}$\newline${\pm4}$       & ${288}$\newline${\pm4}$       & ${290}$\newline${\pm4}$       & ${301}$\newline${\pm4}$       & ${307}$\newline${\pm4}$       & ${313}$\newline${\pm5}$       & ${356}$\newline${\pm4}$       & ${397}$\newline${\pm6}$       & ${439}$\newline${\pm6}$       \\ \midrule
        ${\lambda_\text{ions}}$ (+1532${\mkern1.5mu}$nm) & 0.6297                        & 0.6275                        & 0.6256                        & 0.6237                        & 0.6221                        & 0.6197                        & 0.6177                        & 0.6168                        & 0.6162                        & 0.6168                        & 0.6176                        & 0.6192                        \\\bottomrule[1.3pt]
\end{tabularx}}

\newcommand*\footnoteZeroFieldTM{{\footnote{From a separate 2PPE measurement.\label{foo:ZeroFieldTM}}}}
\newcommand*\footnoteCaWOCTE{{\footnote{Calculated with lattice parameter at 100-295${\unitSpacing{}}$K from Ref.~\cite{Senyshyn2011}.\label{foo:CaWO-CTE}}}}
\newcommand*\footnoteCaWOYoungMod{{\footnote{The maximum value in the ${{a}{b}}$-plane was taken to estimate the upper bound of the thermally-induced multilayer stress.\label{foo:CaWO-YoungMod}}}}
\newcommand*\extendedDataTableThermalAnalysis{
    \begin{tabularx}{0.9\linewidth}{l@{\hspace{3.0em}}s@{\hspace{0.5em}}s@{\hspace{0.5em}}s@{\hspace{0.5em}}s@{\hspace{0.5em}}s@{\hspace{0.5em}}s} \toprule[1.3pt]
        Material                                         & Si                          & \textQuartz{}          & LN,e                               & LN,o                               & \textCaWO{},${c}$                              & \textCaWO{},${a}$                                   \\ \midrule \\ [-2.0ex]
        ${\mathscr{E}}$ (GPa)                            & ${130}$ \cite{Hopcroft2010} & ${75}$ \cite{Zhao1999} & ${199}$ \cite{Weigel2017,Weis1985} & ${173}$ \cite{Weigel2017,Weis1985} & ${108}$ \cite{Gorodtsov2021}                   & ${137}$\footnoteCaWOYoungMod{} \cite{Gorodtsov2021} \\ \midrule
        ${\alpha}$ (${10^{-6}\unitSpacing{}}$K${}^{-1}$) & ${2.6}$ \cite{Swenson1983}  & ${1}$ \cite{Zhao1999}  & ${3.4}$ \cite{Pignatiello2007}     & ${13.4}$ \cite{Pignatiello2007}    & ${13.4}$\footnoteCaWOCTE{} \cite{Senyshyn2011} & ${7.9}$\footref{foo:CaWO-CTE} \cite{Senyshyn2011}   \\ \bottomrule[1.3pt]
\end{tabularx}}%

\newcommand*\extendedDataTableTriExp{
    \begin{tabularx}{0.8\linewidth}{s@{\hspace{0.0em}}s@{\hspace{1.0em}}s@{\hspace{1.0em}}s@{\hspace{1.0em}}s@{\hspace{1.0em}}s} \toprule[1.3pt]
        ${\absB{}}$             & ${p_1}$           & ${p_{\text{Z}}}$  & ${p_{\text{W}}}$  & ${T_{\text{Z}}}$            & ${T_{\text{W}}}$              \\ \midrule
        0.15${\unitSpacing{}}$T & ${0.360\pm0.009}$ & ${0.276\pm0.008}$ & ${0.364\pm0.006}$ & ${126\pm8\unitSpacing{}}$ms & ${1070\pm129\unitSpacing{}}$s \\ \midrule
        0.2${\unitSpacing{}}$T  & ${0.578\pm0.024}$ & ${0.205\pm0.020}$ & ${0.217\pm0.017}$ & ${45\pm7\unitSpacing{}}$ms  & ${1056\pm539\unitSpacing{}}$s \\ \bottomrule[1.3pt]
    \end{tabularx}
}

\newcommand*\footnoteCeff{{\footnote{Effective cooperativity that governs their atomic frequency comb storage efficiency.\label{foo:Ceff}}}}
\newcommand*\footnoteLyasota{{\footnote{The authors infer a sample temperature $\gtrsim$0.5\,K from the absence of Zeeman-branch depopulation at 400\,mT.\label{foo:Lyasota}}}}
\newcommand*\footnoteSampleTemperature{{\footnote{Base temperature of dilution refrigerator; sample temperature was not discussed.\label{foo:SampleTemperature}}}}
\newcommand*\extendedDataTableLiteratureMap{
    \begin{tabularx}{0.99\linewidth}{@{}X c c c X X @{}}\toprule[1.3pt]
        \textbf{Platform}                                                                              & ${\boldsymbol{C}}$            & ${\boldsymbol{T}}$                    & ${\boldsymbol{B}}$ & ${\boldsymbol{\Gamma_{\mkern-2mu\text{\textbf{eff}}}}}$ & \textbf{Spectral diffusion}                                                                                \\ \midrule \\ [-2.0ex]
        \multicolumn{6}{l}{\textbf{Heterogeneously Integrated}}                                                                                                                                                                                                                                                                                                            \\ \midrule
        \textErCaWO{} + Si resonator (single ion)~\cite{ourariIndistinguishableTelecomBand2023a}       & ---                           & 0.47\,K                               & 0.06\,T            & 31\,kHz (${T_2=10.2\,}$µs)                              & 63\,kHz long-term                                                                                          \\
        \textEr{}:TiO\textsubscript{2} on SiN nanophotonic waveguides~\cite{Gupta2025}                 & ---                           & 100\,mK                               & 0.433\,T           & {5\,kHz} (${T_2=64\,}$µs)                               & 27\,kHz within 4\,ms                                                                                       \\
        \textEr{}:CeO\textsubscript{2} epitaxial on Si~\cite{Zhang2024}                                & ---                           & 3.6\,K                                & 0.102\,T           & 440\,kHz (${T_2=0.72\,}$µs)                             & Not quantified                                                                                             \\
        \textEr{}:\textYSO{} on TFLN ring~\cite{yangPhotonicIntegrationEr2021}                         & $0.36$                        & 4\,K                                  & ---                & ---                                                     & ---                                                                                                        \\
        ${\alpha}$-Si ring + \textEr{}:\textYSO{}~\cite{Miyazono2017}                                  & $0.54$                        & 4.5\,K                                & None               & ---                                                     & ---                                                                                                        \\
        \textEr{}:\textYSO{} on SiC ring~\cite{Kolar2026}                                              & $1.9$                         & 7\,mK\footnoteSampleTemperature{}     & 1.238\,T           & ---                                                     & ---                                                                                                        \\

        \midrule
        \ours{\textbf{This work}}: \ours{\textEr{}:\textCaWO{} bonded to TFLN} \ours{(\textSampleA{})} & \ours{\textbf{${4.9 - 6.7}$}} & \ours{75\,mK}                         & \ours{0.2\,T}      & \ours{\textbf{289\,$\pm$\,5\,Hz}}                      & \ours{\textbf{${\boldsymbol{\GammaSD{}=1.5}}$\,kHz}, ${\boldsymbol{R=86}}$\,\textbf{Hz}, verified to 2\,s} \\ \midrule
        \multicolumn{6}{l}{\textbf{Implanted and doped waveguides}}                                                                                                                                                                                                                                                                                                        \\ \midrule
        \textEr{}:LiNbO\textsubscript{3} SmartCut, ridge waveguides + microrings~\cite{wangErLiNb2022} & ---                           & 20\,mK                                & 0.5\,T             & {1.8\,kHz} (${T_2=180\,}$µs)                            & Present with stretched exponential, not quantified (2PPE only)                                             \\
        Er implanted, SiC-on-insulator film~\cite{Lyasota2026}                                         & ---                           & $>$0.5\,K                             & None               & {441\,kHz}                                              & Not quantified                                                                                             \\
        Er implanted, Si waveguides~\cite{gritschNarrowOpticalTransitions2022}                         & ---                           & 2\,K                                  & None               & 9.3\,kHz                                                & Not quantified                                                                                             \\
        Er implanted, commercial Si waveguides~\cite{Rinner2023}                                       & ---                           & $<$2\,K                               & 9\,T               & $\leq$28.1\,kHz                                         & Not quantified                                                                                             \\
        \textEr{}:LiNbO\textsubscript{3} SmartCut~\cite{Barya2026}                                     & ---                           & 30\,mK\footref{foo:SampleTemperature} & 1.5\,T             & 3.5\,kHz (${T_2=90\,}$µs)                               & Present with stretched exponential, not quantified                                                         \\
        \textsuperscript{167}\textEr{}:LiNbO\textsubscript{3} SmartCut~\cite{TFLNprog2026}             & $0.39$\footnoteCeff{}         & 260\,mK                               & 2\,T               & {3.4\,kHz} (${T_2=93\,}$µs)                             & Present with stretched exponential, not quantified                                                         \\ \midrule
        \multicolumn{6}{l}{\textbf{Bulk}}                                                                                                                                                                                                                                                                                                                                  \\ \midrule
        Er in isotopically enriched Si~\cite{berkmanLongOpticalElectron2025}                           & ---                           & $<$20\,mK                             & 0.011\,T           & $<$70\,kHz                                              & ${\GammaSD{}<70}$\,kHz over 50\,$\upmu$s                                                                   \\
        \textsuperscript{167}\textEr{}:\textYSO{}~\cite{Guo2026}                                       & ---                           & 100\,mK                               & 0.2\,T             & 100\,Hz                                                 & ${\GammaSD{}=880}$~--~1010\,Hz                                                                             \\
        \textEr{}:\textYSO{}~\cite{Boettger2009}                                                       & ---                           & 1.5\,K                                & 7\,T               & 73\,Hz                                                  & Negligible                                                                                                 \\
        \textErYO{}~\cite{Fukumori2020}                                                                & ---                           & $<$100\,mK                            & 0.7\,T             & 580\,Hz                                                 & $\sim$2\,kHz over 1\,ms                                                                                    \\
        Er-doped silica fibre~\cite{Rasekh2026}                                                        & ---                           & 7\,mK                                 & 0.09\,T            & $\sim$8\,kHz                                            & ${\GammaSD{}=38\,}$kHz, ${R=1\,}$kHz                                                                       \\ \bottomrule[1.3pt]
    \end{tabularx}
}
\clearpage{}%

\title{Supplementary Information: Millisecond optical coherence and strong collective coupling in an integrated telecom rare-earth photonic platform}
\author{Kah Jen Wo\,\orcidlink{0000-0003-3806-9233}}\email{kjwo@u.nus.edu}
\affiliation{Centre for Quantum Technologies, Queenstown 117543, Singapore}

\author{Pavel A. Dmitriev\,\orcidlink{0009-0006-9681-9409}}\email{pavel.a.dmitriev@nus.edu.sg}
\affiliation{Centre for Quantum Technologies, Queenstown 117543, Singapore}
\affiliation{National University of Singapore, Department of Materials Science and Engineering, Singapore}

\author{Karthik Dasigi\,\orcidlink{0009-0007-8127-2292}}%
\affiliation{Centre for Quantum Technologies, Queenstown 117543, Singapore}

\author{Fumiya Hanamura\,\orcidlink{0000-0002-2382-4593}}%
\affiliation{Centre for Quantum Technologies, Queenstown 117543, Singapore}

\author{Steven Touzard\,\orcidlink{0000-0002-3475-2839}}\email{steven.touzard@nus.edu.sg}
\affiliation{Centre for Quantum Technologies, Queenstown 117543, Singapore}
\affiliation{National University of Singapore, Department of Physics, Singapore}
 
\maketitle
\tableofcontents

\clearpage
\section{Sample preparation: Photonic circuit design considerations}\label{sec:lnoi-design}

We designed our thin film lithium niobate (TFLN) photonic circuits for high-${Q}$, single-mode operation while bonded to calcium~tungstate~(\textCaWO{}). Since adding \textCaWO{} on top of the TFLN waveguide decreases the index contrast and reduces mode confinement, radiative losses due to bending is a major limiting factor to the geometry of the resonator. Our optimised design for the ring resonator has a width of 4${\unitSpacing{}}$µm and radius of 250${\unitSpacing{}}$µm, which, while supporting many modes before bonding, is effectively single-mode after bonding to \textCaWO{}.

Outside of the ring resonator and the pulley coupler, all waveguide bends were defined as \textit{Euler} bends~\cite{bahadori2019universal} to minimise mode mismatch and bending losses.

Fabrication limitations were taken into account when designing the ring to bus waveguide coupler and grating couplers for the optical fibres.
Lithium niobate waveguide fabrication is generally limited by sidewall roughness causing scattering losses, so when designing resonators, care needs to be taken when choosing between a deeper etch for better mode confinement but higher sidewall losses, or a shallower etch with lower sidewall losses and lower confinement (higher radiative bending losses)~\cite{gao2023compact}.
Our best results were with a 250${\unitSpacing{}}$nm deep etch of a 500${\unitSpacing{}}$nm lithium niobate film.
Because our fabrication process is limited to sidewall angles of about 60\textdegree{}, we are limited to a minimum spacing between features of 250${\unitSpacing{}}$nm, which requires a further buffer owing to redeposition affecting nearby structures.

For coupling the ring to the bus waveguide, we opted for a pulley coupler design, where coupling strength can be controlled by both the gap between the ring and bus waveguides and by the wrapping angle of the pulley coupler.
Our chosen gap for best control over coupling within fabrication limits was 1.1${\unitSpacing{}}$µm.
For our wrapping angle of the pulley coupler, we chose 30\textdegree{}.
For best control of the coupling, we decreased the width of the bus waveguide relative to the ring resonator waveguide to match the group velocity of the fundamental TE\textsubscript{0} mode in the resonator and in the bus waveguide~\cite{dai2011novel}. This also provides additional mode selectivity by decreasing coupling to all other modes. With these design decisions and nanofabrication optimisations described in \cref{sec:lnoi-fab}, we were able to consistently fabricate ring resonators with high quality factors. The measured intrinsic and extrinsic quality factors of the device from the main text, \textSampleA{}, are shown in \cref{fig:quality-factor-comparison-fits}.

For the grating coupler, optimising for maximum transmission at the \textErCaWO{} ${\ket{\downarrow}_{Z_1}\!\leftrightarrow\!\ket{\downarrow}_{Y_1}}$ optical transition near 1532.63${\unitSpacing{}}$nm yielded a grating pitch of 964${\unitSpacing{}}$nm with a fill factor of 21\%, meaning 202${\unitSpacing{}}$nm wide ridges with a gap of 764${\unitSpacing{}}$nm. The shape of the grating was chosen to linearly expand over 600${\unitSpacing{}}$µm from the bus waveguide width to 15${\unitSpacing{}}$µm to preserve the modes in the waveguide (i.e., minimise mode hybridisation) while expanding them to better match the modes of the optical fibres.

When designing the electrodes for electro-optic tuning of the resonator, we needed to take several factors into account~\cite{li2020lithium, wang2018nanophotonic, guarino2007electro}.
First, the electrodes need to be aligned to the $r_{33}$ electro-optic component of lithium niobate, in-plane in our X-cut TFLN samples.
We limit the pulley coupler wrapping angle and align the electrodes so that we maximise the electro-optic tuning.
Next, we maximised the length of the resonator covered by electrodes, without affecting the coupling region.
The electrodes are 15${\unitSpacing{}}$µm wide, and we chose a comparatively large edge-to-edge gap of 4${\unitSpacing{}}$µm between each electrode and the waveguide to limit electrode-induced optical losses, at the cost of a weaker field strength, while leaving enough leeway for fabrication imperfections.
Lastly, we chose an electrode thickness of 150${\unitSpacing{}}$nm to ensure most of the electric field passes through the rib waveguide, while still being compatible with our bonding process.
Since our measurement process did not involve fast tuning of the resonator, capacitance of the electrodes was not considered in the design process. Adding electrodes to the \textSampleA{} ring resonator caused the intrinsic quality factor to decrease by a factor of 2 (\cref{fig:quality-factor-comparison-fits}b).

\begin{figure}[!h]
    \centering
    \includegraphics{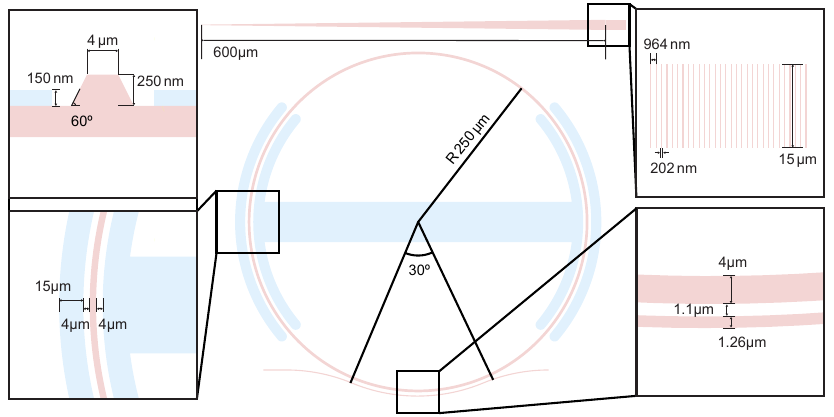}
    \caption{
        \textbf{Dimensions of photonic circuit.}
        Schematic top view of ring resonator, pulley coupler, electrodes and tapered grating coupler. Insets, clockwise from top left: Cross-sectional view of resonator and electrodes, close up view of grating coupler, close up view of ring resonator and pulley coupler, close up view of electrodes and ring resonator.
    }
    \label{fig:lnoi-design}
\end{figure}%

\section{Sample preparation: Thin-film lithium niobate fabrication}\label{sec:lnoi-fab}

Samples were fabricated out of 500${\unitSpacing{}}$nm X-cut thin-film lithium niobate on insulator (LNOI) chips, purchased from~\texttt{NANOLN}.
Patterning was done using hydrogen silsesquioxane (HSQ) resist with electron beam lithography (EBL) for the ring resonator and waveguides and photolithography for the electrodes and bonding mesa.
A total of four lithography steps were used to create the final chip.
To align the lithography masks, we first added marks at the corners of the chip using a bilayer \texttt{AZ1512}~/~\texttt{LOR-5A} resist stack, e-beam deposition of a 100${\unitSpacing{}}$nm Ti layer and subsequent lift-off in \texttt{Remover PG}.
Titanium was chosen because it provides enough contrast for EBL alignment, yet can be easily removed with hydrofluoric acid (HF), which is used during one of the final cleaning steps in the fabrication process.

Most of the complexity of the nanofabrication process lay in the development of lithium niobate (LN) patterning and etching.

The final step of the fabrication process involves \textit{direct bonding} of an erbium-doped \textErCaWO{} to the finished LNOI photonic chip (\cref{sec:direct-bonding}).
Therefore, it is important to retain as much unetched LN as possible to increase the bonding area and bond strength. We grew a 600${\unitSpacing{}}$nm layer of \textQuartz{} using plasma-enhanced chemical vapour deposition (PECVD) to use as a hard mask during the subsequent etching of LN.
Next, we defined trenches in the \textQuartz{} for the waveguides and electrodes.
We wrote the trench pattern using \texttt{AZ1512} photoresist with an \texttt{AR 300-80 new} adhesion promotion layer, exposed by 405${\unitSpacing{}}$nm light and developed by \texttt{MIF319} developer and then transferred the pattern to the \textQuartz{} by a wet chemical etch using 1\% w/w HF, taking care not to overetch to prevent damage to the LN layer by the HF~\cite{yu2024poling}.

Next, we wrote the waveguides and rings with EBL using HSQ (\texttt{FOx-16}) resist.
A 600${\unitSpacing{}}$nm-thick layer of HSQ was spin coated onto the chips to create a sufficiently thick mask for reactive ion etching (RIE).
Exposure was done using EBL with a total dose of 2000${\unitSpacing{}}$µC/cm$^2$ for the rings and waveguides, and 1800${\unitSpacing{}}$µC/cm$^2$ for the denser grating coupler features.
The pattern was developed using 25\% w/w Tetramethylammonium Hydroxide (TMAH) to ensure high contrast and well-defined nanogaps especially in the grating couplers and ring to bus waveguide coupling region~\cite{deotare_nanobeam_2012,choy2013nanophotonic}.
Special care was taken when writing the rings themselves in EBL: we made sure that the rings fit into one exposure field to eliminate any stitching errors that would contribute to scattering losses.
Additionally, the whole exposure was split into 4 passes with slight offsets (physical stage movement compensated by beam deflection within the write field) to smooth out any staircasing caused by the discrete nature of the EBL patterning process, especially when approximating smooth curved lines.
Moreover, proximity effect correction was crucial as it accounts for electron backscattering due to the thick resist layer and the largely non-conductive LNOI chip.

Following the patterning, we etched the LN using inductively coupled plasma reactive ion etching (ICP-RIE).
LN is notoriously difficult to etch using RIE because standard fluorine- or chlorine-based plasmas either leave behind insoluble residues or drastically increase surface roughness, making pure argon plasma the best option for high-quality etching~\cite{shen2021comparative}.
A pure argon plasma etch, being a pure physical etch, leads to redeposition of the etched material~\cite{kumar2024optimization}.
The physical nature of the etch process also limits the etch selectivity, in our case never going above 1.2:1~(LN : \textQuartz{}), hence the requirements for thick masks.
Optimisation of the RIE power and pressure can decrease the amount of redeposited material~\cite{https://doi.org/10.4218/etrij.2024-0137, 10.1515/nanoph-2022-0676, yang2021low}, but cannot completely solve the problem. Thus, the remaining redeposited material needs to be removed with a post-etch cleaning process. SC-1 solution (De-ionised water : Ammonium Hydroxide : Hydrogen Peroxide) is currently one of the most commonly used cleaning agents to remove redeposited LN from the chip~\cite{https://doi.org/10.4218/etrij.2024-0137, kumar2024optimization, ulliac_argon_2016, holzgrafe_cavity_2022}. Varying concentrations, temperatures and durations have been proposed for cleaning, and in our case a 2:1:1 v/v concentration at room temperature for 1~hour yielded the best results in removing redeposited material.
Care must be taken not to use overly aggressive SC-1 conditions, since it will also attack the unetched LN itself -- and has been used as a stand-alone wet etchant of LN~\cite{zhuang2023high}.
Additional steps in the cleaning process involved using HF to strip the oxide hardmasks and piranha solution to clear any organic contaminants.
Another consideration for RIE etch parameters is the effect of RIE power on the resulting sidewall angles, where higher RIE power generally produces more vertical sidewalls~\cite{yang2021low}, but in our case would lead to the deterioration of the hard mask and worse overall etch outcomes.
Our optimal conditions (150${\unitSpacing{}}$W RF and 1500${\unitSpacing{}}$W ICP, 40${\unitSpacing{}}$sccm Ar at 2${\unitSpacing{}}$mTorr) resulted in 60\textdegree{} sidewalls with a 50${\unitSpacing{}}$nm/min etch rate and we etched our waveguides to a depth of 250${\unitSpacing{}}$nm (controlled by ellipsometric measurements before and after the etch process) in a single continuous etch.

The \texttt{NANOLN} fabrication process to create LNOI wafers itself introduces defects into the LN layer and additional defects are introduced during the argon etch and post-etch wet chemical processing. These defects can be repaired by an annealing process. After transferring the ring patterns to the LN and cleaning any redeposition residues, we anneal our chips for 1~hour at 520\textdegree{}C in a pure oxygen atmosphere -- repairing any deep defects from the helium implantation during the \texttt{NANOLN} fabrication process and new surface defects introduced by our own processing, while the oxygen atmosphere prevents lithium depletion in the surface layer~\cite{shams_ansari_integrated_2022,holzgrafe_cavity_2022,leidinger_influence_2016,shams-ansari_reduced_2022,zhang_electronically_2019,desiatov_ultra-low-loss_2019,shams-ansari_thin-film_2022,sosunov2024impact}.

Finally, we added electrodes to the rings for electro-optic tuneability again using a bilayer \texttt{AZ1512} / \texttt{LOR-5A} resist stack, e-beam deposition of a Ti/Al (10${\unitSpacing{}}$nm / 150${\unitSpacing{}}$nm) electrode layer and subsequent lift-off in \texttt{Remover PG}. We chose aluminium over gold for the electrode material to prevent formation of intermetallic compounds with the aluminium wires used for wirebonding during the final packaging step, which might affect the structural stability of the bonds during thermal cycling~\cite{xu2024evolution}.

\begin{figure}[!h]
    \centering
    \includegraphics{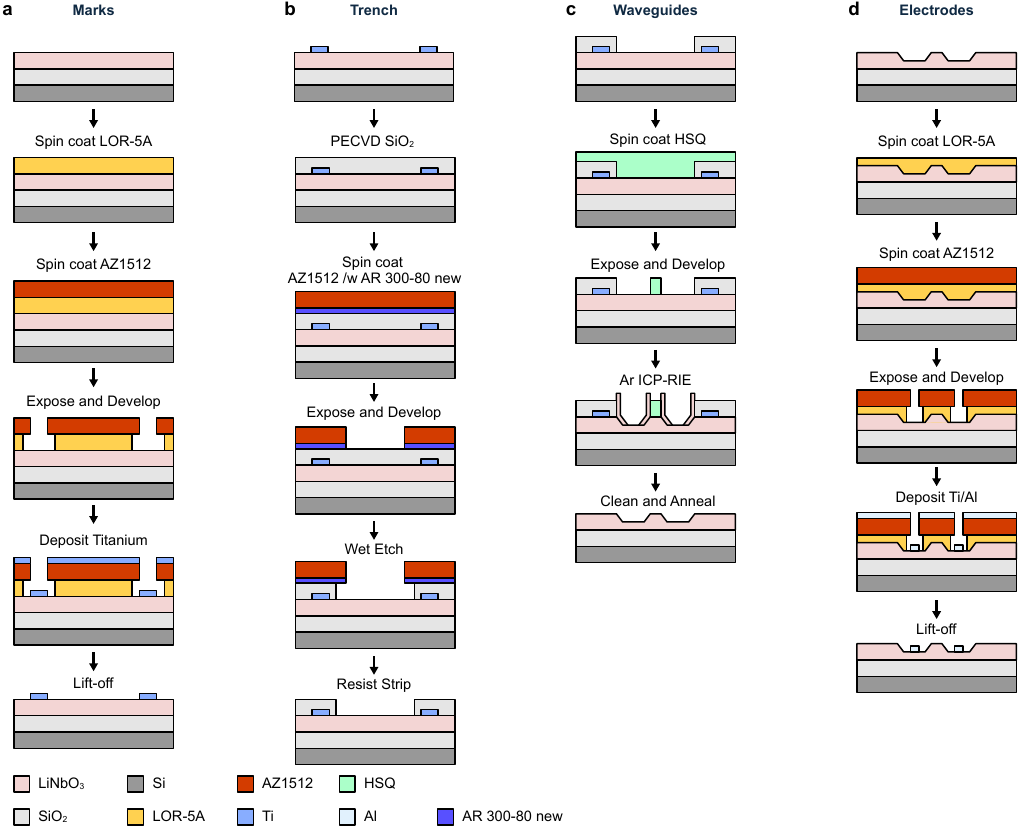}
    \caption{
        \textbf{Thin-film lithium niobate nanofabrication flow schematic.}
        Schematic cross-sectional illustration of the four lithography steps used to fabricate the TFLN photonic chip.
        \textbf{a},~Alignment marks: a bilayer LOR-5A/AZ1512 resist stack is patterned via 405${\unitSpacing{}}$nm photolithography, followed by e-beam evaporation of 100${\unitSpacing{}}$nm Ti and lift-off.
        \textbf{b},~Bonding mesa and trench definition: a 600${\unitSpacing{}}$nm PECVD \textQuartz{} hard mask layer is deposited and then defined by photolithography and wet etching in 1\% HF to define trenches for the waveguides and electrodes.
        \textbf{c},~Waveguide and ring resonator patterning: HSQ (FOx-16) resist is exposed by electron beam lithography, developed in 25\% TMAH, and the pattern is transferred into the LN layer by Ar ICP-RIE dry etching. Post-etch cleaning (SC-1 and piranha) removes redeposited material, followed by annealing at 520\textdegree{}C in O\textsubscript{2} to repair etch-induced and implantation defects.
        \textbf{d},~Electrode fabrication: a bilayer LOR-5A/AZ1512 resist stack is patterned via 405${\unitSpacing{}}$nm photolithography, followed by e-beam evaporation of Ti/Al (10${\unitSpacing{}}$nm/150${\unitSpacing{}}$nm) and lift-off to form the electrodes for electro-optic tuning.
    }
    \label{fig:lnoi-fab-process}
\end{figure}

The detailed fabrication process flows for the samples are presented in \cref{tab:lnoi-fab-flow}. Corresponding schematics are shown in \cref{fig:lnoi-fab-process}.
\begin{table}[!h]
    \centering
    \begin{tabularx}{0.97\linewidth}{X@{\hspace{0.01\textwidth}}X}
        \begin{definition}\label{fabflow:alignment-marks}
            \textbf{1. Alignment marks}
            \begin{enumerate}
                    \itemsep-0.12em
                \item Wash (Acetone / Isopropyl Alcohol (IPA) / De-ionised water (DI)).
                \item Bake for 5 minutes at 140\textdegree{}C.
                \item Spin coat LOR-5A, 60${\unitSpacing{}}$s, 4000~RPM.
                \item Bake for 2 minutes at 180\textdegree{}C.
                \item Spin coat AZ1512, 60${\unitSpacing{}}$s, 4000~RPM.
                \item Bake for 60${\unitSpacing{}}$s at 100\textdegree{}C.
                \item Expose, 405${\unitSpacing{}}$nm with a dose of 50${\unitSpacing{}}$µC/cm$^2$.
                \item Develop in MIF319 (2.5\% TMAH) for 90${\unitSpacing{}}$s at room temperature (RT).
                \item DI stop.
                \item Deposit 100${\unitSpacing{}}$nm Ti using e-Beam evaporator.
                \item Lift-off using Remover PG at RT for 24~h.
                \item DI stop.
                \item Wash (Acetone / IPA / DI).
            \end{enumerate}
        \end{definition} &
        \begin{definition}\label{fabflow:ring-resonator}
            \textbf{3. Ring resonator}
            \begin{enumerate}
                    \itemsep-0.12em
                \item Wash (Acetone / IPA / DI / MIF319 / DI).
                \item Bake for 5 minutes at 140\textdegree{}C.
                \item Spin coat FOx-16 HSQ resist, 60${\unitSpacing{}}$s, 2500~RPM.
                \item Bake for 5~min at 80\textdegree{}C.
                \item Expose, EBL at $2$~nA with a total dose of 2000${\unitSpacing{}}$µC/cm$^2$.
                \item Develop 25\% w/w TMAH for $17$~s.
                \item Develop MIF319 for 60${\unitSpacing{}}$s.
                \item DI stop.
                \item Etch using ICP-RIE, Argon 40${\unitSpacing{}}$sccm, 2${\unitSpacing{}}$mTorr, 1500${\unitSpacing{}}$W ICP / 150${\unitSpacing{}}$W RF.
                \item Strip Ti, HSQ and \textQuartz{} in 1\% HF for 10~min.
                \item Clean using 3:1 v/v Piranha solution (96\% Sulphuric Acid : 30\% Hydrogen Peroxide) for 30~min, no external heating.
                \item Clean using 2:1:1 v/v/v SC-1 solution (DI : 30\% Ammonium Hydroxide : 30\% Hydrogen Peroxide) for 60~min, no external heating.
                \item DI stop.
                \item Anneal at 520\textdegree{}C in a pure Oxygen atmosphere for 1~h.
            \end{enumerate}
        \end{definition} %
    \end{tabularx}
    \begin{tabularx}{0.97\linewidth}{X@{\hspace{0.01\textwidth}}X}
        \begin{definition}\label{fabflow:bonding-mesa-and-trench}
            \textbf{2. Bonding mesa}
            \begin{enumerate}
                    \itemsep-0.12em
                \item Wash (Acetone / IPA / DI).
                \item Bake for 5 minutes at 140\textdegree{}C.
                \item Deposit 600${\unitSpacing{}}$nm of \textQuartz{} using PECVD, 300\textdegree{}C, N$_2$O 460${\unitSpacing{}}$sccm, SiH$_4$ 3${\unitSpacing{}}$sccm, 20${\unitSpacing{}}$W RF.
                \item Spin coat AR 300-80 new, 60${\unitSpacing{}}$s, 6000~RPM.
                \item Bake for 2 minutes at 180\textdegree{}C.
                \item Spin coat AZ1512, 60${\unitSpacing{}}$s, 4000~RPM.
                \item Bake 90${\unitSpacing{}}$s at 115\textdegree{}C.
                \item Expose, 405${\unitSpacing{}}$nm with a dose of 75${\unitSpacing{}}$µC/cm$^2$.
                \item Develop in MIF319 for 30${\unitSpacing{}}$s at RT.
                \item Etch \textQuartz{} in 1\% w/w Hydrofluoric Acid (HF) for 10~min.
                \item DI stop.
                \item Strip resist in Remover PG at RT for 24~h.
            \end{enumerate}
        \end{definition} &
        \begin{definition}\label{fabflow:electrodes}
            \textbf{4. Electrodes}
            \begin{enumerate}
                    \itemsep-0.12em
                \item Wash (Acetone / Isopropyl Alcohol (IPA) / De-ionised water (DI)).
                \item Bake for 5 minutes at 140\textdegree{}C.
                \item Spin coat LOR-5A, 60${\unitSpacing{}}$s, 4000~RPM.
                \item Bake for 2 minutes at 180\textdegree{}C.
                \item Spin coat AZ1512, 60${\unitSpacing{}}$s, 4000~RPM.
                \item Bake for 60${\unitSpacing{}}$s at 100\textdegree{}C.
                \item Expose, 405${\unitSpacing{}}$nm with a dose of 50${\unitSpacing{}}$µC/cm$^2$.
                \item Develop in MIF319 (2.5\% TMAH) for 90${\unitSpacing{}}$s at room temperature (RT).
                \item DI stop.
                \item Deposit 10${\unitSpacing{}}$nm Ti and 150${\unitSpacing{}}$nm Al using e-Beam evaporator.
                \item Lift-off using Remover PG at RT for 24~h.
                \item DI stop.
                \item Wash (Acetone / IPA / DI).
                \item Bake for 5 minutes at 180\textdegree{}C.
            \end{enumerate}
        \end{definition}%
    \end{tabularx}%
    \caption{
        \textbf{Thin-film lithium niobate nanofabrication flows.}
        Summary of the thin-film lithium niobate fabrication workflow, divided into four modules: alignment mark definition, bonding mesa preparation, ring resonator patterning and electrode deposition.
    }
    \label{tab:lnoi-fab-flow}
\end{table}%
\clearpage

\section{Sample preparation: Direct bonding}\label{sec:direct-bonding}

\begin{figure}[!b]
    \centering
    \includegraphics{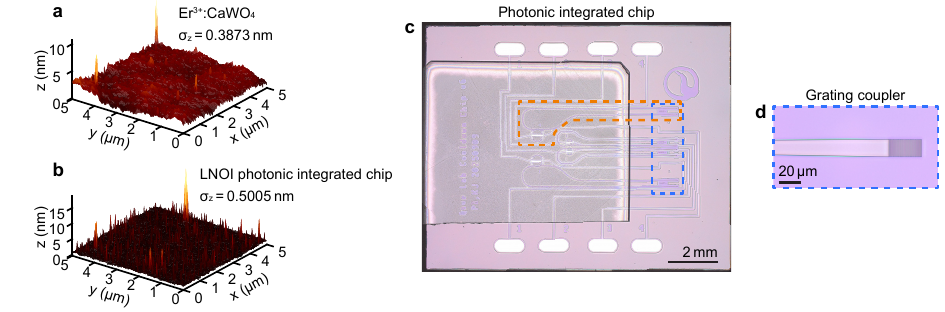}
    \caption{
        \textbf{Bonding surface characterisation and bonded photonic integrated chip.}
        \textbf{a},
        \textbf{b},~Atomic force microscope scans of a \textErCaWO{} crystal surface (root-mean-square roughness ${\sigma_\text{z}=0.3873\unitSpacing{}}$nm) and the bonding interface of a patterned thin-film lithium niobate-on-insulator photonic integrated chip (${\sigma_\text{z}=0.5005\unitSpacing{}}$nm)
        respectively.
        Both values satisfy the sub-nanometre roughness requirement for direct bonding.
        \textbf{c},~Top-down optical microscope image of the thin-film lithium niobate photonic integrated chip bonded with \textErCaWO{}.
        The nanophotonic structures corresponding to the ring resonator characterised in the main text (\textSampleA{}) are outlined in dashed orange lines, and the grating couplers in dashed blue lines.
        \textbf{d},~Zoomed-in optical microscope image of a grating coupler on the same chip.
    }
    \label{fig:optical-microscope-image}
\end{figure}

Successful direct bonding requires that the root-mean-square (RMS) surface roughness of both bonding interfaces remains below 1${\unitSpacing{}}$nm.
To meet this requirement, we used an 8.5$\times$8.5$\times$0.5$\unitSpacing{}$mm$^3$ 50${\unitSpacing{}}$ppm \textErCaWO{} chip (\texttt{SurfaceNet~GmbH}) that was epi-polished on both sides.
While only a single polished surface is required for bonding, this dual-sided polish provides optical transparency through the chip, a feature critical for verifying bond integrity and microring resonator coverage via optical microscopy.
The surface of a representative, as-received sample was characterised with a \texttt{Bruker~Dimension~Icon} atomic force microscope (AFM), yielding an RMS surface roughness below 1${\unitSpacing{}}$nm (see \cref{fig:optical-microscope-image}a).
Similarly, a representative patterned LNOI chip was characterised and also found to have an RMS surface roughness below 1${\unitSpacing{}}$nm (see \cref{fig:optical-microscope-image}b).
As the characterisation was performed in a standard laboratory setting rather than a cleanroom, this value represents a conservative upper bound on the intrinsic material roughness, inclusive of any potential surface particulates.

To allow the crystal host to fit onto the patterned photonic chip (see \cref{fig:optical-microscope-image}c), the crystal host was mechanically diced to $\sim$8.5$\times$6.5$\unitSpacing{}$mm$^2$ with an \texttt{Accretech~SS20} dicer.
\begin{figure}[!h]
    \centering
    \includegraphics{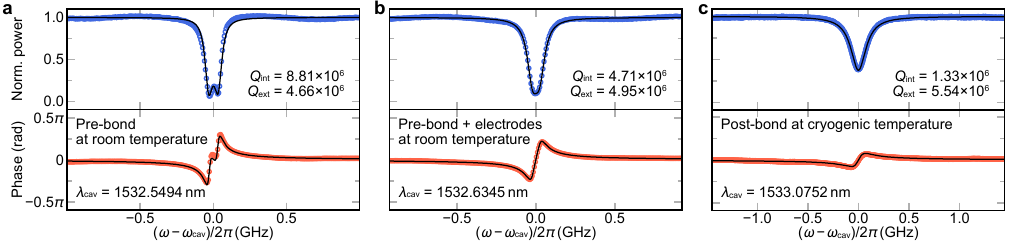}
    \caption{
        \textbf{Complex transmission spectrum of the characterised microring resonator.}
        ~Normalised transmitted power and phase response of the \textSampleA{} ring resonator device measured with our heterodyne setup (see \cref{sec:setup-heterodyne}) along with the complex fits for
        \textbf{a}, before bonding and without electrodes,
        \textbf{b}, before bonding and with electrodes, and
        \textbf{c}, after bonding. The measurements before bonding were done at room temperature and atmospheric conditions. The measurement post-bonding was done at cryogenic temperature and high vacuum.
        The intrinsic and extrinsic quality factors ${Q_\text{int}=\omega_\text{cav}/\kappa_\text{int}}$ and ${Q_\text{ext}=\omega_\text{cav}/\kappa_\text{ext}}$ were obtained by fitting the complex transmission to the coupled-mode model of Ref.~\cite{Puckett2021}, which includes the mode splitting from coupling between the forward- and backward-propagating modes.
    }
    \label{fig:quality-factor-comparison-fits}
\end{figure}
The diced \textErCaWO{} together with the patterned photonic chip were then cleaned with ultrasonication in acetone for approximately 5 minutes, followed by another 5 minutes of ultrasonication in isopropyl alcohol in a class-10~cleanroom.
Contact with the bonding interfaces of the \textErCaWO{} chip and the photonic chip was minimised by using a polytetrafluoroethylene (PTFE) wafer rack that allows the chips to be slotted in vertically during the ultrasonication bath.
After the ultrasonication cleaning, the chips were rinsed with de-ionised water and blow dried with a nitrogen gun.

We prepared two samples using different bonding agents: \textSampleA{} and \textSampleB{}.
The \textSampleA{}, whose results are presented in the main text, was bonded using de-ionised (DI) water, while \textSampleB{} was bonded using 30\% (w/w) ammonium hydroxide (\textAmmonia{}) in water. This surface-activating method is based on Ref.~\cite{Kang2024}.
Apart from the choice of bonding agent, the bonding procedure was identical for both samples and proceeded as follows.

The photonic chip was placed with the bonding interface facing upwards on a flat surface in a class-100 cleanroom.
A microdroplet of the bonding agent (DI water or \textAmmonia{}, as appropriate) was placed near the ring resonators by eye using a PTFE tweezer that had been dipped into the corresponding liquid.
The tweezer itself was cleaned with isopropyl alcohol ultrasonication prior to this step, and during the placement of the microdroplet, the tweezer never physically touched the photonic chip.
After the droplet was placed, the \textErCaWO{} was flipped and placed gently onto the droplet.
Without any externally applied force, the liquid was allowed to evaporate, and direct bonding between the \textErCaWO{} and the photonic chip formed.
The chips were handled exclusively using polyetheretherketone (PEEK) tweezers, which we found to minimise chipping of the \textErCaWO{} crystal compared to ceramic or metal tweezers.

The bonded stack was stored at room temperature in a nitrogen desiccator for 24 hours to allow for full evaporation of the bonding agent and for the bonds to strengthen.
The bonded stack was then annealed on a hotplate at 100\textdegree{}C under atmospheric conditions.
\textSampleB{} was annealed for approximately 12 hours and loaded first into the dilution refrigerator, serving as the pathfinder device on which the measurement protocols and thermalisation procedures were developed and optimised.
\textSampleA{} remained on the hotplate during this period, accumulating an anneal time of approximately 1~month; shear-test measurements on a control sample and thermal cycling of \textSampleB{} confirmed that the shorter anneal produces bond strengths sufficient for repeated cooldowns to millikelvin temperatures (see \cref{sec:shear-test}).
The experience gained with \textSampleB{} informed the thermalisation and mounting procedures used for \textSampleA{}, which is the primary device reported in the main text.
After passively cooling back down to room temperature, \textSampleA{} was ready to be loaded into our Bluefors LD400 dilution refrigerator (see \cref{sec:dilution-fridge-setup}).
For a quantitative study of the bonding strength using shear test, see \cref{sec:shear-test}.
The optical microscope image in \cref{fig:optical-microscope-image}c corresponds to \textSampleA{}.

We characterised the ring resonator using our heterodyne setup (see \cref{sec:setup-heterodyne}), and observed that the intrinsic quality factor $Q_{\text{int}}$ decreased after bonding.
For the \textSampleA{} ring resonator near the ${\ket{\downarrow}_{Z_1}\!\leftrightarrow\!\ket{\downarrow}_{Y_1}}$ optical transition, $Q_\text{int}$ decreased from ${8.8\times10^{6}}$ to ${4.7\times10^{6}}$ after adding the electrodes, and to ${1.3\times10^{6}}$ after bonding, measured at cryogenic temperature (\cref{fig:quality-factor-comparison-fits}).
Mode simulations predict a negligible absorption loss from the electrodes at the chosen gap (see \cref{sec:lnoi-design}), so we attribute the reduction upon adding the electrodes to the electrode fabrication process.
We attribute the decrease of the intrinsic quality factor after bonding to increased participation of the optical mode with defects near the bonding interface or in the crystal, as well as mechanical stress arising from thermal-expansion mismatch, leading to additional losses.
In addition, over the course of cooling both \textSampleA{} and \textSampleB{} down to $\sim$4$\unitSpacing{}$K in our Bluefors LD400 dilution refrigerator, we observed a non-monotonic behaviour in the intrinsic quality factor of the bare ring resonator, which we attribute to slippage of the \textErCaWO{} that is on top of the ring resonator (see \cref{sec:shear-test}).
Despite the decrease in intrinsic quality factor, we show that in \cref{sec:lumerical-filling-factor} we can still achieve collective cooperativity $C>1$. At temperatures lower than $\sim$4$\unitSpacing{}$K, the intrinsic quality factor stabilises at $\sim$1 million for \textSampleA{}.

\section{Experimental setup in the dilution refrigerator}\label{sec:dilution-fridge-setup}
\begin{figure}[!h]
    \centering
    \includegraphics{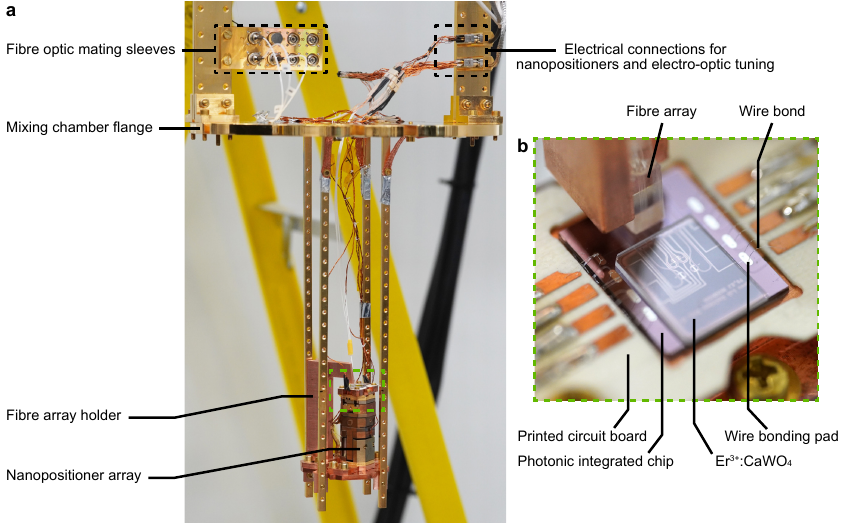}
    \caption{
        \textbf{Experimental setup in a dilution refrigerator.}
        \textbf{a},~Photograph of the photonic integrated chip at the mixing chamber flange of a Bluefors LD400 dilution refrigerator, showing the fibre array holder, nanopositioner array, fibre optic mating sleeves, and electrical connections.
        \textbf{b},~Close-up photograph of the photonic integrated chip mounted on a printed circuit board, showing the fibre array for optical coupling to photonic integrated chip, the bonded \textErCaWO{} crystal, and aluminium wire bonds for electro-optic tuning.
    }
    \label{fig:dilution-fridge-setup}
\end{figure}

To couple telecom C-band light into our Bluefors LD400 dilution refrigerator, we employed SMF-28 single mode fibres (Corning).
These were connected via mating sleeves at the mixing chamber to a fibre array unit for coupling to the on-chip grating couplers (see \cref{fig:dilution-fridge-setup}).
To minimise the passive thermal load on the mixing chamber and enable operation at base temperature, the pigtails of the fibre array were shortened to the required length by splicing with a \texttt{Fujikura~45S} splicer.
The fibre array was then bonded to a copper holder using the low-outgassing UV-cured \texttt{Norland~Optical~Adhesive~NOA~88}.
To enable \textit{in situ} optimal fibre-to-chip alignment at cryogenic temperatures, the chip was mounted on an \texttt{Attocube} piezo-based nanopositioner array \texttt{ANPx101} (2$\times$), and \texttt{ANPz102} (1$\times$).
The integrity of the fibre array assembly has been validated through five full thermal cycles from room temperature to the cryostat's base temperature, without degradation in alignment stability or coupling efficiency (${\sim}$15.3\% efficiency per facet).

To fix the integrated photonic chip in place so that it does not slip during position adjustments, we applied \texttt{Apiezon~N~grease} to the copper sample holder that is on the nanopositioner array before affixing the integrated photonic chip onto it. We chose this grease due to its ability to maintain good thermal conductivity in cryogenic environments.
To control the electro-optic tuning of the ring resonator, we used a \texttt{Westbond~7674E} to wire bond the chip to a \texttt{Rogers~RO4350B} printed circuit board (PCB) with aluminium wires. This PCB was chosen for its high thermal conductivity.
The footprint of the setup was designed such that it fits within the bore of an \texttt{American~Magnetics~Inc.} 3-dimensional vector magnet. All copper used in our experimental setup is oxygen-free (C101) to maximise thermal conductivity and minimise outgassing.

\section{Electro-optic tuning}\label{sec:eo-tuning}

\begin{figure}[!h]
    \centering
    \includegraphics{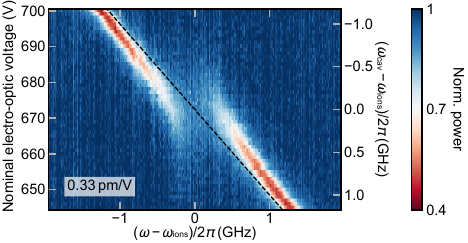}
    \caption{
        \textbf{Electro-optic tuning and fit.}
        Measured transmitted power of \textSampleA{} at ${\absB{}=1\unitSpacing{}}$T, normalised to the off-resonant
        background, versus probe detuning from the erbium ${{\ketdownz}{\leftrightarrow}{\ketdowny}}$ transition and the nominal voltage applied to the electrodes.
        Here, ${\omega/2\pi}$ is the probe frequency, and ${\omega_\text{ions}/2\pi}$ is the ${{\ketdownz}{\leftrightarrow}{\ketdowny}}$ transition frequency.
        Dashed line: fitted linear dependence of resonator frequency ${\omega_\text{cav}/2\pi}$ against nominal voltage. Its slope is the electro-optic tuning coefficient, ${\approx}0.33\unitSpacing{}$pm/V, equivalently ${\approx}42\unitSpacing{}$MHz/V, referred to the nominal voltage; the voltage across the ring resonator was not measured.
        This measurement was performed at a nominal mixing chamber temperature ${\sim}$8${\unitSpacing{}}$mK.
    }
    \label{fig:electro-optic-coefficient-plot}
\end{figure}

The central electrode inside the ring is electrically floating, so that each side of the ring experiences half of the applied voltage.
The largest nominal voltage was 700${\unitSpacing{}}$V, corresponding to a tuning range of about 29${\unitSpacing{}}$GHz (\cref{fig:electro-optic-coefficient-plot}), about one third of the 86${\unitSpacing{}}$GHz free spectral range of the ring; higher voltages are limited by the dielectric breakdown of the DC lines of the dilution refrigerator.
Voltages of this magnitude were applied only at cryogenic temperature in high vacuum.
The voltage was ramped at 0.5${\unitSpacing{}}$V/s after amplification, corresponding to about 21${\unitSpacing{}}$MHz/s, and passed through a resistor--capacitor low-pass filter (56${\unitSpacing{}}$k$\Omega$, 66${\unitSpacing{}}$nF) with a cutoff of 43${\unitSpacing{}}$Hz.
After a change of voltage, the resonance drifted slowly over hours, and returning to the starting voltage showed hysteresis.
We therefore swept the voltage in one direction only and acquired each spectrum, which takes about a minute; no drift was observed on this timescale.
The probe power of 0.03${\unitSpacing{}}$µW at the cryostat input is about 37${\unitSpacing{}}$dB below a power at which we observed no photorefractive effects in a ring resonator with a radius of 40${\unitSpacing{}}$µm.

\section{Measurement setups}\label{sec:setup-heterodyne}

\begin{figure}[!h]
    \centering
    \includegraphics{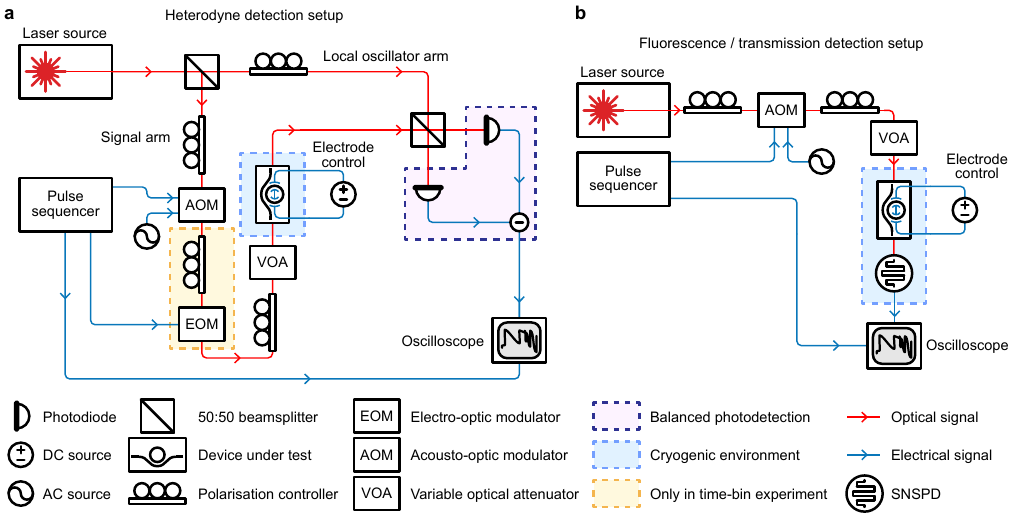}
    \caption{
        \textbf{Optical measurement setups.}
        \textbf{a}, Heterodyne detection setup based on a free-running Mach-Zehnder interferometer for optical vector analysis. The laser output is split by a 50:50 beamsplitter into a local oscillator (LO) arm and a signal arm.
        The signal arm is frequency-shifted by an acousto-optic modulator (AOM) and attenuated by a variable optical attenuator (VOA) before passing through the device under test (DUT) inside the cryogenic environment.
        The transmitted signal is recombined with the LO arm at a balanced photodetector, and the resulting electronic signal is recorded on an oscilloscope.
        An electro-optic modulator (EOM) and polarisation controllers are included in the signal path for the coherent storage and retrieval experiment.
        \textbf{b}, Fluorescence and transmission detection setup. The laser output passes through an AOM and VOA before being routed to the DUT in the cryogenic environment. Transmitted or fluorescence signals are detected by superconducting nanowire single-photon detectors (SNSPDs) and read out on an oscilloscope.
        A pulse sequencer controls the timing of optical pulse gating and data acquisition in both configurations.
        Electrode control provides electrical biasing of the DUT via a DC source.
    }
    \label{fig:detection-schemes}
\end{figure}

We perform optical vector analysis using a free-running Mach-Zehnder interferometer setup based on Ref.~\cite{Dasigi2026} which enables phase-sensitive detection, as shown in \cref{fig:detection-schemes}a.
To acquire the complex transmission over a wavelength range, which is required for collective cooperativity extraction, we use \texttt{Santec~TSL-570} as our laser source, which enables fast wavelength sweep.
A wavelength sweep rate of 100${\unitSpacing{}}$nm/s is chosen such that phase fluctuations due to vibrations and thermal fluctuations in the optical fibres are negligible over a wavelength range of interest.
This signal is then split 50:50 into the local oscillator (LO) arm and signal arm.
The signal in signal arm is detuned by an acousto-optic modulator (AOM) at +200${\unitSpacing{}}$MHz, then passes through the device under test (DUT) before being detected at the balanced photodetector along with the LO arm. The AOM used is \texttt{SGTF200-1550-1T-1D}.
To control the input power, we use a \texttt{Santec~OVA-100} variable optical attenuator (VOA) to attenuate the signal going into the DUT.
The balanced photodetector (\texttt{Wieserlabs~WL-BPD1GA}) converts optical signal into electronic signal, which is then acquired through a \texttt{TeledyneLecroy~610Z} oscilloscope.
This electronic signal can then be demodulated either in software or in hardware, to extract the amplitude and phase components of the complex transmission.
Polarisation controllers were added in between components to match polarisation and maximise overall coupling efficiency.
For \textEr{} coherence, spectral diffusion, and inversion echo measurements, instead of a scanning laser, the laser source is replaced by an ultra-low-noise, narrow-linewidth, single-frequency \texttt{Precilaser~FL-SF-1533-S} laser.
Its linewidth as reported by Precilaser is ${<}{2}{\unitSpacing{}}$kHz at 100${\unitSpacing{}}$µs integration. The pulse sequencer~(\texttt{ArTiQ~2238~MCX-TTL}) gates the AOM to form optical pulses required for the three-pulse photon echo (3PPE) measurement.
To shift the phase of individual AOM-gated pulses in our coherent interference experiment, an \texttt{exail~MPX-LN-0.1} electro-optic modulator (EOM) is used.

To measure the fluorescence of the \textEr{} ion ensemble, we excite the ensemble via an AOM-gated optical pulse, and measure the fluorescence using a \texttt{Single~Quantum~SSPD-1550-70-80} superconducting nanowire single photon detector (SNSPD) mounted on the 4K flange of our Bluefors LD400 dilution refrigerator.
The SNSPD signals were counted and averaged on a \texttt{TeledyneLecroy~610Z} oscilloscope.
The measurement schematic is shown in \cref{fig:detection-schemes}b.
The extracted optical ${T_1}$ is later validated with inversion-recovery echo measurement.

\section{Thermal expansion contrast analysis and bond strength characterisation}\label{sec:shear-test}
To determine the appropriate \textCaWO{} crystal orientation for integration with LNOI that is robust to temperature changes due to post-bond annealing, we employ a numerical analysis using the multilayer strain framework from~Ref.~\cite{Weigel2017}, which captures three distinct contributions to the thermally induced strain: (i) the intrinsic thermal contraction of each individual material, (ii) the in-plane constraint imposed by coefficient of thermal expansion (CTE)-mismatched layers bonded in a common stack, and (iii) the additional deformation arising from curvature of the composite structure due to asymmetric thermal stresses.
Using this model, in which each layer has a thickness ${\ell}$ and width $w$, the strain in the $i$-th layer is given by~${\varepsilon_i=\alpha_i\Delta T+\tfrac{F_i}{\mathscr{E}_i \mathscr{A}_i}+\tfrac{\ell_i}{2\mathscr{R}}}$, where for each layer, ${\alpha_i}$ is the CTE, $F_i$ the thermally-induced axial force, ${\mathscr{E}_i}$ the Young's modulus, and ${\mathscr{A}_i=\ell_i w_i}$ the cross-sectional area.
${\Delta T}$ is the temperature difference and ${\mathscr{R}}$ the radius of common curvature.
We note that this model does not take into account any slippage of layers, and that elastic material response is obeyed throughout, with stress and strain related through the Hooke's law ${\sigma=\mathscr{E}\varepsilon}$.
By solving for the following system of linear equations for 4 layers: \textCaWO{}--\textLN{}--\textQuartz{}--Si
\begin{align}
    \textstyle{\sum}_{i=1}^{4}F_i    & =0, \label{eqn:stress-1}                                                                                                        \\
    \textstyle{\sum}_{i=1}^{4}h_iF_i & =\frac{\textstyle{\sum}_{i=1}^{4}\mathscr{E}_i{I}_i}{\mathscr{R}},                                         \label{eqn:stress-2} \\
    \alpha_j\Delta T+\frac{F_j}{\mathscr{E}_j \mathscr{A}_j}+\frac{\ell_j}{2\mathscr{R}}
    & = \alpha_{j+1}\Delta T+\frac{F_{j+1}}{\mathscr{E}_{j+1} \mathscr{A}_{j+1}}-\frac{\ell_{j+1}}{2\mathscr{R}},\label{eqn:stress-3}
\end{align}
we determine the total strain ${\varepsilon_i}$ in each layer.
When no external force is assumed, the sum of thermally-induced axial forces is given by \cref{eqn:stress-1}.
The moments due to those forces follow \cref{eqn:stress-2}, where ${I_i=\ell_{i}^{3} w_i/12}$ is the area moment of inertia, and ${h_i}$ is the distance from the bottom of the stack to the centre of the ${i}$-th layer.
The continuity of strain at each interface ${j=1,...,3}$ yields \cref{eqn:stress-3}.
The thicknesses in our bonded stack are ${\ell_{\text{\textCaWO{}}}=500\unitSpacing{}}$µm, ${\ell_{\text{LN}}=500\unitSpacing{}}$nm, ${\ell_{\text{\textQuartz{}}}=4.7\unitSpacing{}}$µm, and ${\ell_{\text{Si}}=525\unitSpacing{}}$µm.
The Young's moduli and CTEs are given in \cref{tab:thermal-analysis}.
For simplicity, the widths in the numerical analysis were set to 8${\unitSpacing{}}$mm.
\begin{table}[!b]
    \centering
    \extendedDataTableThermalAnalysis{}%
    \caption{
        \textbf{Mechanical and thermal parameters used in the numerical analysis of thermally induced stress.}
        Here, ${\mathscr{E}}$ denotes Young's modulus, and ${\alpha}$ denotes the coefficient of thermal expansion (CTE).
        (e,o) denotes the parameter values parallel to the extraordinary- and ordinary-axes of lithium~niobate.
        (${a}$,${c}$) denotes the parameter values parallel to the $a$- and $c$-axes of calcium~tungstate.
    }
    \label{tab:thermal-analysis}
\end{table}%

\begin{figure}[!h]
    \centering
    \includegraphics{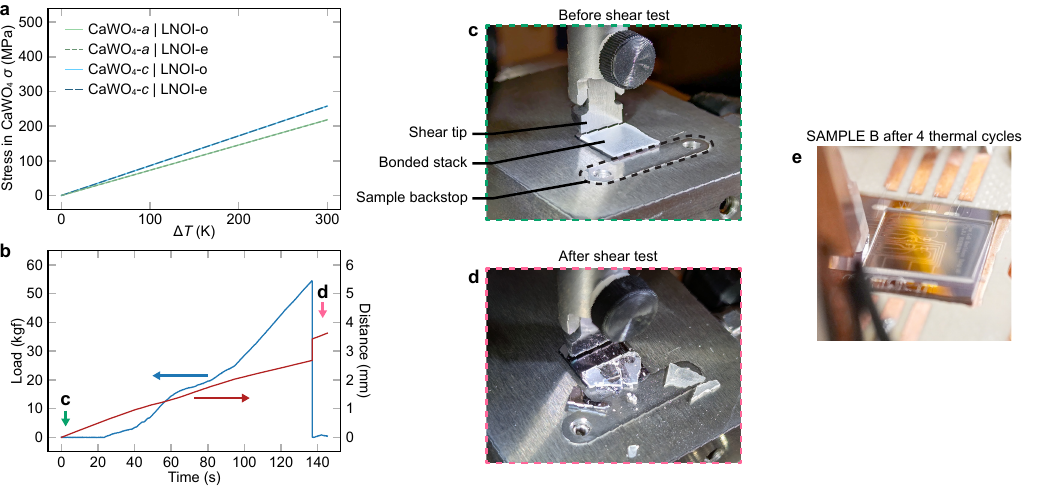}
    \caption{
        \textbf{Theoretical thermal stress and experimental shear testing of bonded \textCaWO{} on LNOI.}
        \textbf{a},~Calculated thermal stress, ${\sigma}$, in the \textCaWO{} layer as a function of temperature change ${\Delta T}$ for different crystal cut combinations. Note that the theoretical curves for X-cut and Z-cut LNOI overlap almost perfectly for both cuts of the \textCaWO{} orientations.
        \textbf{b},~Experimental load and displacement curves as a function of time during a destructive shear test of an undoped \textCaWO{} chip bonded to an unpatterned LNOI substrate. The green and pink arrows (labelled \textbf{c} and \textbf{d}) indicate the time points corresponding to the photographs in the subsequent panels.
        \textbf{c},
        \textbf{d},~Photographs of the experimental setup \textbf{c}, before and \textbf{d}, after the shear test. To provide a visual scale, the undoped \textCaWO{} measures 10${\times}$10$\unitSpacing{}$mm$^2$. The sample backstop ensures the LNOI substrate remains stationary under high lateral loads.
        \textbf{e},~Photograph of \textSampleB{} after 4 thermal cycles, showing that the \textErCaWO{} is still bonded to the underlying photonic integrated chip.
    }
    \label{fig:thermal-analysis}
\end{figure}

The stress in the \textCaWO{} layer is shown in \cref{fig:thermal-analysis}a.
We observed no significant difference of stress in \textCaWO{} along the extraordinary versus ordinary axes of \textLN{}, which indicates that the stress contribution is dominated by the much thicker \textQuartz{}+Si substrate.
Our analysis shows that by taking the ${c}$-cut \textCaWO{} (i.e., ${a}$- and ${b}$-axes are in-plane), we minimise the stress that is thermally-induced in it.

To experimentally gauge the strength of our bonded stack comprising an undoped 10${\times}$10${\unitSpacing{}}$mm${^2}$ $c$-cut 1-side epi-polished \textCaWO{} crystal bonded to an unpatterned LNOI chip, we performed a destructive shear test on it at room temperature using a \texttt{Nordson~DAGE~4000} bond tester, as shown in \cref{fig:thermal-analysis}b-d.
The bond tester uses a shear tip positioned above the LNOI chip but below the \textCaWO{} chip.
The shear tip moves at a set rate of 25${\unitSpacing{}}$µm/s to push only against the \textCaWO{}, while the whole bonded stack is held in place by the sample backstop (\cref{fig:thermal-analysis}c).
Our test reveals that the maximum load sustained is ${54.54\unitSpacing{}}$kgf, which converts to a maximum shear stress of ${7.9\unitSpacing{}}$MPa.

\section{Collective cooperativity estimation}\label{sec:lumerical-filling-factor}

\begin{figure}[!h]
    \centering
    \includegraphics{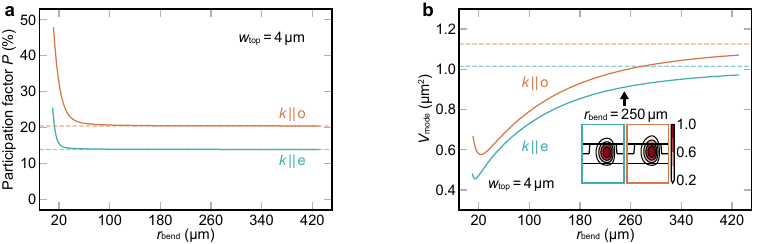}
    \caption{
        \textbf{Simulated \textErCaWO{} participation and mode volume per unit propagation length.}
        \textbf{a}, The simulated participation factor ${P}$ of the fundamental TE\textsubscript{0} mode as a function of waveguide bend radius ${r_\text{bend}}$ for top width ${w_\text{top}=4\unitSpacing{}}$µm. Mode propagation along the extraordinary and ordinary axes of the lithium niobate are simulated, denoted by ${{k}{\parallel}{\text{e}}}$ and ${{k}{\parallel}{\text{o}}}$, respectively.
        \textbf{b}, The corresponding simulated mode volume per unit propagation length ${V_\text{mode}}$.
        The dashed lines indicate corresponding values for straight waveguide. The insets show the normalised energy density mode profiles for ${r_\text{bend}=250\unitSpacing{}}$µm.
    }
    \label{fig:filling-factor}
\end{figure}

To estimate the collective cooperativity of our \textErCaWO{}-on-TFLN platform, we first evaluate the participation $P$ of electric field energy of the waveguide's optical TE\textsubscript{0} mode in the \textErCaWO{} crystal, defined as~\cite{yangPhotonicIntegrationEr2021}
\begin{equation}
    P=\frac
    {\int_{\text{\textCaWO{}}}
        \mathbf{E}^{*}(\vec{r})\cdot \overset{\text{\tiny$\leftrightarrow$}}{\epsilon}(\vec{r}) \cdot \mathbf{E}(\vec{r})
    \,\mathrm{d}V}
    {\int\mathbf{E}^{*}(\vec{r})\cdot \overset{\text{\tiny$\leftrightarrow$}}{\epsilon}(\vec{r}) \cdot \mathbf{E}(\vec{r})\,\mathrm{d}V},
\end{equation}
where ${\overset{\text{\tiny$\leftrightarrow$}}{\epsilon}}$ is the permittivity tensor.
\Cref{fig:filling-factor}a shows the simulated $P$ as a function of the bend radius ${r_\text{bend}}$ for 50\% etched 500$\unitSpacing{}$nm X-cut LNOI waveguide with top width ${w_\text{top}=4\unitSpacing{}}$µm and 60\textdegree{} sidewall angle, corresponding to the ring resonator geometry of the device characterised in the main text, i.e., \textSampleA{}.
Directly above the waveguide in the simulation geometry is a ${c}$-cut \textCaWO{}.
A 4${\unitSpacing{}}$µm top width ensures fundamental-mode TE\textsubscript{0} operation by suppressing higher-order modes in the \textErCaWO{}-on-TFLN ring resonator.
Mode propagation along the extraordinary and ordinary axes of the \textLN{} were simulated separately, denoted by ${k}{\parallel}{\text{e}}$ and ${k}{\parallel}{\text{o}}$, respectively.
For the simulation, the refractive index values ${n_{\text{\textCaWO{}},a}=1.8842}$, ${n_{\text{\textCaWO{}},c}=1.8984}$, ${n_{\text{\textLN{}},\text{o}}=2.2117}$, ${n_{\text{\textLN{}},\text{e}}=2.1381}$, and ${n_{\text{\textQuartz{}}}}=1.4442$ were used at simulation wavelength ${\lambda_\text{sim}=1532.63\unitSpacing{}}$nm.

To provide physical intuition for the spatial scale of the light-matter coupling, we additionally plot the mode volume per unit propagation length ${V_\text{mode}=\tfrac{\int\mathbf{E}^{*}(\vec{r})\cdot \overset{\text{\tiny$\leftrightarrow$}}{\epsilon}(\vec{r}) \cdot \mathbf{E}(\vec{r})\,\mathrm{d}A}{\max{\mkern-4mu(\mathbf{E}^{*}(\vec{r})\cdot \overset{\text{\tiny$\leftrightarrow$}}{\epsilon}(\vec{r}) \cdot \mathbf{E}(\vec{r}))}}}$, where the integral runs over the waveguide cross-section, in \cref{fig:filling-factor}b.
This shows that ${V_\text{mode}}$ remains under 1${\unitSpacing{}}$µm$^2$, confirming that the guided mode is tightly confined compared to free space approaches.
This tight confinement, combined with the strong participation in the \textCaWO{}, ensures efficient coupling between the ring resonator photons and the near-surface ensemble of \textEr{} ions, which directly enters the collective cooperativity through the collective coupling strength defined as \cite{yangPhotonicIntegrationEr2021}
\begin{equation}\label{eqn:big-g}
    G=\frac{\mu}{n_\text{mode}}\sqrt{\frac{P\rho\omega}{2\epsilon_0\hbar}},
\end{equation}
where $\mu$ is the transition dipole moment, $\rho$ the \textEr{} ion density, $\epsilon_0$ the vacuum permittivity, $\hbar$ the reduced Planck constant, and ${n_\text{mode}}$ the effective refractive index of the optical resonator mode.

To estimate the transition dipole moment ${\mu}$ for the ${{Z_1}{\leftrightarrow}{Y_1}}$ transition, we first calculate the oscillator strength given by \cite{Bartolo1968,Henderson2006}
\begin{equation}\label{eqn:oscillator-strength}
    \oscillatorStrength{},
\end{equation}
where ${\alpha(\nu)}$ is the absorption coefficient as a function of frequency ${\nu=\omega/2\pi}$, $e$ the electron charge, $m_e$ the electron mass, and $n$ is the refractive index of the material in which the \textEr{} ion ensemble resides.
In our case, ${n=n_{\text{\textCaWO{}},a}=1.8842}$.
The real cavity model has been found suitable for substitutional ions \cite{Dolgaleva2012}, and therefore we take ${\chi_L}=3n^2/(2n^2+1)$.
Given the 50${\unitSpacing{}}$ppm dopant concentration of the \textErCaWO{}, its ion density is calculated to be ${\rho=6.4\times 10^{23}\unitSpacing{}}$m${^{-3}}$.

The absorption coefficient here is given by ${\alpha(\nu)=-2.3\log{(|A(\nu)|^2/\max{|A(\nu)|^2})}/(PL)}$~\citep[\S{}${\unitSpacing{}}$8.2.2.4]{Liu2005}, where ${L=7850\unitSpacing{}}$µm is the optical length of the bus waveguide which is estimated based on the bonding area in \cref{fig:optical-microscope-image}c, and $A(\nu)$ is the complex transmission.
The participation factors of the bus waveguide are simulated to ${P_{{k}{\parallel}{\text{e}}}=14.8\%}$ and ${P_{{k}{\parallel}{\text{o}}}=23.5\%}$ for ${r_\text{bend}\rightarrow\infty}$, which is a good approximation since the bend radii of the bus waveguide are large enough.
By taking the weighted average of these participation factors based on the bus waveguide geometry, we approximate ${P={(L_{{k}{\parallel}{\text{o}}}/L})P_{{k}{\parallel}{\text{o}}}+(1-{L_{{k}{\parallel}{\text{o}}}/L})P_{{k}{\parallel}{\text{e}}}=15.6\%}$, with the length parallel to TFLN's ordinary axis taken to be $L_{{k}{\parallel}{\text{o}}}=680\unitSpacing{}$µm.
Taking $\alpha(\nu)$ at zero field, as shown in \cref{fig:magnetic-field-dependence}a, we obtained ${f=2.2\times10^{\smash{-7}}}$.
The transition dipole moment is related to the oscillator strength through ${\transitionDipoleMoment{}}$ \cite{Bartolo1968,Henderson2006,Zhong2018}. Substituting the obtained oscillator strength into the expression we get ${\mu=1.6\times10^{\smash{-32}}\unitSpacing{}}$C${\cdot}$m.

Having established the transition dipole moment value, we now proceed to estimating the collective cooperativity of \textEr{} ions coupled to the ring resonator.
Since the geometry of the resonator is a ring, the effective participation factor can be approximated as the average between ${k}{\parallel}{\text{e}}$ and ${k}{\parallel}{\text{o}}$ values such that ${{P}=(P_{{k}{\parallel}{\text{e}}} + P_{{k}{\parallel}{\text{o}}})/2}$.
The resonator characterised in the main text has a bend radius ${r_{\text{bend}}=250\unitSpacing{}}$µm, which corresponds to ${P=17.2\unitSpacing{}\%}$.
The corresponding effective indices of the TE\textsubscript{0} mode found from the simulation are ${n_{\text{mode},{k}{\parallel}{\text{e}}}=2.0311}$ and ${n_{\text{mode},{k}{\parallel}{\text{o}}}=1.9722}$. Taking its average, we obtain $n_{\text{mode}}=2.0017$.
Finally, by substituting above values into \cref{eqn:big-g}, we obtain ${G_\text{estimated}/2\pi=349\unitSpacing{}}$MHz, which leads to collective cooperativity ${C_\text{estimated}=7.4\pm0.6}$ for $\kappa_\text{tot}$ and $\Gammainh{}$ values at ${|\mathbf{B}|=0\unitSpacing{}}$T in \cref{tab:collective-cooperativity}.
Note that the eigenmode simulations in this section were done with a commercial software: \texttt{Ansys~Lumerical}.

To determine the experimental collective cooperativity, we approximate the inhomogeneous optical transition of the \textEr{} ensemble by a Lorentzian spectral distribution.
We compared this choice against a Gaussian profile and found that the Lorentzian profile consistently provided better agreement with the measured transmission data.
We also found no significant difference in the collective cooperativities using the Gaussian inhomogeneous broadening model.
The choice of Lorentzian profile is also consistent with previous rare-earth ion (REI) spectroscopy \cite{Boettger2006a}, REI resonator-ensemble modelling \cite{yangPhotonicIntegrationEr2021}, and is expected in rare-earth crystals with dilute defects \cite{Thiel2011,Orth1994}.
Using the Lorentzian model, the complex transmission of the ring resonator at angular frequency $\omega_\text{cav}$ in the presence of \textEr{} ions at frequency $\omega_\text{ions}$ is given by~\cite{Diniz2011}
\begin{equation}\label{eqn:lorentzian_inhomogeneous_transmission}
    A=1-\frac{\iu{}\kappa_\text{ext}}{(\omega-\omega_\text{cav})+\iu{}\kappa_\text{tot}/2-\tfrac{G^2}{\omega-\omega_\text{ions}+\iu{}\Gammainh{}/2}},
\end{equation}
where $\omega$ is the probe angular frequency, ${\kappa_\text{tot}=\kappa_\text{int}+\kappa_\text{ext}}$ the total linewidth of the resonator and $\Gammainh{}$ the inhomogeneous broadening linewidth of the \textEr{} ions.
By fitting the complex transmission to \cref{eqn:lorentzian_inhomogeneous_transmission}, we extract collective cooperativities through $\kappa_\text{tot}$, ${\Gammainh{}}$ and ${G}$ for various magnetic fields (\cref{tab:collective-cooperativity}).
To reliably extract the experimental collective cooperativity, we separately determine the intrinsic $\kappa_\text{int}$ and extrinsic linewidths $\kappa_\text{ext}$ of the ring resonator with detuning ${\gg}{\Gammainh{}}{,}{G}$.
This allows us to fix ${\kappa_{\text{tot}}}$ and fit to \cref{eqn:lorentzian_inhomogeneous_transmission} with only two free parameters: ${\Gammainh{}}$ and $G$.
Because the laser is swept repeatedly with an interval between sweeps shorter than the shelving-state lifetimes, the ensemble is probed in an optically pumped steady state whose depletion depends on magnetic field through both the shelving branching ratios and their lifetimes; the extracted ${G}$ is therefore a lower bound on its intrinsic value.
For more information about the inhomogeneous broadening of the \textEr{} ions coupled to the ring resonator versus bus waveguide, see \cref{sec:magnetic-field-dependence-gamma-inh}.
The power at the cryostat input is 0.03${\unitSpacing{}}$µW, and for each resonator-ensemble detuning we average over 10 wavelength sweeps.

\begin{table}[!b]
    \centering
    \extendedDataTableAvoidedCrossing{}%
    \caption{
        \textbf{Fitted and measured parameters of the resonator-coupled \textEr{} ion ensemble.}
        The table details parameters for the \textSampleA{} at various magnetic field strengths and includes comparative data from \textSampleB{} at zero field.
        The centre wavelength $\lambda_\text{ions}$ corresponds to ${\ketdownz}{\leftrightarrow}{\ketdowny}$ transition frequency of the resonator-coupled ions, which is additionally shifted relative to the bus waveguide-coupled ions due to the DC electric field applied for electro-optic tuning (up to 700${\unitSpacing{}}$V).
        The resonator linewidths $\kappa_\text{int}$ and $\kappa_\text{ext}$ were determined independently with the resonator detuned by ${\gg}{G}$ from the \textEr{} transition and held fixed during subsequent fits.
        The inhomogeneous broadening linewidth ${\Gammainh{}}$ and collective coupling strength $G$ were obtained as free parameters by fitting the complex transmission to the coupled-resonator model \cref{eqn:lorentzian_inhomogeneous_transmission}, which assumes a Lorentzian inhomogeneous profile. The collective cooperativity $C$ is derived from the fitted parameters with standard error propagation.
        All uncertainties represent one-standard-deviation confidence intervals from least-squares fits.
        Measurements for \textSampleA{} were performed at a nominal mixing chamber temperature ${\sim}$8${\unitSpacing{}}$mK.
    }
    \label{tab:collective-cooperativity}
\end{table}%
We now compare the experimentally determined collective cooperativity with our theoretical estimate. From the fit to our \textSampleA{} zero-field transmission data \cref{tab:collective-cooperativity}, we extract a collective cooperativity $C=6.66\pm0.42$.
This value is comparable with the predicted value of ${C_\text{estimated}=7.4\pm0.6}$, which was calculated based on the simulated mode properties and the oscillator strength derived from bus waveguide absorption.

To validate the reproducibility of our platform, we also characterised a pathfinder device, \textSampleB{}, which featured a ring with a smaller bend radius (${r_\text{bend}=140\unitSpacing{}}$µm).
The following data were obtained during its initial thermal cycles.
By fitting the complex transmission to \cref{eqn:lorentzian_inhomogeneous_transmission}, we determined a collective cooperativity of ${C=9.06\pm0.96}$. The parameters extracted from this fit are detailed in \cref{tab:collective-cooperativity}. The low inhomogeneous broadening observed for this device $\Gammainh{}$ is consistent with the restricted sampling with reduced strain variation (see \cref{sec:magnetic-field-dependence-gamma-inh}).

\section{Magnetic field dependence of inhomogeneous broadening}\label{sec:magnetic-field-dependence-gamma-inh}
\begin{figure}[!h]
    \centering
    \includegraphics{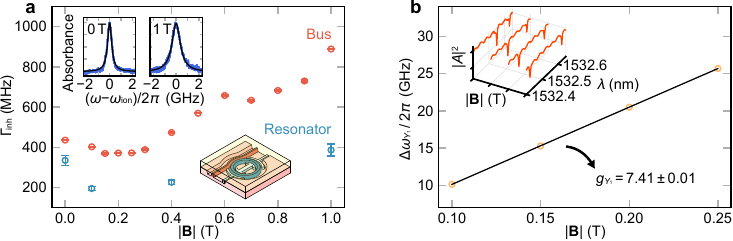}
    \caption{
        \textbf{Magnetic field dependence of inhomogeneous broadening linewidth and Zeeman splitting of ${\boldsymbol{Y_1}}$ manifold.}
        \textbf{a},~Inhomogeneous broadening linewidth ${\Gammainh{}}$ as a function of magnetic field strength ${\absB{}}$ for \textEr{} ions coupled to the bus waveguide (orange) and to the ring resonator (blue).
        The inset plots show absorbance spectra at ${\absB{}=0\unitSpacing{}}$T and 1${\unitSpacing{}}$T for bus waveguide coupled ions.
        \textbf{b},~Zeeman splitting ${\Delta\omega_{Y_1}/2\pi}$ of the ${Y_1}$ excited-state Kramers doublet as a function of ${\absB{}}$, with a linear fit yielding an effective $g$-factor ${g_{Y_1}=7.41\pm0.01}$.
        Inset: transmission spectra at corresponding magnetic fields showing the progressive Zeeman splitting.
        The error bars represent one-standard-deviation confidence intervals from least-squares fits.
    }
    \label{fig:magnetic-field-dependence}
\end{figure}

The absorbance of the ${\ket{\downarrow}_{Z_1}}{\leftrightarrow}{\ket{\downarrow}_{Y_1}}$ transition of the \textEr{} ions coupled to the bus waveguide was acquired using our heterodyne detection setup (see \cref{sec:setup-heterodyne}), and fitted to a Lorentzian profile to extract the inhomogeneous broadening linewidth ${\Gammainh{}}$ for various magnetic field strengths ${|\mathbf{B}|}$, as shown by the orange markers in \cref{fig:magnetic-field-dependence}a.
We overlay in blue markers the inhomogeneous broadening linewidths of \textEr{} ions coupled only to the ring resonator, found from fitting our measured complex transmission to \cref{eqn:lorentzian_inhomogeneous_transmission}.
At higher magnetic field strengths, the inhomogeneous broadening linewidth increases, consistent with a spatial variation of the Zeeman shift across the crystal, which can arise from inhomogeneity of the field of the vector magnet (see \cref{sec:dilution-fridge-setup}) and from strain-induced variations of the effective $g$-factor.
The overlaid resonator data reveals that at ${|\mathbf{B}|=0\unitSpacing{}}$T, there is already a decrease in inhomogeneous broadening by coupling to \textEr{} ions in a relatively small region compared to the ions coupled to the bus waveguide.
At non-zero magnetic fields, the suppression effect is magnified further, confirming a spatially varying local environment experienced by the \textEr{} ion ensemble.
These differences in inhomogeneous broadening reflect static variations of strain and of the Zeeman shift across the crystal.
In contrast, the homogeneous linewidth and spectral diffusion (\cref{sec:spectral-diffusion}) are set by the fluctuating environment of the ions, namely the \textEr{} electron-spin bath, the \textTungstenNuclear{} nuclear spins and phonons, which we expect to be common to the ions coupled to the bus waveguide and to the ring resonator, since both lie in the same bonded \textErCaWO{} crystal.

To characterise the $g$-factor corresponding to the electronic transition ${\ket{\downarrow}_{Y_1}}{\leftrightarrow}{{\ket{\uparrow}_{Y_1}}}$ which we denote as $g_{Y_1}$, we perform a linear fit to the frequency difference $\Delta\omega_{Y_1}$ between the
${\ket{\downarrow}_{Z_1}}{\leftrightarrow}{{\ket{\downarrow}_{Y_1}}}$
and
${\ket{\downarrow}_{Z_1}}{\leftrightarrow}{{\ket{\uparrow}_{Y_1}}}$ transitions as shown in \cref{fig:magnetic-field-dependence}b.
From this we extract ${g_{Y_1}=7.41\pm0.01}$, which agrees well with Ref.~\cite{Becker2026}.

\section{Spectral diffusion and coherence}\label{sec:spectral-diffusion}
To investigate spectral diffusion of \textEr{} ions, we employ the 3-pulse photon echo (3PPE) or \textit{stimulated photon echo} experiment, where the delay between the first two pulses is ${\tau_{12}}$, and the delay between the second and third pulses is ${\tau_{23}}$.
The delays are measured from the centre of each pulse.
For this measurement, all optical pulses had a duration of 150${\unitSpacing{}}$ns and a power of 0.3${\unitSpacing{}}$mW at the cryostat input, and the resulting echoes were averaged at the intensity level over 50~--~170~acquisitions with repetition time 100${\unitSpacing{}}$ms.
The pulse duration and power were set empirically to maximise the echo signal.
The stimulated echo intensity is given by \cite{Boettger2006}
\begin{equation}\label{eqn:3ppe_intensity}
    I_\text{3PPE}=I_0\exp{(-4\pi\tau_{12}\theoGammaeff{})},
\end{equation}
where the $\tau_{23}$ dependence of the echo amplitude is absorbed into the constant $I_0$, and $\theoGammaeff{}$ is the effective linewidth. It is modelled as a function of ${\tau_{23}}$ through the following relation \cite{Boettger2006}
\begin{equation}\label{eqn:theoretical_effective_linewidth}
    \theoGammaeffExpression{},
\end{equation}
where $\GammaZ{}$ is the homogeneous linewidth as if there were no spectral diffusion, $R$ the rate of spectral diffusion, and $\GammaSD{}$ the linewidth of spectral diffusion.
To extract the effective linewidth ${\theoGammaeff{}}$ as a function of ${\tau_{23}}$ as shown in the main text, we fitted the echo decay against $\tau_{12}$ with \cref{eqn:3ppe_intensity} at each ${\tau_{23}}$.

\begin{figure}[h!]
    \centering
    \includegraphics{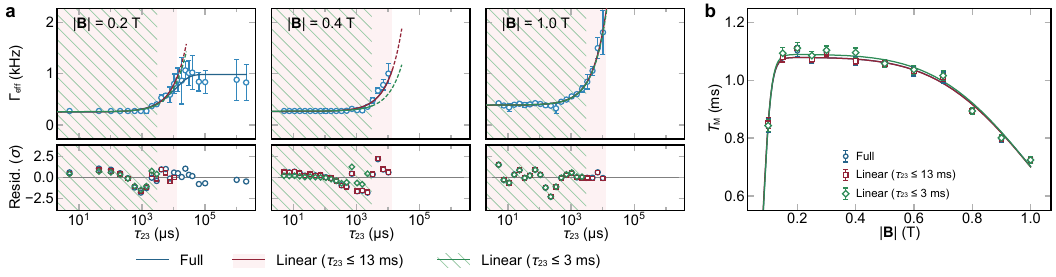}
    \caption{
        \textbf{Effect of the linearisation of the spectral diffusion model on the empirical coherence time.}
        \textbf{a},~Effective linewidth ${\theoGammaeff{}}$ against waiting time $\tau_{23}$ (top panels) at
        ${\absB{}=0.2}$, 0.4 and 1.0${\unitSpacing{}}$T, with the full expression
        ${\theoGammaeff{}=\GammaZ{}+\tfrac{1}{2}\GammaSD{}(1-\exp{(-R\tau_{23})})}$ (blue) and the linearised model
        ${\theoGammaeff{}\approxeq\GammaZ{}+\tfrac{1}{2}\GammaSD{}R\tau_{23}}$ fitted to ${\tau_{23}}{\leq}{13}{\unitSpacing{}}$ms
        (red; shaded region; the fit used in the main text's Fig.~2e) and to ${\tau_{23}}{\leq}{3}{\unitSpacing{}}$ms (green; hatched
        region). The linearised fits are solid within their fitting window and dashed beyond it. The bottom panels show the
        normalised residuals ${(\theoGammaeff{}-\text{model})/\sigma}$ of the full (circles) and linearised
        (${\tau_{23}}{\leq}{13}{\unitSpacing{}}$ms, squares; ${\tau_{23}}{\leq}{3}{\unitSpacing{}}$ms, diamonds) fits.
        \textbf{b},~Empirical coherence time ${T_\text{M}}$ against magnetic field obtained from the three fits in \textbf{a},
        each with a fit of the field-dependence model of \cref{eqn:gamma_eff_vs_magnetic_field} (lines).
        The fitted parameters at all magnetic fields are given in \cref{tab:spectral-diffusion}, and those of the field-dependence fits in \cref{tab:field-dependence}.
    }
    \label{fig:linear-vs-full-3ppe}
\end{figure}

\begin{figure}[h!]
    \centering
    \includegraphics{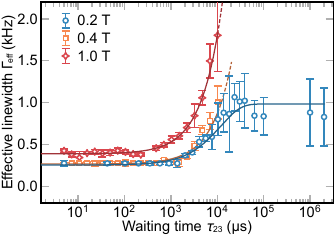}
    \caption{
        \textbf{Effective linewidth against waiting time at different magnetic fields.}
        Effective linewidth ${\theoGammaeff{}}$ against waiting time $\tau_{23}$ at ${\absB{}=0.2}$, 0.4 and 1.0${\unitSpacing{}}$T, with fits of the full expression
        ${\theoGammaeff{}=\GammaZ{}+\tfrac{1}{2}\GammaSD{}(1-\exp{(-R\tau_{23})})}$ (lines).
        At 0.2${\unitSpacing{}}$T, the saturation of the spectral diffusion is resolved (${R}{=}{86\pm18}{\unitSpacing{}}$Hz).
        At 0.4${\unitSpacing{}}$T and 1.0${\unitSpacing{}}$T, it is not resolved within the measured range of $\tau_{23}$, and only the product ${\GammaSD{}R}$ is determined (\cref{tab:spectral-diffusion}); the dashed lines extrapolate these fits beyond the last data point.
        Error bars represent one-standard-deviation uncertainties from the fits of the echo decays with \cref{eqn:3ppe_intensity}.
    }
    \label{fig:3ppe-for-other-fields}
\end{figure}

To study the effective coherence time of the \textEr{} ions in our system, it is instructive to consider ${\tau_{23}=0}$ where the 3PPE measurement becomes equivalent to the two-pulse photon echo (2PPE) or \textit{Hahn echo} measurement. The empirical effective coherence time ${T_{\text{M}}}$ at ${\tau_{23}=0}$ extracted from the spectral diffusion parameters can be written as \cite{Boettger2006,Thiel2011}
\begin{equation}\label{eqn:empirical_effective_coherence_time}
    T_\text{M}=\frac{2\GammaZ{}}{\GammaSD{}R}
    \Bigg(\mkern-5mu
        -1+\sqrt{1+\frac{\GammaSD{}R}{\pi\GammaZ^{\mkern-5mu\smash{2}}}}
    \Bigg).
\end{equation}
The corresponding empirical effective homogeneous linewidth is then
\begin{equation}\label{eqn:empirical_effective_linewidth}
    \empiGammaeff{}=\frac{1}{\pi T_\text{M}}.
\end{equation}
To obtain each data point in the empirical coherence time versus magnetic field strength plot in the main text, we substitute into \cref{eqn:empirical_effective_coherence_time} the ${\GammaZ{}}$ and ${\GammaSD{}R}$ values obtained from fitting the linearised model ${\theoGammaeff{}\approxeq\GammaZ{}+\tfrac{1}{2}{\GammaSD{}R}\tau_{23}}$, which follows from ${1-\exp{(-R\tau_{23})}\approx R\tau_{23}}$ for ${R\tau_{23}\ll1}$, to ${\tau_{23}}{\leq}{2T_1}$ (\cref{tab:spectral-diffusion}).
At ${\absB{}=0.2\unitSpacing{}}$T, where the saturation of the spectral diffusion is resolved (${R}{=}{86\pm18}{\unitSpacing{}}$Hz), $R\tau_{23}$ reaches ${\approx}0.9$ at the last fitted point (${\tau_{23}=10.3\unitSpacing{}}$ms), so the condition ${R\tau_{23}\ll1}$ is not strictly satisfied.
We therefore verified that $T_\text{M}$ is insensitive to the approximation (\cref{fig:linear-vs-full-3ppe,tab:spectral-diffusion}).
Fitting the full expression, \cref{eqn:theoretical_effective_linewidth}, changes $T_\text{M}$ by at most 0.9\,\% at any field.
Restricting the linearised fit to ${\tau_{23}}{\leq}{3}{\unitSpacing{}}$ms, the largest window for which ${R\tau_{23}}{\leq}{0.25}$ at every field where $R$ is constrained by the data, changes $T_\text{M}$ by at most 2.7\,\%, within 1.5 standard deviations.
The fitted slope ${\GammaSD{}R}$ of the linearised model should be regarded as an effective initial rate: at ${\absB{}=0.2\unitSpacing{}}$T it is ${\approx}17\,\%$ lower than ${\GammaSD{}R}$ from the full expression.
The values of ${R}$ and ${\GammaSD{}}$ quoted separately in the main text are obtained from the full expression, fitted over the entire measured range of ${\tau_{23}}$ at ${\absB{}=0.2\unitSpacing{}}$T.
In \cref{fig:3ppe-for-other-fields}, we compare these full-expression fits at ${\absB{}=0.2}$, 0.4 and 1.0${\unitSpacing{}}$T; at the higher fields, the saturation of the effective linewidth is not resolved within the measured range of ${\tau_{23}}$.
The empirical effective coherence times are then converted into homogeneous linewidths through \cref{eqn:empirical_effective_linewidth}.

As a magnetic field is introduced, the effective homogeneous linewidth of the \textEr{} ions drops due to the suppression of electronic spin flip-flop interactions, which contributes to the homogeneous linewidth according to the Boltzmann distribution~\cite{Boettger2006}
\begin{equation}\label{eqn:gamma_flip_flop}
    \Delta\Gammaff{}=\alpha_\text{ff}\exp{\Big(-\frac{\geff{}\bohrM{}B}{2\kB{}T}\Big)}\sech{\Big(\frac{\geff{}\bohrM{}B}{2\kB{}T}\Big)},
\end{equation}
where ${\alpha_\text{ff}}$ is a coefficient describing the strength of the spin flip-flop processes,
${\geff{}}$ is the effective $g$-factor of the electronic spin state,
${\bohrM{}}$ the Bohr magneton, ${\kB{}}$ the Boltzmann constant, and $T$ the effective temperature.
In addition to the suppression of flip-flop interactions, we observed another mechanism causing an increase in homogeneous linewidth at higher magnetic field strengths.
We attribute this increase in homogeneous linewidth to the one-phonon direct process, which involves the absorption/emission of a single phonon from the resonant electronic spin transition. This linewidth contribution due to the direct process has the form~\cite{Abragam1970}
\begin{equation}\label{eqn:gamma_direct_process}
    \Delta\GammaD{}=\alpha_\text{D}{\geff{}^{\mkern-6mu3}}B^5\coth{\Big(\frac{\geff{}\bohrM{}B}{2\kB{}T}\Big)},
\end{equation}
where ${\alpha_{\text{D}}}$ describes the strength of the one-phonon direct process.
The complete contribution to the effective linewidth as a function of magnetic field ${\absB{}}$ is then
\begin{equation}\label{eqn:gamma_eff_vs_magnetic_field}
    \empiGammaeff{}=\GammaeffZ{}+\Delta\Gammaff{}+\Delta\GammaD{},
\end{equation}
where $\GammaeffZ{}$ is the effective linewidth from magnetic field-independent contributions.
In our fitting, we let ${\alpha_\text{ff}=\empiGammaeff{}|_{\absB{}=0}-\GammaeffZ{}}$, where ${\empiGammaeff{}|_{\absB{}=0}}$ is the empirical linewidth at zero field, measured separately using 2PPE to be ${T_{\text{M}}=4.49\pm0.15\unitSpacing{}}$µs, fitted with a single exponential.
Using this model (\cref{eqn:gamma_eff_vs_magnetic_field}), we find good agreement, with
${\GammaeffZ{}=295\pm1\unitSpacing{}}$Hz,
${\alpha_{\text{ff}}=71\pm2\unitSpacing{}}$kHz,
${\alpha_{\text{D}}{\geff{}^{\mkern-2mu3}}=160\pm5\unitSpacing{}}$Hz${\unitSpacing{}}$T${}^{\smash{-5}}$,
and
${g_\text{eff}\bohrM{}/\kB{}T=75\pm1\unitSpacing{}}$T${^{-1}}$.
By taking $g$-factor of the ${Z_1}$ manifold to be ${g_{Z_1}=8.38}$~\cite{LeDantec2021}, we find that the effective spin temperature of our system is ${T=75\pm1\unitSpacing{}}$mK.
Repeating the fit with $T_\text{M}$ from the full expression or from the linearised model restricted to ${\tau_{23}}{\leq}{3}{\unitSpacing{}}$ms gives ${T=75.6\pm1.3\unitSpacing{}}$mK and ${75.9\pm1.1\unitSpacing{}}$mK, respectively, consistent with the value above (\cref{fig:linear-vs-full-3ppe}b and \cref{tab:field-dependence}).
\begin{table}[!h]
    \centering
    \extendedDataTableSpectralDiffusion{}%
    \caption{
        \textbf{Fitted and measured spectral diffusion parameters of the \textEr{} ensemble at various magnetic field strengths.}
        The centre wavelength $\lambda_\text{ions}$ corresponds to ${\ketdownz}{\leftrightarrow}{\ketdowny}$ transition frequency of the bus waveguide-coupled ions.
        The spectral diffusion-independent linewidth ${\GammaZ{}}$ and spectral diffusion product ${\GammaSD{}R}$ were obtained as free parameters by fitting \cref{eqn:theoretical_effective_linewidth} in three ways:
        with the full expression ${\theoGammaeff{}=\GammaZ{}+\tfrac{1}{2}\GammaSD{}(1-\exp{(-R\tau_{23})})}$ (Full), and with the linearised model ${\theoGammaeff{}\approxeq\GammaZ{}+\tfrac{1}{2}\GammaSD{}R\tau_{23}}$, valid for ${R\tau_{23}}{\ll}{1}$, restricted to ${\tau_{23}}{\leq}{13}{\unitSpacing{}}$ms (equivalent to ${\tau_{23}}{\leq}{2T_1}$; values used in the main text) or to ${\tau_{23}}{\leq}{3}{\unitSpacing{}}$ms.
        The fitting windows and the residuals of the three fits are shown in \cref{fig:linear-vs-full-3ppe}.
        The saturation rate $R$ is from the full expression; where the data do not resolve the saturation, a 95\% upper limit from a $\chi^{\smash{2}}$ profile is given.
        The empirical coherence time $T_\text{M}$ follows from ${\GammaZ{}}$ and ${\GammaSD{}R}$ via \cref{eqn:empirical_effective_coherence_time}, and ${\tilde{\Gamma}_{\mkern-2.5mu\text{eff}}}{=}{1/(\pi T_\text{M})}$.
        All uncertainties represent one-standard-deviation confidence intervals from least-squares fits, propagated to $T_\text{M}$ and ${\tilde{\Gamma}_{\mkern-2.5mu\text{eff}}}$.
        Measurements were performed at a nominal mixing chamber temperature of ${\sim}$8${\unitSpacing{}}$mK.
    }
    \label{tab:spectral-diffusion}
\end{table}%
\begin{table}[!h]
    \centering
    \begin{tabularx}{0.99\linewidth}{l@{\hspace{2em}}X@{\hspace{0.5em}}X@{\hspace{0.5em}}X} \toprule[1.3pt]
        & \textbf{Full}   & \textbf{Linear,} $\boldsymbol{{\tau_{23}\leq13}\unitSpacing{}}$\textbf{ms} & \textbf{Linear,} $\boldsymbol{{\tau_{23}\leq3}\unitSpacing{}}$\textbf{ms} \\ \midrule \\ [-2.0ex]
        ${\GammaeffZ{}}$ (Hz)                                                                 & ${294.7\pm1.6}$ & ${294.7\pm1.4}$                                                            & ${292.0\pm1.6}$                                                           \\ \midrule
        ${\alpha_\text{ff}}$ (kHz)                                                            & ${70.6\pm2.3}$  & ${70.6\pm2.3}$                                                             & ${70.6\pm2.3}$                                                            \\ \midrule
        ${\alpha_{\text{D}}{\geff{}^{\mkern-2mu3}}}$ (Hz${\unitSpacing{}}$T${}^{\smash{-5}}$) & ${159\pm6}$     & ${160\pm5}$                                                                & ${162\pm5}$                                                               \\ \midrule
        ${\geff{}\bohrM{}/\kB{}T}$ (T${^{-1}}$)                                               & ${74.5\pm1.3}$  & ${74.9\pm1.1}$                                                             & ${74.2\pm1.1}$                                                            \\ \midrule
        ${T}$ (mK)                                                                            & ${75.6\pm1.3}$  & ${75.1\pm1.1}$                                                             & ${75.9\pm1.1}$                                                            \\ \bottomrule[1.3pt]
    \end{tabularx}
    \caption{
        \textbf{Magnetic field dependence of the empirical effective linewidth.}
        Parameters of \cref{eqn:gamma_eff_vs_magnetic_field} fitted to ${\empiGammaeff{}}$ obtained from each of the three sets of $T_\text{M}$ in \cref{tab:spectral-diffusion} (shown in \cref{fig:linear-vs-full-3ppe}b).
        The values in the main text are from the linearised model fitted to ${\tau_{23}}{\leq}{13}{\unitSpacing{}}$ms.
        The temperature $T$ assumes ${\geff{}=g_{Z_1}=8.38}$~\cite{LeDantec2021}.
        All uncertainties represent one-standard-deviation confidence intervals from least-squares fits; that of ${\alpha_\text{ff}}$ also includes the uncertainty of ${\empiGammaeff{}|_{\absB{}=0}}$, which was separately measured with 2PPE.
    }
    \label{tab:field-dependence}
\end{table}%

\section{Cooperativity and coherence at common magnetic fields}\label{sec:common-field}
\Cref{tab:common-field} collects the collective cooperativity and the empirical coherence time of the \textSampleA{} at the magnetic fields where both were measured.
The cooperativity is obtained from the ions coupled to the ring resonator (\cref{sec:lumerical-filling-factor}), and the coherence time from the ions coupled to the bus waveguide (\cref{sec:spectral-diffusion}); both ensembles lie in the same \textErCaWO{} crystal and were measured in the same cooldown.
At 0.4${\unitSpacing{}}$T, the device provides ${C=4.93\pm0.20}$ together with ${T_\text{M}=1068\pm13\unitSpacing{}}$µs.
The largest cooperativity, ${C=6.66\pm0.42}$, is reached at zero field, where electronic spin flip-flops limit the coherence time to ${T_\text{M}=4.49\pm0.15\unitSpacing{}}$µs (\cref{sec:spectral-diffusion}).
\begin{table}[!h]
    \centering
    \begin{tabularx}{0.99\linewidth}{l@{\hspace{2em}}X@{\hspace{0.5em}}X@{\hspace{0.5em}}X@{\hspace{0.5em}}X} \toprule[1.3pt]
        ${\absB{}}$ (T) & ${C}$                          & ${G/2\pi}$ (MHz)                & ${\Gamma_{\mkern-2.5mu\text{inh}}/2\pi}$ (MHz) & ${T_\text{M}}$ (µs)            \\ \midrule \\ [-2.0ex]
        0.0             & ${6.66\pm0.42}$~\ringMarker{} & ${330.57\pm5.67}$~\ringMarker{} & ${331.79\pm20.94}$~\ringMarker{}               & ${4.49\pm0.15}$\footnoteZeroFieldTM{}~\busMarker{}  \\ \midrule
        0.1             & ${4.95\pm0.25}$~\ringMarker{} & ${211.00\pm2.60}$~\ringMarker{} & ${195.00\pm10.22}$~\ringMarker{}               & ${852\pm19}$~\busMarker{}     \\ \midrule
        0.4             & ${4.93\pm0.20}$~\ringMarker{} & ${231.34\pm2.56}$~\ringMarker{} & ${226.22\pm10.75}$~\ringMarker{}               & ${1068\pm13}$~\busMarker{}    \\ \midrule
        1.0             & ${5.77\pm0.77}$~\ringMarker{} & ${353.49\pm25.87}$~\ringMarker{} & ${386.58\pm29.49}$~\ringMarker{}              & ${724\pm10}$~\busMarker{}     \\ \bottomrule[1.3pt]
    \end{tabularx}
    \caption{
        \textbf{Collective cooperativity and empirical coherence time of the \textSampleA{} at common magnetic fields.}
        Entries marked \ringMarker{} are from the ions coupled to the ring resonator (\cref{tab:collective-cooperativity}), and entries marked \busMarker{} from the ions coupled to the bus waveguide.
        ${T_\text{M}}$ is from the linearised spectral diffusion fit used in the main text (\cref{tab:spectral-diffusion}).
        All uncertainties represent one-standard-deviation confidence intervals from least-squares fits.
        Measurements were performed in the same cooldown at a nominal mixing chamber temperature of ${\sim}$8${\unitSpacing{}}$mK.
    }
    \label{tab:common-field}
\end{table}%

\section{Optical lifetime ${\boldsymbol{T_1}}$}\label{sec:purcell}
\begin{figure}[!h]
    \centering
    \includegraphics{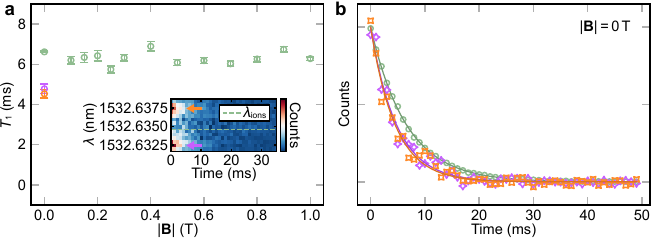}
    \caption{
        \textbf{Relaxation time ${\boldsymbol{T_1}}$ of \textEr{} ions.}
        \textbf{a}, Optical lifetime ${T_1}$ as a function of applied magnetic field strength ${\absB{}}$ for bare \textEr{} ions not coupled to the ring resonator (green) and for ions coupled to the ring resonator at the lower (orange) and upper (purple) polariton branches.
        The bare ions lifetime shows no systematic dependence on ${\absB{}}$.
        Inset: two-dimensional fluorescence map showing the polariton mode structure near the ${\ketdownz{}}{\leftrightarrow}{\ketdowny{}}$ transition.
        \textbf{b}, Fluorescence decay curves at ${\absB{}=0\unitSpacing{}}$T for the same lower and upper polaritons as in \textbf{a}, together with single-exponential fits (solid lines).
        The error bars represent one-standard-deviation confidence intervals from least-squares fits.
    }
    \label{fig:t1-purcell-effect}
\end{figure}

The bare relaxation time $T_1$ of the \textEr{} ions not coupled to the ring resonator shows no systematic dependence on the applied magnetic field strength over the range investigated ${|\mathbf{B}|\in[0\unitSpacing{}\text{T},1\unitSpacing{}\text{T}]}$,
as shown by the green markers in \cref{fig:t1-purcell-effect}a.
We calculate the average bare relaxation time ${T_1=6.47\pm0.03\unitSpacing{}}$ms across all measured magnetic field strengths.
At the two fields used for the inversion-recovery measurements (\cref{sec:inversion-echo}), the individually fitted values are ${T_1=6.34\pm0.25\unitSpacing{}}$ms at 0.15${\unitSpacing{}}$T and ${T_1=6.42\pm0.25\unitSpacing{}}$ms at 0.2${\unitSpacing{}}$T.
When the ring resonator is brought to the ${Z_1}{\leftrightarrow}{Y_1}$ transition frequency at zero field, the coupled system forms polaritons (hybridised light-matter states) as shown in the inset of \cref{fig:t1-purcell-effect}a at wavelengths ${1532.6324\unitSpacing{}}$nm and ${1532.6374\unitSpacing{}}$nm.
These polaritons are each detuned from \textEr{} centre wavelength by ${\sim}{G}$.
The fluorescence decay data corresponding to the lower and upper polaritons are shown in orange and purple colours, respectively, in \cref{fig:t1-purcell-effect}b.
Single-exponential fits yield ${T_1=4.53\pm0.21\unitSpacing{}}$ms for the lower polariton and ${T_1=4.78\pm0.24\unitSpacing{}}$ms for the upper polariton.

\section{Inversion-recovery echo measurement}\label{sec:inversion-echo}
An initial pulse of length ${600\unitSpacing{}}$ns excites the population; after a variable delay $\tau_\text{inv}$, a Hahn echo sequence with delay ${\tau_\text{H}=42\unitSpacing{}}$µs and pulse length ${150\unitSpacing{}}$ns reads out the population remaining at the initial excitation frequency.
We average over 50~--~100 shots for each data point.
The fractional stretching factor of 0.5 used in the stretched exponential for ${T_\text{W}}$ can be seen as a distribution of characteristic relaxation rates~\cite{BerberanSantos2005,Wang2025}.
\begin{table}[!h]
    \centering
    \extendedDataTableTriExp{}%
    \caption{
        \textbf{Inversion echo fitted parameters and lifetimes.} The parameters were obtained by fitting to ${\triExp{}}$. The optical ${T_1}$ was fixed to values obtained from an independent fluorescence measurement. Note that $p_\text{W}$ does not reflect the true branching factor because no clear plateau of the intensity of the recovered echo at measured time delays was observed.
    }
    \label{tab:tri-exp}
\end{table}%

\section{Optical phase storage pulse parameters}\label{sec:phase-retrieval-and-storage}
The power of the optical pulses used here is 24${\unitSpacing{}}$µW at the cryostat input, and the resulting echoes were averaged at the intensity level over 600 acquisitions with repetition time 200${\unitSpacing{}}$ms.
All pulses have a duration of 150${\unitSpacing{}}$ns and the same power, so that the two write pulses and the two read pulses have equal pulse areas, as required for full interference visibility.
The corresponding excitation bandwidth of ${\sim}6\unitSpacing{}$MHz is much narrower than the inhomogeneous linewidth and much wider than the grating period ${1/\tau_{12}\approx0.75\unitSpacing{}}$MHz, where ${\tau_{12}=1.33\unitSpacing{}}$µs.
The delays $\tau_{12}$ and ${\tau_\text{d}=6.15\unitSpacing{}}$µs are much shorter than the ${\sim}160\unitSpacing{}$µs coherence time of the laser, and $\tau_\text{d}$ exceeds $\tau_{12}$ so that no echo coincides with a read pulse.

\section{Resonator-limited quantum memory efficiency}\label{sec:memory-efficiency}
To quantify how the resonator loss limits a future cavity-assisted echo memory, we consider the revival of silenced echo (ROSE) protocol~\cite{Damon2011} in a resonator.
Adapting the efficiency of Ref.~\cite{Hanamura2026} to a resonator with non-zero intrinsic loss gives
\begin{equation}\label{eqn:rose-efficiency}
    \eta_\text{ROSE}=\Big(\frac{\kappa_\text{ext}}{\kappa_\text{tot}}\,\frac{4C}{(1+C)^2}\Big)^{2},
\end{equation}
where ${C=4G^2/(\kappa_\text{tot}\Gammainh{})}$ and ${\kappa_\text{tot}=\kappa_\text{ext}+\kappa_\text{int}}$.
The collective coupling strength $G$ (\cref{eqn:big-g}) does not depend on the resonator loss, which therefore enters only through $\kappa_\text{tot}$.
For a given $\kappa_\text{int}$, \cref{eqn:rose-efficiency} is maximised at ${\kappa_\text{tot}=4G^2/\Gammainh{}+2\kappa_\text{int}}$, which gives
\begin{equation}\label{eqn:rose-efficiency-max}
    \eta_\text{ROSE}^\text{max}=\Big(\frac{C_\text{int}}{1+C_\text{int}}\Big)^{2},\qquad C_\text{int}=\frac{4G^2}{\kappa_\text{int}\Gammainh{}}.
\end{equation}
For the zero-field parameters of the \textSampleA{} (\cref{tab:collective-cooperativity}), ${C_\text{int}=8.3}$ and ${\eta_\text{ROSE}^\text{max}\approx80\%}$, reached at ${\kappa_\text{ext}/2\pi\approx1.5\unitSpacing{}}$GHz.
With the intrinsic quality factor before bonding, ${Q_\text{int}\approx5\times10^{6}}$ (${\kappa_\text{int}/2\pi\approx39\unitSpacing{}}$MHz), these become ${C_\text{int}\approx34}$ and ${\eta_\text{ROSE}^\text{max}\approx94\%}$.
These values account only for the resonator coupling and loss; dephasing during storage and imperfect rephasing reduce the efficiency further.

\clearpage{}
\section{Erbium coherence and coupling in literature}
\begin{table}[!h]
    \centering
    \extendedDataTableLiteratureMap{}%
    \caption{
        \textbf{Optical coherence and cooperativity across erbium photonic platforms.}
    }
    \label{tab:literature-map}
\end{table}%

\stoptoc

\def\bibsection{
\section*{SUPPLEMENTARY \refname}}